\documentclass[reqno,centertags]{amsart}

\usepackage {graphicx}
\usepackage {psfrag}
\usepackage{pstricks,multido}
\usepackage{amsmath,amsthm,amscd,amssymb}
\usepackage{latexsym}
\newcommand{\C}{\mathbb C}

\newcommand{\Z}{\mathbb Z}
\newcommand{\N}{\mathbb N}
\newcommand{\R}{\mathbb R}

\newcommand{\cal}{\mathcal}

\newtheorem{Thm}{Theorem}

\newtheorem{Lem}[Thm]{Lemma}
\newtheorem{Cor}[Thm]{Corollary}

\theoremstyle{definition}

\newtheorem{Def}[Thm]{Definition}
\newtheorem{Rem}[Thm]{Remark}

\begin{document}

\title[Absolutely Continuous Spectrum]{Absolutely Continuous Spectrum  of a
Polyharmonic Operator with a Limit Periodic Potential  in Dimension Two.}
\author[Y.~Karpeshina and Y.-R.~Lee]{Yulia Karpeshina and Young-Ran~Lee}

%Dr. Karpeshina's address
\address{Department of Mathematics, Campbell Hall, University of Alabama at Birmingham,
1300 University Boulevard, Birmingham, AL 35294.}
\email{karpeshi@math.uab.edu}%

%Young-Ran's address
\address{Department of Mathematical Sciences, Korea Advanced Institute of Science and Technology
333 Gwahangno, Yuseong-gu, Daejeon 305-701, Republic of Korea.}%
\email{youngranlee@kaist.ac.kr}

\thanks{Research partially supported by USNSF Grant DMS-0201383}
%\thanks{\copyright 2007 by the authors. Faithful reproduction of this article,
%       in its entirety, by any means is permitted for non-commercial purposes}
%\keywords{Non-linear Schr\"odinger equation, decay of eigenfunctions}
%\subjclass[2000]{35B20, 35B40, 35P30 }
\date{\today, AC091007.tex }

\begin{abstract}
We consider a polyharmonic operator
$H=(-\Delta)^l+V(x)$ in  dimension two with $l\geq 6$, $l$ being an
integer,  and  a limit-periodic potential $V(x)$. We prove that the
spectrum contains a semiaxis of absolutely continuous spectrum.
\end{abstract}

\maketitle

\section{Main Results.}

We study  an operator
    \begin{equation}
    H=(-\Delta)^l+V(x) \label{limper}
    \end{equation}
    in two dimensions, where $l\geq 6$ is an integer and $V(x)$ is a limit-periodic
potential
    \begin{equation}\label{V}
    V(x)=\sum _{r=1}^{\infty}V_r(x);
    \end{equation}
here $\{V_r\}_{r=1}^{\infty}$ is a family of periodic potentials
with doubling periods and decreasing $L_{\infty}$-norms, namely,
$V_r$ has orthogonal periods $2^{r-1}\vec \beta _1,\ 2^{r-1}\vec
\beta _2$ and
    \begin{equation}
    \|V_r\|_{\infty}<\hat Cexp(-2^{\eta r}) \label{r}
    \end{equation}
for some $\eta> 2+64/(2l-11)$. Without loss of generality, we assume that $\hat C=1$ and $\int
_{Q_r}V_r(x)dx=0$, $Q_r$ being the elementary cell of periods
corresponding to $V_r(x)$.

The one-dimensional analog of (\ref{limper}), (\ref{V}) with $l=1$
is already thoroughly investigated. It is proven in
\cite{1}--\cite{9} that the spectrum of the operator $H_1u=-u''+Vu$
is generically a Cantor type set. It has positive Lebesgue measure
\cite{1,8}. The spectrum is absolutely continuous \cite{1,3},
\cite{6}--\cite{7}. Generalized eigenfunctions can be represented in
the form of $e^{ikx}u(x)$, $u(x)$ being limit-periodic \cite{6,8,9}.
The case of a complex-valued potential is studied in \cite{10}.
Integrated density of states is investigated in
\cite{11}--\cite{14}. Properties of eigenfunctions of discrete
multidimensional limit-periodic Schr\"odinger operators are studied
in \cite{15}. As to the continuum multidimensional case, it is
proved \cite{14} that the integrated density of states for
(\ref{limper}) is the limit of densities of states for periodic
operators. A particular case of a periodic operator ($V_r=0$ when
$r\geq 2$) for  dimensions $d\geq 2$ and different $l$ is already
studied well, e.g., see \cite{9r} -- \cite{30}.
 Here we  prove that the spectrum of (\ref{limper}),
(\ref{V}) contains a semiaxis of absolutely continuous spectrum. This
paper is based on \cite{KL}. We proved the following results for the
case $d=2$, $l\geq 6$ in \cite{KL}.
   \begin{enumerate}
    \item The spectrum of the operator (\ref{limper}), (\ref{V})
    contains a semiaxis. A proof of the analogous result by different
    means can be found in \cite{20}.  The more general
    case $8l>d+3$, $d\neq 1(\mbox{mod}4)$, is considered in \cite{20},  however,
    under the additional restriction on the potential:
       the lattices of periods  of all periodic potentials
      $V_r$ have to contain a nonzero vector $\vec \gamma$ in common,
      i.e., $V(x)$ is periodic in one direction.

    \item There are generalized eigenfunctions $\Psi_{\infty }(\vec k, \vec
    x)$,
    corresponding to the semiaxis, which are    close to plane waves:
    for every $\vec k $ in a %extensive (for definition, see \eqref{16b})
    subset $\cal{G} _{\infty }$ of $\R^2$, there is
    a solution $\Psi_{\infty }(\vec k, \vec x)$ of the  equation
    $H\Psi _{\infty }=\lambda _{\infty }\Psi _{\infty }$ which can be described by
    the formula
    \begin{equation}
    \Psi_{\infty }(\vec k, \vec x)
    =e^{i\langle \vec k, \vec x \rangle}\left(1+u_{\infty}(\vec k, \vec
    x)\right), \label{aplane}
    \end{equation}
    \begin{equation}
    \|u_{\infty}\|_{L_{\infty }(\R^2)}\underset{|\vec k| \rightarrow
     \infty}{=}O\left(|\vec k|^{-\gamma _1}\right),\ \ \ \gamma _1>0,
    \label{aplane1}
    \end{equation}
    where $u_{\infty}(\vec k, \vec x)$ is a limit-periodic
    function
    \begin{equation}
    u_{\infty}(\vec k, \vec x)=\sum_{r=1}^{\infty}  u_r(\vec k, \vec
    x),\label{aplane2}
    \end{equation}
    $u_r(\vec k, \vec x)$ being periodic with periods $2^{r-1} \vec \beta _1,\ 2^{r-1} \vec \beta _2$.
    The  eigenvalue $\lambda _{\infty }(\vec k)$ corresponding to
    $\Psi_{\infty }(\vec k, \vec x)$ is close to $|\vec k|^{2l}$:
    \begin{equation}
    \lambda _{\infty }(\vec k)\underset{|\vec k| \rightarrow
     \infty}{=}|\vec k|^{2l}+
    O\left(|\vec k|^{-\gamma _2}\right),\ \ \ \gamma _2>0. \label{16a}
    \end{equation}
     The ``non-resonance" set $\cal{G} _{\infty }$ of
       vectors $\vec k$, for which (\ref{aplane}) -- (\ref{16a}) hold, is
       a Cantor type set $\cal{G} _{\infty }=\bigcap _{n=1}^{\infty }\cal{G}_n$,
       where $\{\cal{G} _n\}_{n=1}^{\infty}$ is a decreasing sequence of
       sets in $\R^2$. Each $\cal{G} _n$ has a finite number of holes in each bounded
       region. More and more holes appear as $n$ increases;
       however, holes added at each step are of smaller and smaller size.
       The set $\cal{G} _{\infty }$ satisfies the estimate
       \begin{equation}
       \left|\cal{G} _{\infty }\cap
        \bf B_R\right|\underset{R \rightarrow
     \infty}{=}|{\bf B_R}| \bigl(1+O(R^{-\gamma _3})\bigr),\ \ \ \gamma _3>0,\label{full}
       \end{equation}
       where $\bf B_R$ is the disk of radius $R$ centered at the
       origin and $|\cdot |$ is Lebesgue measure in $\R^2$.

     \item The set $\cal{D}_{\infty}(\lambda)$, defined as a level
     (isoenergetic) set for $\lambda _{\infty }(\vec k)$,
     $$ {\cal D} _{\infty}(\lambda)=\left\{ \vec k \in \cal{G} _{\infty }
     :\lambda _{\infty }(\vec k)=\lambda \right\},$$
     is shown to be a slightly distorted circle with an infinite number of holes. It can be
    described by  the formula
    \begin{equation}
    {\cal D}_{\infty}(\lambda)=\left\{\vec k:\vec k=\varkappa _{\infty}(\lambda, \vec{\nu})\vec{\nu},
    \ \vec{\nu} \in {\cal B}_{\infty}(\lambda)\right\}, \label{D}
    \end{equation}
    where ${\cal B}_{\infty }(\lambda )$ is a subset of the unit circle $S_1$.
    The set ${\cal B}_{\infty }(\lambda )$ can be interpreted as the set of possible
    directions of propagation for  almost plane waves (\ref{aplane}).
    The set ${\cal B}_{\infty }(\lambda )$ has a Cantor type structure
    and an asymptotically full measure on $S_1$ as
    $\lambda \to \infty $:
    \begin{equation}
    L\bigl({\cal B}_{\infty }(\lambda )\bigr)\underset{\lambda \rightarrow
     \infty}{=}2\pi +O\left(\lambda^{-\gamma _3/2l}\right),
    \label{B}
    \end{equation}
    here and below $L(\cdot)$ is a length of a curve. The value
    $\varkappa _{\infty }(\lambda ,\vec \nu )$ in (\ref{D}) is the
    ``radius" of ${\cal D}_{\infty}(\lambda)$ in a direction $\vec \nu$.
    The function $\varkappa _{\infty }(\lambda ,\vec \nu)-\lambda^{1/2l}$
    describes the deviation of ${\cal D}_{\infty}(\lambda)$
    from the perfect circle of the radius $\lambda^{1/2l}$. It is shown that the deviation is small
    \begin{equation}
    \varkappa _{\infty }(\lambda ,\vec \nu )\underset{\lambda
    \rightarrow
    \infty}{=}\lambda^{1/2l}+O\left(\lambda^{-\gamma _4 }\right), \ \ \ \gamma _4>0. \label{h}
    \end{equation}

\end{enumerate}

In this paper, we use the technique of \cite{KL} to prove absolute
continuity of the branch of the spectrum (the semiaxis)  corresponding to $\Psi_{\infty }(\vec k, \vec x)$.

\begin{figure}
\begin{minipage}[t]{6.6cm}
\centering
\includegraphics[totalheight=.25\textheight]{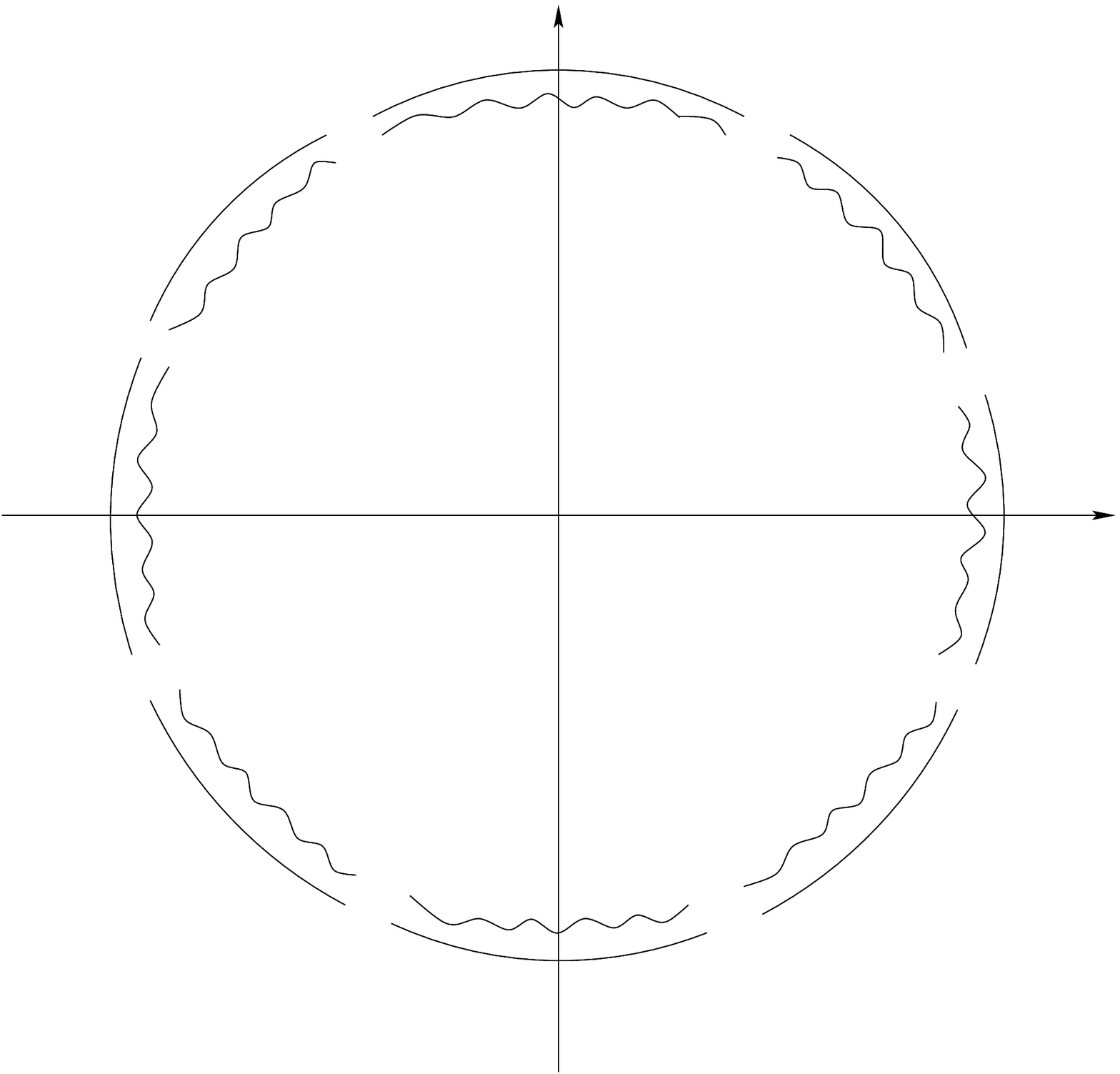}
\caption{Distorted circle with holes,
$\cal{D}_1(\lambda)$}\label{F:1}
\end{minipage}
\hfill
\begin{minipage}[t]{6.6cm}
\centering
\includegraphics[totalheight=.25\textheight]{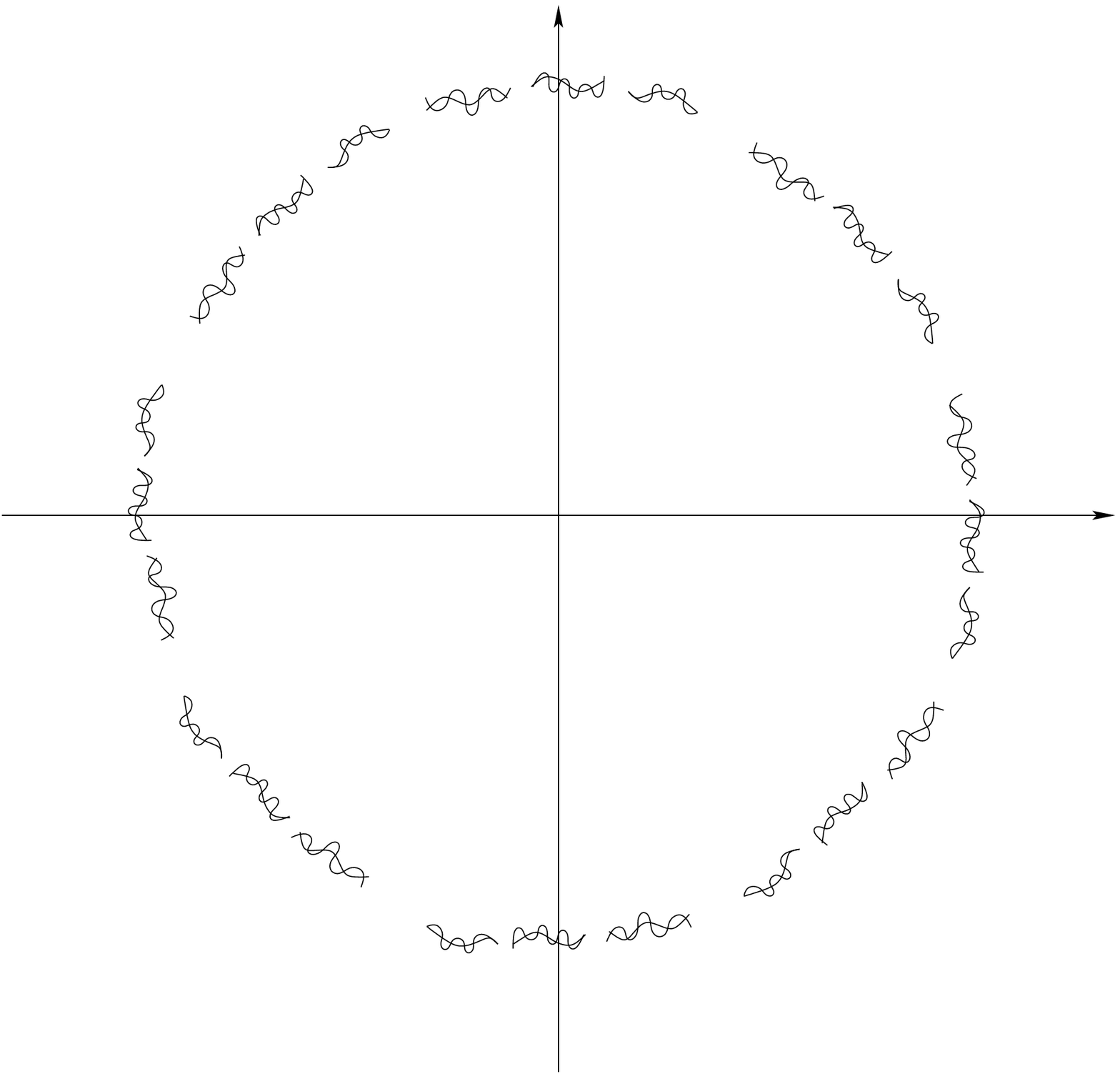}
\caption{Distorted circle with holes,
$\cal{D}_2(\lambda)$}\label{F:2}
\end{minipage}
\hfill
\end{figure}

In \cite{KL}, we  develop  a modification of the
Kolmogorov-Arnold-Moser (KAM) method  to prove the results listed
above. The paper \cite{KL} is inspired by \cite{21,22,24}, where the
method is used for periodic problems. In \cite{21}, KAM method is
applied to classical Hamiltonian systems. In \cite{22,24}, the
technique developed in \cite{21}  is  applied for semiclassical
approximation for multidimensional periodic Schr\"{o}dinger
operators at high energies. In  \cite{KL}, we consider  a sequence
of operators
    $$ H_0=(-\Delta )^l, \ \ \ \ \ \
H^{(n)}=H_0+\sum_{r=1}^{M_n} V_r,\ \ \ n\geq 1, \ M_n \to \infty
\mbox{ as } n \to \infty .$$ Obviously, $\|H-H^{(n)}\|\to 0$ as
$n\to \infty $ and $H^{(n)}=H^{(n-1)}+W_n$, where
$W_n=\sum_{r=M_{n-1}+1}^{M_n} V_r$. We treat each operator
$H^{(n)}$, $n\geq 1$, as a perturbation of the previous operator
 $H^{(n-1)}$. Each operator $H^{(n)}$ is periodic; however, the
periods go to infinity as $n \to \infty$. We show that there exists
$\lambda_*=\lambda_*(V)$ such that the semiaxis $[\lambda _*, \infty
)$ is contained in the spectra of all operators $H^{(n)}$. For every
operator $H^{(n)}$, there is a set of eigenfunctions (corresponding
to the semiaxis)  close to plane waves: for every $\vec k $ in an
extensive subset $\cal{G} _n$ of $\R^2$, there is a solution
$\Psi_{n}(\vec k, \vec x)$ of the differential equation $H^{(n)}\Psi
_n=\lambda ^{(n)}\Psi _n$, which can be represented by the formula
    \begin{equation}
    \Psi_n (\vec k, \vec x)
    =e^{i\langle \vec k, \vec x \rangle}\left(1+\tilde u_{n}(\vec k, \vec
    x)\right),\ \ \
    \|\tilde u_{n}\|=_{|\vec k| \to \infty}O(|\vec k|^{-\gamma _1}),\ \
    \ \gamma _1>0, \label{na}
    \end{equation}
where $\tilde u_{n}(\vec k, \vec x)$ has periods $2^{M_n-1}\vec \beta _1,
2^{M_n-1}\vec \beta _2$.\footnote{Obviously, $\tilde u_{n}(\vec k, \vec x)$
is simply related to functions $u_{r}(\vec k, \vec x)$
used in (\ref{aplane2}): $\tilde u_{n}(\vec k, \vec x)=\sum
_{r=M_{n-1}+1}^{M_n} u_{r}(\vec k, \vec x)$. } The corresponding
eigenvalue $\lambda ^{(n)}(\vec k)$ is close to $|\vec k|^{2l}$:
    $$ \lambda ^{(n)}(\vec k)\underset{|\vec k| \rightarrow
     \infty}{=}|\vec k|^{2l}+
    O\left(|\vec k|^{-\gamma _2}\right),\ \ \ \gamma _2>0.$$
The non-resonance set $\cal{G} _{n}$ is shown to be extensive in
$\R^2$:
       \begin{equation}
       \left|\cal{G} _{n}\cap
       \bf B_R\right|\underset{R \rightarrow
     \infty}{=}|{\bf B_R}|\bigl(1+O(R^{-\gamma _3})\bigr). \label{16b}
       \end{equation}
Estimates (\ref{na}) -- (\ref{16b}) are uniform in $n$.
The set ${\cal D}_{n}(\lambda)$ is defined as the level
(isoenergetic) set for non-resonant eigenvalue $\lambda ^{(n)}(\vec
k)$:
    \begin{equation}
     {\cal D} _{n}(\lambda)=\left\{ \vec k \in
    \cal{G} _n:\lambda ^{(n)}(\vec k)=\lambda \right\}.\label{Dn*}
    \end{equation}
This set is shown to be a slightly distorted circle with a finite
number of holes (see Figs. \ref{F:1}, \ref{F:2}), the set ${\cal D}
_{1}(\lambda)$ being strictly inside the circle of the radius
$\lambda^{1/2l}$ for sufficiently large $\lambda $. The set ${\cal
D} _{n}(\lambda)$ can be described by the formula
    \begin{equation}
    {\cal D}_{n}(\lambda)=\left\{\vec k:\vec k=
    \varkappa_{n}(\lambda, \vec{\nu})\vec{\nu},
    \ \vec{\nu} \in {\cal B}_{n}(\lambda)\right\}, \label{Dn}
    \end{equation}
where ${\cal B}_{n}(\lambda )$ is a subset  of the unit circle
$S_1$. The set ${\cal B}_{n}(\lambda )$ can be interpreted as the
set of possible directions of propagation for  almost plane waves
(\ref{na}). It has an asymptotically full measure on $S_1$ as
$\lambda \to \infty $:
    \begin{equation}
    L\bigl({\cal B}_{n}(\lambda )\bigr)\underset{\lambda \to \infty
    }{=}2\pi +O\left(\lambda^{-\gamma _3/2l}\right). \label{Bn}
    \end{equation}
The set ${\cal B}_{n}(\lambda)$ has only a finite number of holes;
however, their number grows with $n$. More and more holes of a
smaller and smaller size are added at each step. The value
$\varkappa_{n}(\lambda ,\vec \nu )-\lambda^{1/2l}$ gives the
deviation of ${\cal D}_{n}(\lambda)$ from the circle of the radius
$\lambda^{1/2l}$  in the direction $\vec \nu $. It is shown that the
deviation is asymptotically small:
    \begin{equation}
    \varkappa_{n}(\lambda ,\vec \nu)
    =\lambda^{1/2l}+O\left(\lambda^{-\gamma _4 }\right),\ \ \ \
    \frac{\partial \varkappa_{n}(\lambda ,\vec \nu)}{\partial \varphi
    }=O\left(\lambda^{-\gamma _5 }\right), \quad \gamma _4, \gamma _5>0,
    \label{Dec9a}
    \end{equation}
$\varphi $ being an angle variable, $\vec \nu =(\cos \varphi ,\sin
\varphi )$.  Estimates (\ref{Bn}), (\ref{Dec9a}) are uniform in $n$.

% It is shown that ${\cal D}_1(\lambda )$ is strictly
%inside the perfect circle, here and below we assume without loss of
%generality that $\int V_r=0$, for all $r$, integral being taken over
%the elementary cell of periods.

At each step, more and more points are excluded from the
non-resonance sets $\cal{G} _n$ ; thus, $\{ \cal{G} _n
\}_{n=1}^{\infty }$ is a decreasing sequence of sets. The set
$\cal{G} _\infty $ is defined as the limit set $\cal{G}
_\infty=\bigcap _{n=1}^{\infty }\cal{G} _n $. It has an infinite
number of holes, but nevertheless satisfies the relation
(\ref{full}). For every $\vec k \in \cal{G} _\infty $ and every $n$,
there is a generalized eigenfunction of $H^{(n)}$ of the type
(\ref{na}). It is shown that  the sequence $\Psi _n(\vec k, \vec x)$
has a limit in $L_{\infty }(\R^2)$ when $\vec k \in \cal{G} _\infty
$. The function $\Psi _{\infty }(\vec k, \vec x) =\lim _{n\to \infty
}\Psi _n(\vec k, \vec x)$ is a generalized eigenfunction of $H$. It
can be written in the form (\ref{aplane}) -- (\ref{aplane2}).
Naturally, the corresponding eigenvalue $\lambda _{\infty }(\vec k)
$ is the limit of $\lambda ^{(n )}(\vec k )$ as $n \to \infty $.

It is shown that $\{{\cal B}_n(\lambda)\}_{n=1}^{\infty }$ is a
decreasing sequence of sets at each step more and more directions
being excluded. We consider the limit ${\cal B}_{\infty}(\lambda)$
of ${\cal B}_n(\lambda)$,
    $${\cal B}_{\infty}(\lambda)=\bigcap_{n=1}^{\infty} {\cal B}_n(\lambda).$$
This set has a Cantor type structure on the unit circle. It is shown
that ${\cal B}_{\infty}(\lambda)$ has asymptotically full measure on
the unit circle (see (\ref{B})). We prove that the sequence
$\varkappa _n(\lambda ,\vec \nu )$, $n=1,2,... $, describing the
isoenergetic curves $\cal{D}_n$, converges rapidly (super
exponentially) as $n\to \infty$. Hence,
 ${\cal D}_{\infty}(\lambda)$ can be described as the limit of  ${\cal
D}_n(\lambda)$ in the sense (\ref{D}), where $\varkappa
_{\infty}(\lambda, \vec{\nu})=\lim _{n \to \infty} \varkappa
_n(\lambda, \vec{\nu})$ for every $\vec{\nu} \in {\cal
B}_{\infty}(\lambda)$. It is shown that the derivatives of the
functions $\varkappa _n(\lambda, \vec{\nu})$ (with respect to the
angle variable on the unit circle) have a limit as $n\to \infty $
for every $\vec{\nu} \in {\cal B}_{\infty}(\lambda)$. We denote this
limit by $\frac{\partial \varkappa_{\infty}(\lambda ,\vec
\nu)}{\partial \varphi }$. Using (\ref{Dec9a}), we prove  that
    \begin{equation}
    \frac{\partial \varkappa_{\infty}(\lambda ,\vec \nu)}{\partial
    \varphi }=O\left(\lambda^{-\gamma _5 }\right).
    \label{Dec9a*}
    \end{equation}
Thus, the limit curve ${\cal D}_{\infty}(\lambda)$ has a
tangent vector in spite of its Cantor type structure, the tangent
vector being the limit of corresponding tangent vectors for ${\cal
D}_n(\lambda)$ as $n\to \infty $.  The curve  ${\cal
D}_{\infty}(\lambda)$ looks like a slightly distorted circle with
infinite number of holes.

The main technical difficulty overcome in \cite{KL} is the
construction of non-resonance sets $\cal{B} _n(\lambda)$ for every
fixed sufficiently large $\lambda $, $\lambda >\lambda_*(V)$, where
$\lambda_*$ is the same for all $n$. The set $\cal{B} _n(\lambda)$
is obtained  by deleting a ``resonant" part from
$\cal{B}_{n-1}(\lambda)$. The definition of  $\cal{B} _{n-1}
(\lambda)\setminus \cal{B}_{n}(\lambda)$ includes Bloch eigenvalues
of $H^{(n-1)}$. To describe  $\cal{B} _{n-1}(\lambda)\setminus
\cal{B}_{n}(\lambda)$, one has to use not only non-resonant
eigenvalues of the type (\ref{16a}) but also resonant eigenvalues,
for which no suitable  formulas are known. The absence of formulas
causes difficulties in estimating the size of $\cal{B}
_{n-1}(\lambda)\setminus \cal{B}_n(\lambda)$. To deal with this
problem, we start by introducing an angle variable $\varphi \in
[0,2\pi)$,  $\vec \nu = (\cos \varphi , \sin \varphi )\in S_1$ and
consider sets $\cal{B}_n(\lambda)$ in terms of this variable. Next,
we show that the resonant set  $\cal{B} _{n-1}(\lambda)\setminus
\cal{B}_{n}(\lambda)$ can be described as the set of zeros of
determinants of the type $\det (I+A_{n-1}(\varphi))$,
$A_{n-1}(\varphi )$ being a trace type operator,
    $$I+A_{n-1}(\varphi )=\bigl(H^{(n-1)}\bigl(\vec \varkappa _{n-1}(\varphi )
       +\vec b\bigr)-\lambda-\epsilon \bigr) \bigl(H_{0}\bigl(\vec \varkappa _{n-1}(\varphi )
       +\vec b\bigr)+\lambda\bigr)^{-1},$$
where $\vec \varkappa _{n-1}(\varphi )$ is a vector-function
describing $\cal{D}_{n-1}(\lambda)$ : $\vec \varkappa
_{n-1}(\varphi)=\varkappa_{n-1}(\lambda, \vec \nu)\vec \nu$. To
obtain $\cal{B} _{n-1}(\lambda)\setminus \cal{B}_{n}(\lambda)$, we
take all values of $\epsilon $ in a small interval and vectors $\vec
b$ in a finite set, $\vec b\neq 0$. Further, we extend our
considerations to a complex neighborhood $\varPhi _0$ of $[0,2\pi)$.
We show that the determinants are analytic functions of $\varphi $
in $\varPhi _0$, and thus reduce the problem of estimating the size
of the resonance set to a problem in complex analysis. We use
theorems for analytic functions to count the zeros of the
determinants and to investigate how far zeros move when $\epsilon $
changes. This enables us to estimate the size of the zero set of the
determinants and hence the size of the non-resonance set $\varPhi
_n\subset \varPhi _0$,  which is defined as a non-zero set for the
determinants. Proving that the non-resonance set $\varPhi _n$ is
sufficiently large, we obtain estimates (\ref{16b}) for $\cal{G} _n$
and (\ref{Bn}) for $\cal{B}_n$, the set  $\cal{B}_n$ being the
intersecton of $\varPhi _n$ with the real line. To obtain $\varPhi
_n$ we delete  from $\varPhi _0$ more and more holes of smaller and
smaller radii at each step. Thus, the non-resonance set $\varPhi
_n\subset \varPhi _0$ has the structure of Swiss Cheese (Fig.
\ref{F:7}, \ref{F:8}). We call deleting  the resonance set from
$\varPhi _0$ at each step of the recurrent procedure the ``Swiss
Cheese Method".  The essential difference of our method from those
applied  earlier in similar situations (see, e.g., \cite{21,22,24})
is that we construct a non-resonance set not only in the whole space
of a parameter ($\vec k\in \R^2$ here) but also on  the isoenergetic
curves ${\cal D}_n(\lambda )$ in the space of parameter when
$\lambda$ is sufficiently large. Estimates for the size of
non-resonant sets on a curve require more subtle technical
considerations than those sufficient for description of a
non-resonant set in the whole space of the parameter.

Here, we use information obtained in \cite{KL} to prove absolute continuity of the branch of the spectrum
(the semiaxis) corresponding to the functions $\Psi _{\infty }(\vec k, \vec x)$,
$\vec k\in \cal{G} _{\infty }$. Absolute continuity follows from the
convergence of the spectral projections corresponding to $\Psi
_{n}(\vec k, \vec x)$, $\vec k\in \cal{G} _{\infty }$, to spectral
projections of $H$ (in the strong sense uniformly in $\lambda $) and
properties of the level curves $\cal{D}_{\infty }(\lambda )$,
$\lambda >\lambda _*$. Roughly speaking, the area between isoenergetic curves
$\cal{D}_{\infty }(\lambda +\epsilon
)$ and $\cal{D}_{\infty }(\lambda )$ (integrated density of states)
is proportional to $\epsilon$.

Note that generalization of  results from the case $l\geq 6$, $l$
being an integer, to the case of rational $l$ satisfying the same
inequality is relatively simple; it requires just slightly more
careful technical considerations. The restriction $l\geq 6$  is also
technical, though it is more difficult to lift. The condition $l\geq
6$ is needed only for the first two steps of the recurrent procedure
in \cite{KL}. The requirement for super exponential decay of
$\|V_r\|$ as $r \to \infty $ is more essential than $l\geq 6$ since
it is needed to ensure convergence of the recurrent procedure. It is
not essential that  potentials $V_r$ have doubling periods; periods
of the type $q^{r-1}\vec \beta _1,\ q^{r-1}\vec \beta _2$, $q \in
\N,$ can be treated in the same way.

The periodic case ($V_r=0$, when $r\geq 2$) is
already carefully investigated  for dimensions $d\geq 2$ and  different $l$  \cite{9r}--\cite{30}.
For briefness,
we mention here only results for dimension two. Absolute continuity of the whole spectrum
is proven in \cite{9r} for $l=1$, however the proof can be
extended for higher integers $l$. Bethe-Sommerfeld conjecture is first
proved for $d=2$, $l=1$ in \cite{11r}, \cite{19r} and for
$l\geq 1$ in \cite{29r}. The perturbation formulas for eigenvalues are constructed in
\cite{28r}. The formulas for eigenfunctions and
the corresponding isoenergetic surfaces are obtained in \cite{29r}.

The plan of the paper is the following. In Section 2,  we sketch  main
steps of the recurrent procedure and the ``Swiss cheese method" developed  in \cite{KL}.
Section 3 describes
eigenfunctions and isoenergetic surfaces of $H$. The proof of the absolute continuity is in Section 4
using the results in Sections 2 and 3.

%%%%%%\vspace{5mm} \noindent {\bf Acknowledgement} The authors are very
%%%%%grateful to Prof. G. Stolz and Prof. G. Gallavotti for useful
%%%%%%%%55discussions.

\section{Recurrent Procedure.}
\subsection{The First Approximation.}\label{chapt3}

%%%%%%\spacing{1.66}

\subsubsection{The Main Operator $H$ and the First Operator $H^{(1)}$.}

We introduce the first operator $H^{(1)}$, which corresponds to a
partial sum in the series (\ref{V})
    \begin{equation}\label{2.1}
     H^{(1)}=(-\Delta )^l+W_1,
      \qquad
    W_1=\sum_{r=1}^{M_1}V_r,
     \end{equation}
where $M_1$ is chosen in such a way that $
     2^{M_1}\approx k^{s_1}$ \footnote{We write $a(k)\approx b(k)$
when the inequalities $\frac{1}{2}b(k)<a(k)<2b(k)
%%%%%\ c_1\neq c_1(k),\c_2\neq c_2(k)
$ hold.} for a $k>1$, $s_1=(2l-11)/32$. For simplicity, we let the
potentials $V_r$ have periods directed along the axes, i.e., the periods of $V_r$
are $2^{r-1}\vec \beta_1=2^{r-1}(\beta_1,0)$
and $2^{r-1}\vec \beta_2=2^{r-1}(0,\beta_2)$. Then, obviously, the periods of
$W_1$ are $(a_1,0)=2^{M_1-1} (\beta _1,0)$ and $(0,a_2)=2^{M_1-1}
(0,\beta _2)$, and
 $a_1\approx k^{s_1}\beta _1/2,$ $a_2\approx k^{s_1}\beta _2/2$.
     Note that
    $$\|W_1\|_{\infty} \leq \sum_{n=1}^{ M_1}
    \|V_n\|_{\infty}
    %%%%%%%%%%%\leq \sum_{n=1}^{M_1}\exp(-2^{\eta n})
    =O(1)\ \text{as}\  k \to \infty. $$
It is well-known (see, e.g., \cite{RS}) that
spectral analysis of a periodic operator $H^{(1)}$ can be reduced
to  analysis of a family of operators $H^{(1)}(t)$,
$t\in K_1$, where $K_1$ is the elementary cell of the dual lattice,
$K_1=[0,2\pi a_1^{-1})\times [0,2\pi a_2^{-1}).$ The vector $t$ is
called  $quasimomentum$. An operator $H^{(1)}(t)$, $t\in K_1$, acts
in $L_2(Q_1)$, $Q_1$ being the elementary cell of the periods of the
potential, $Q_1=[0,a_1]\times [0,a_2].$ The operator $H^{(1)}(t)$ is
described by formula (\ref{2.1}) and the quasiperiodic boundary conditions
for a function and its derivatives:
    \begin{equation} \label{quasiperiodic}
    \begin{array}{ll}u(a_1,x_{2})= \exp (it_1a_1)u(0,x_{2}),& u(x_1,a_{2})= \exp
    (it_2a_2)u(x_{1},0),\\
u_{x_1}^{(j)}(a_1,x_{2})=\exp (it_1a_1)u_{x_1}^{(j)}(0,x_{2}),&
u_{x_2}^{(j)}(x_1,a_{2})=  \exp
    (it_2a_2)u_{x_2}^{(j)}(x_{1},0),\end{array}
    \end{equation}
$0<j<2l$. Each operator $H^{(1)}(t)$, $t\in K_1$, has a discrete
bounded below spectrum $\Lambda ^{(1)}(t)$
    $$\Lambda ^{(1)}(t)=\bigcup _{n=1}^{\infty }\lambda _n^{(1)}(t),\
    \lambda^{(1)} _n(t)\to _{n\to \infty }\infty .$$
The spectrum $\Lambda ^{(1)}$ of the operator $H^{(1)}$ is the union
of the spectra of the operators $H^{(1)}(t)$ over $t \in K_1$,
$\Lambda ^{(1)}=\cup _{t\in K_1}\Lambda ^{(1)}(t)=\bigcup _{n\in
\N,t\in K_1}\lambda _n^{(1)}(t).$ The functions $\lambda
_n^{(1)}(t)$ are continuous in $t$, so $\Lambda ^{(1)}$ has a band
structure
    \begin{equation}\label{bands}
    \Lambda ^{(1)}=\bigcup _{n=1}^{\infty }[q_n^{(1)},Q_n^{(1)}],\ q_n^{(1)}
    =\min _{t\in K_1}\lambda _n^{(1)}(t),\
     Q_n^{(1)}=\max _{t\in K_1}\lambda ^{(1)}_n(t).
     \end{equation}

The eigenfunctions of $H^{(1)}(t)$ and $H^{(1)}$ are simply related.
Extending all the eigenfunctions of the operators $H^{(1)}(t)$
quasiperiodically (see (\ref{quasiperiodic})) to $\R^2$, we obtain a
complete system of generalized eigenfunctions of $H^{(1)}$.

Let $H_0^{(1)}$ be the operator (\ref{limper}) corresponding to  $V=0$.
We consider that it has periods $(a_1,0),(0,a_2)$ and that
operators $H_0^{(1)}(t)$, $t\in K_1$ are defined in $L_2(Q_1)$. The eigenfunctions of the
operator $H_0^{(1)}(t)$, $t\in K_1$, are   plane waves satisfying (\ref{quasi}).
They are naturally indexed by   points of $\Z^2$
    $$\Psi _j^0(t,x)=|Q_1|^{-1/2}\exp
    i \langle \vec{p}_j(t),x \rangle, \ \ j\in \Z^2,$$
the  eigenvalue corresponding to $\Psi _j^0(t,x)$ being equal to $p_j^{2l}(t)$, where
here and below
    $$\vec{p}_j(t)=2\pi j/a+t, \ \ 2\pi j/a=(2\pi j_1/a_1, 2\pi j_2/a_2), \ j\in
    \Z^2,\ \ |Q_1|=a_1a_2,\ \ p_j^{2l}(t)=|\vec{p}_j(t)|^{2l}.    $$

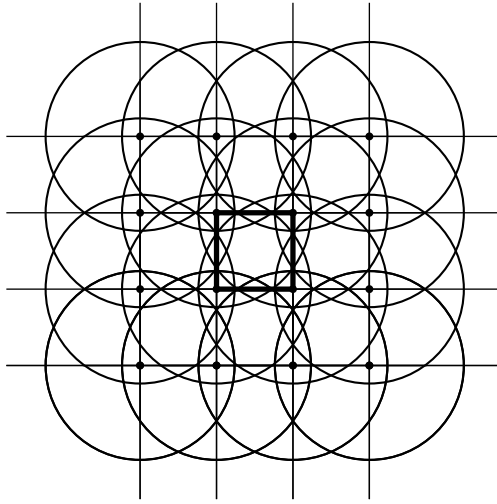
\begin{figure}
\centering

\begin{pspicture}(0,0)(7,7)
\psset{unit=1in}

\def\4row{
\pscircle(0.5,0.5){0.5} \qdisk(0.5,0.5){1.5pt}
\psline[linewidth=0.5pt](-0.2,0.5)(1.2,0.5)
\psline[linewidth=0.5pt](0.5,-0.2)(0.5,1.2)}

\multips(0,0)(0.4,0.0){4}{\4row}

\multido{\n=0+0.4}{4} {\multips(\n,0)(0.0,0.4){4}{\4row}}

%%%%%%% square in the middle %%%%%%%%%%
\psline[linewidth=2pt](0.9,0.9)(1.3,0.9)
\psline[linewidth=2pt](0.9,0.9)(0.9,1.3)
\psline[linewidth=2pt](0.9,1.3)(1.3,1.3)
\psline[linewidth=2pt](1.3,0.9)(1.3,1.3)
%%%%%%%%%%%%%%%%%%%%%%%%%%%%%%%%%%%%%%%

\end{pspicture}

\caption{The isoenergetic surface $S_0(\lambda)$ of the free
operator $H_0^{(1)}$}\label{F:3}
\end{figure}

Next, we introduce an isoenergetic surface\footnote{``surface" is a traditional term.
In our case, it is a curve.} $S_0(\lambda)$ of the free
operator $H_0^{(1)}$. A point $t \in K_1$ belongs to $S_0(\lambda)$ if and
only if $H_0^{(1)}(t)$ has an eigenvalue equal to $\lambda$, i.e.,
there exists $j \in \Z^2$ such that $p_j^{2l}(t)=\lambda$. This
surface can be obtained as follows: the circle of radius $k=\lambda
^{1/(2l)}$ centered at the origin is divided into pieces by
the dual lattice $\{\vec p_q(0)\}_{q \in \Z^2}$, and then all pieces
are translated in a parallel manner into the cell $K_1$ of the dual
lattice. We also can  get $S_0(\lambda)$ by drawing sufficiently
many circles of radii $k$ centered at the dual lattice $\{\vec
p_q(0)\}_{q \in \Z^2}$ and by looking at the figure in the cell
$K_1$. As the result of either of these two procedures we obtain a circle
of radius $k$ ``packed into the bag $K_1$" as shown in the
Fig.~\ref{F:3}. Note that each piece of $S_0(\lambda )$ can be
described by an equation $p_j^{2l}(t)=\lambda$ for a fixed $j$. If
$t\in S_0(\lambda )$, then $j$ can be uniquely defined from the last
equation, unless $t$ is not a point of self-intersection of the
isoenergetic surface. A point $t$ is a self-intersection of
$S_0(\lambda )$ if and only if
\begin{equation}
p_q^{2l}(t)=p_j^{2l}(t)=k^{2l}  \label{Apr2} \end{equation}
for at
least on pair of indices $q,j$, $q\neq j$.

 Note that any vector $\vec \varkappa $ in $\R ^2$ can be uniquely represented
 in the form $\vec \varkappa =\vec{p}_j(t)$, where $j\in \Z ^2$ and
$t\in K_1$. Let $\cal K_1$ be the parallel shift into $K_1$: $${\cal
K_1}:\R ^2\to K_1, \ \ {\cal K_1}\bigl(\vec{p}_j(t)\bigr)=t.$$
Obviously, ${\cal K_1}S_k=S_0(\lambda )$ and $L\bigl(S_0(\lambda
)\bigr)=L\bigl(S_k\bigr)=2\pi k$, $k=\lambda ^{1/(2l)}$, $S_k$
being the circle of radius $k$ centered at the origin.

The operator $H^{(1)}(t)$, $t\in K_1$, has the following matrix
representation in the basis of plane waves $\Psi _j^0(t,x)$:
    $$H^{(1)}(t)_{mq}=p_m^{2l}(t)\delta _{mq}+
    w_{m-q},\  \ m,q\in \Z^2,$$
here and below $\delta_{mq}$ is the Kronecker symbol, $w_{m-q}$ are Fourier coefficients of $W_1$, the
coefficient $w_0$ being equal to zero. The matrix
$H^{(1)}(t)_{mq}$ also describes an operator in the space $l_2$ of
square summable sequences with indices in $\Z^2$, the operator in
$l_2$ being unitarily equivalent to $H^{(1)}(t)$ in $L_2(Q_1)$.  From now on, we denote the operator in $l_2$
also by $H^{(1)}(t)$. Note that
the canonical basis in $l_2$ does not depend on $t$, all
dependence on $t$ being in the matrix. Thus, the matrix
$H^{(1)}(t)_{mq}$ and hence the operator $H^{(1)}(t):l_2\to
l_2$ can be analytically extended in $t$ from $K_1$ to $\C^2$. We consider
$H^{(1)}(t):l_2\to l_2$ for real and complex $t$. Further, when
we refer to $H^{(1)}(t)$ for $t\in \C^2$, we mean the operator in $l_2$.

\subsubsection{Perturbation Formulas.}
\begin{figure}%[htp]
    \centering
    \includegraphics[totalheight=.2\textheight]{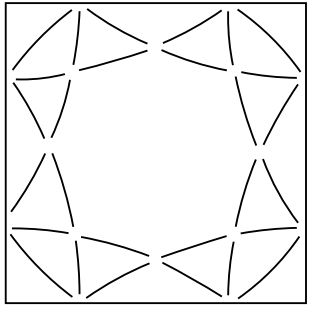}
    \caption{The first non-resonance set $\chi_1(\lambda)$}\label{F:4}
    \end{figure}

In this section, we consider
the operator $H^{(1)}(t)$ as a perturbation of the free operator $H^{(1)}_0(t)$. We show
that for every sufficiently large $\lambda $, there is a
``non-resonant" subset $\chi _1(\lambda)$ of $S_0(\lambda )$ such that
perturbation series for an eigenvalue and a spectral projection of
$H^{(1)}(t)$ converge when $t\in \chi _1(\lambda)$. The set $\chi _1(\lambda)$ is
obtained by deleting small neighborhoods of self-intersections of
$S_0(\lambda )$; see Fig. \ref{F:4}. The self-intersections are
described by (\ref{Apr2}) and correspond to degenerated eigenvalues
of $H^{(1)}_0(t)$. The size of the neighborhood is
$k^{-1-4s_1-\delta }$, $k=\lambda ^{1/(2l)}$, where $\delta $ is
a small positive number. The set $\chi _1(\lambda)$ is
sufficiently large: its relative measure with respect to
$S_0(\lambda )$ tends to 1 as $\lambda \to \infty $. The precise
formulation of these results is given in the next lemma, proved
by elementary geometric considerations in  \cite{19} \footnote{More precisely, Lemma
\ref{L:2.1} corresponds to Lemma 2.1 on page 26 in \cite{19}. There
is a slight difference between two lemmas. Lemma 2.1 in \cite{19} is
proved for the case of fixed periods $\beta _1,\beta _2$. In Lemma
\ref{L:2.1} here, we consider the periods $a_1\approx k^{s_1}\beta
_1/2, a_2\approx k^{s_1}\beta _2/2$. However, the proofs are
completely analogous.}.

\begin{Lem}[Geometric Lemma]\label{L:2.1}
For an arbitrarily small positive $\delta $, $2\delta <2l-2-4s_1,$
and sufficiently large $\lambda $, $\lambda
>\lambda _0(V,\delta)$, there exists a non-resonance set
 $\chi _1(\lambda,\delta )\subset S_0(\lambda )$
 such that the following hold.

\begin{enumerate}
\item For any point $t\in \chi_1(\lambda)$,
    \begin{enumerate}
    \item there exists a unique $j\in \Z ^2$ such that $p_j(t)=k$, $k=\lambda ^{1/(2l)}$;
    \item
        \begin{equation}\label{2.5}
        \min _{i\neq j}|p_j^{2}(t)-p_i^{2}(t)|>2k^{-4s_1-\delta }.
        \end{equation}
    \end{enumerate}

\item For any $t$ in the $(2k^{-1-4s_1-2\delta })$-neighborhood of
the non-resonance set in $\C^2$, there exists a unique $j\in \Z
^2$ such that
    \begin{equation}\label{2.6}
    |p_j^2(t)-k^2| < 5k^{-4s_1-2\delta }
    \end{equation}
and (\ref{2.5}) holds.

\item The  non-resonance set $\chi_1(\lambda)$ has an asymptotically full
measure on $S_0(\lambda)$ in the following sense that
    \begin{equation}\label{2.7}
    \frac{L(S_0(\lambda )\setminus \chi _1(\lambda,\delta
    ))}{L(S_0(\lambda))} \underset {\lambda\to \infty }{=}O(k^{- \delta /2}).
    \end{equation}
 \end{enumerate}
\end{Lem}

\begin{Cor}\label{C:2.2}
If $t$ belongs to the $(2k^{-1-4s_1-2\delta })$-neighborhood of the
non-resonance set $\chi_1(\lambda,\delta )$ in $\C^2$, then, for any
$z\in \C^2$ lying on the circle $C_1=\{ z: |z-k^{2l}|=k^{2l-2-4s_1-\delta }\}$
%%%%%%%%%%\begin{equation} C_1=\{ z: |z-k^{2l}| =k^{2l-2-4s_1-\delta }\}
%%%%%%%%%%\\label{Apr3} \end{equation}
 and any $i$ in $ \Z^2$, the
inequality
%
%%%\begin{equation}\label{2.8}
$2| p_i^{2l}(t)-z| >k^{2l-2-4s_1-\delta }$
%%%\end{equation}
%
holds.
\end{Cor}
Let $E_j(t)$ be the spectral projection of the free operator,
corresponding to the eigenvalue $p_j^{2l}(t):$ $(E_j)_{rm}=\delta
_{jr}\delta _{jm} $. In the $(2k^{-1-4s_1-2\delta})$-neighborhood of
$\chi _1(\lambda ,\delta )$, we define functions $g_r^{(1)}(k,t)$
and operator-valued functions $G_r^{(1)}(k,t)$, $r=1, 2, \cdots$ as
follows:
\begin{equation}\label{2.11}
 g_r^{(1)}(k,t)=\frac{(-1)^r}{2\pi ir}\mbox{Tr}\oint
 _{C_1}((H_0^{(1)}(t)-z)^{-1}W_1)^rdz,
 \end{equation}
\begin{equation}\label{2.12}
G_r^{(1)}(k,t)=\frac{(-1)^{r+1}}{2\pi i}\oint
_{C_1}((H_0^{(1)}(t)-z)^{-1}W_1)^r (H_0^{(1)}(t)-z)^{-1}dz.
 \end{equation}
 To find
$g_r^{(1)}(k,t)$ and $G_r^{(1)}(k,t)$, it is necessary to compute
the residues of a rational function of a simple structure, whose
numerator does not depend on $z$, while the denominator is a
product of factors of the type $(p_i^{2l}(t)-z)$. For all $t$ in
the non-resonance set within $C_1$, the integrand has a single pole at the point
$z=k^{2l}=p_j^{2l}(t).$
%%%%%%%In the sums, arising in the construction of
%%%%%%%the integrand operators and their traces there are only a finite
%%%%%%%number of terms, having a singularity at the point $z=k^{2l}$.
By computing the residue at this point, we obtain explicit expressions
for $g_r^{(1)}(k,t)$ and $G_r^{(1)}(k,t)$. For example,
$g_1^{(1)}(k,t)=0$,
   \begin{equation}
    \label{2.13}
    g_2^{(1)}(k,t)
    =\sum _{q\in \Z^2,q\neq 0}| w_q| ^2(p_j^{2l}(t)-
        p_{j+q}^{2l}(t))^{-1}
    =\sum _{q\in \Z^2,q\neq 0} \frac{| w_q| ^2(2p_j^{2l}(t)-
        p_{j+q}^{2l}(t)-p_{j-q}^{2l}(t))}{2(p_j^{2l}(t)-
        p_{j+q}^{2l}(t))(p_j^{2l}(t)-
        p_{j-q}^{2l}(t))},
    \end{equation}
    \begin{equation}\label{2.15}
    G_1^{(1)}(k,t)_{rm}=\frac{w_{j-m}}{p_j^{2l}(t)-p_m^{2l}(t)}\delta_{rj}
    + \frac{w_{r-j}}{p_j^{2l}(t)-p_r^{2l}(t)}\delta _{mj}, \mbox{if}\
    \ r\neq m,\quad G_1^{(1)}(k,t)_{jj}=0.
    \end{equation}
%$w_q$ being Fourier coefficients of $W_1$.
It is not difficult to show that $g_2^{(1)}(k,t)>$ for sufficiently large $\lambda$.
For technical reasons, it is convenient to introduce parameter $\alpha $ in front of the
potential $W_1$. Namely, $H_{\alpha }^{(1)}=(-\Delta )^l
+\alpha W_1$, $0\leq \alpha \leq 1$. We denote the operator $H_\alpha ^{(1)}$
with $\alpha =1$ simply by $H^{(1)}.$
\begin{Thm}\label{T:2.3}
 Suppose $t$ belongs to the
$(2k^{-1-4s_1-2\delta })$-neighborhood in $K_1$ of the
non-resonance set $\chi _1(\lambda ,\delta )$, $0<2\delta
<2l-2-4s_1$. Then for sufficiently large $\lambda $, $\lambda
>\lambda _0(V, \delta )$, and for all $\alpha $, $-1\leq \alpha \leq
1$, there exists a single eigenvalue of the operator $H_{\alpha
}^{(1)}(t)$ in the interval $\varepsilon_1 (k,\delta ):=
(k^{2l}-k^{2l-2-4s_1-\delta }, k^{2l}+k^{2l-2-4s_1-\delta })$. It is
given by the series
\begin{equation}\label{2.16}
\lambda_j^{(1)} (\alpha ,t)=p_j^{2l}(t)+\sum _{r=2}^{\infty }\alpha
^rg_r^{(1)}(k,t),
\end{equation}
converging absolutely in the disk $|\alpha|  \leq 1$, where the
index $j$ is  determined  according to Parts 1(a) and 2 of Lemma \ref{L:2.1}.
The spectral projection, corresponding to $\lambda_j^{(1)}
(\alpha ,t),$ is given by the series:
\begin{equation}\label{2.17}
E_j^{(1)} (\alpha ,t)=E_j+\sum _{r=1}^{\infty }\alpha
^rG_r^{(1)}(k,t),
\end{equation}
which converges in the trace class $\mathbf{S_1}$ uniformly with
respect to $\alpha $ in the disk  $| \alpha | \leq 1$.

For the coefficients $g_r^{(1)}(k,t)$,  $G_r^{(1)}(k,t)$ the following
estimates hold:
%\footnote{Here and below $\| \cdot \|_r$,  $r\in N$ is the norm in the class
%$S_r$.}
%
\begin{equation}\label{2.18}
| g_r^{(1)}(k,t) |<k^{2l-2-4s_1-\gamma _0r-\delta },
%\end{equation}
%
%\begin{equation}\label{2.19}
 \| G_r^{(1)}(k,t)\| _1<k^{-\gamma _0r},
\end{equation}
where $\gamma _0=2l-2-4s_1-2\delta $.
%The operator $G_r^{(1)}$ is
%nonzero only on the finite-dimensional subspace $(\sum_{i:| i-j|
%<rR_0}E_i)l_2^2.$
\end{Thm}
\begin{Cor}\label{C:2.4}
For the perturbed eigenvalue and its spectral projection, the
following estimates hold:
\begin{equation}\label{2.20}
\bigl| \lambda_j^{(1)} (\alpha ,t)-p_j^{2l}(t)\bigr| \leq 2\alpha
^2k^{2l-2-4s_1-2\gamma _0-\delta },
\end{equation}
\begin{equation}\label{2.21}
\bigl\|E_j^{(1)} (\alpha ,t)-E_j\bigr\|_1\leq 2 | \alpha |
k^{-\gamma _0}.
\end{equation}
\end{Cor}
%

%%%%%%%%%\begin{Rem} \label{R:May10} The theorem states that $\lambda_j^{(1)} (\alpha ,t)$ is a
%%%%%%%%%single eigenvalue in the interval $\varepsilon_1 (k,\delta )$. This
%%%%%%%%%means that $|\lambda_j^{(1)} (\alpha ,t)-k^{2l}|<k^{2l-2-4s_1-\delta
%%%%%%%%%}$. Formula (\ref{2.20}) provides a stronger estimate on the
%%%%%%%%%location of $\lambda_j^{(1)} (\alpha ,t)$, the right-hand side of
%%%%%%%%%(\ref{2.20}) being smaller than the size of $\varepsilon
%%%%%%%%%_1$.\end{Rem}

%%%%%%%%%%Next, we show that  the series~(\ref{2.16}), ~(\ref{2.17}), can be
%%%%%%%%%extended as holomorphic functions of $t$ in a complex neighborhood
%%%%%%%%%of $\chi _1$; they can be differentiated  with respect to $t$ any
%%%%%%%%%number of times with retaining their asymptotic character.

Let us introduce the notations:
$$T(m):=\frac{\partial ^{|m|}}{\partial t_1^{m_1} \partial t_2^{m_2}},\ \ \
m=(m_1,m_2),\ \ \ |m| :=m_1+m_2,\ \ \  m ! :=m_1 ! m_2 !, \ \ \
 T(0)f  :=f.$$
We show, in \cite{19}, that  the coefficients $g_r^{(1)}(k,t)$ and $G_r^{(1)}(k,t)$
can be extended as holomorphic functions of two variables from the real
$(2k^{-1-4s_1-2\delta})$-neighborhood of the non-resonance set $\chi_1(\lambda, \delta)$
 to its complex neighborhood of the same size and the following estimates hold in the complex neighborhood:
%%%%%%%%\begin{equation}\label{2.43}
$$|T(m)g_r^{(1)}(k,t)| <m!k^{2l-2-4s_1-\delta -\gamma _0r+| m|
(1+4s_1+2\delta )},
%%%%%%%%\end{equation}
%
%
%%%%%%%%\begin{equation}\label{2.44}
\| T(m)G_r^{(1)}(k,t)\|_1 <m!k^{-\gamma _0 r+| m| (1+4s_1+2\delta
)}.$$
%%%%%%%\end{equation}
%%%%%%\end{Lem}
From this, the following theorem follows easily.

\begin{Thm}\label{T:2.5}
%%%%%%Under the conditions of Theorem \ref{T:2.3}
The series~(\ref{2.16}),~(\ref{2.17}) can be extended as
holomorphic functions of two variables to the complex
$(2k^{-1-4s_1-2\delta })$-neighborhood of the non-resonance set
$\chi_1$ from its neighborhood in $K_1$, and the following
estimates hold
in the complex $(2k^{-1-4s_1-2\delta })$-neighborhood of the non-resonance set
$\chi_1$:
%%%%%\end{Thm}
%
%%%%%%\begin{Cor}\label{C:2.6}
%%%%%The following estimates hold  for the perturbed eigenvalue and its
%%%%%spectral projection:
\begin{equation}\label{2.45}
| T(m)(\lambda_j^{(1)} (\alpha ,t)-p_j^{2l}(t))| <2m!\alpha
^2k^{2l-2-4s_1- 2\gamma _0-\delta +| m| (1+4s_1+2\delta )},
\end{equation}
\begin{equation}\label{2.46}
\| T(m)(E_j^{(1)} (\alpha ,t)-E_j)\|_1 <2m!\alpha k^{-\gamma
 _0+|m| (1+4s_1+2\delta )}.
\end{equation}
\end{Thm}
%%%%%%%%%\begin{Cor}
%%%%%%%%%\begin{equation}\label{2.20*}
%%%%%%%%%\nabla \lambda_j^{(1)} (\alpha ,t)=p_j^{2l-1}(t)\vec
%%%%%%%%%p_j(t)\bigl(1+o(1)\bigr).
%%%%%%%%%\end{equation}
%%%%%%%%%\begin{equation}
%%%%%%%%% T(m)\lambda_j^{(1)} (\alpha ,t)<4l^2k^{2l-2},\ \ \ \mbox{if} \ \ \
%%%%%%%%%|m|=2 \label{2.20a}
%%%%%%%%%\end{equation}
%%%%%%%%%\end{Cor}
\subsubsection{Nonresonance Part of Isoenergetic Set of $H^{(1)}$.}

Let $S_1(\lambda )$\footnote{$S_1(\lambda )$ definitely depends on
$\alpha W_1$; however we omit this to keep the notation simple.}
be the isoenergetic set of the operator $H_{\alpha}^{(1)}$, i.e.,
    \begin{equation}\label{2.65.1}
    S_1(\lambda )=\{t \in K_1 : \exists n \in \N: \
    \lambda_n^{(1)}(\alpha,t)=\lambda  \},
    \end{equation}
where $\{\lambda_n^{(1)}(\alpha,t)\}_{n=1}^{\infty }$ is the
complete set of eigenvalues of $H_{\alpha}^{(1)}(t)$.
We construct a ``non-resonance" subset  $\chi _1^*(\lambda )$ of $S_1(\lambda)$, which
corresponds to  non-resonance eigenvalues $\lambda_{j}^{(1)}(\alpha, t)$
given by the perturbation series (\ref{2.16}).
Note that for every $t$ belonging to the non-resonant set
$\chi _1(\lambda,  \delta )$ described by \ref{L:2.1}, there is a single $j\in \Z^2$ such
that $p_j(t)=k$, $k=\lambda ^{1/(2l)}$. This means that the
function $t\to \vec p_j(t)$ maps $\chi _1(\lambda,  \delta )$ into
the circle $S_k$. We denote the image of $\chi
_1(\lambda, \delta )$ in $S_k$  by $\cal{D}_0(\lambda )_{nonres}$.
Obviously,
\begin{equation} \chi _1(\lambda,  \delta )={\cal
K_1}\cal{D}_0(\lambda)_{nonres},\label{p}
\end{equation}
 where
${\cal K_1}$ establishes a one-to-one relation between two sets. Let
$\cal{B}_1(\lambda)$ be a set of unit vectors corresponding to
$\cal{D}_0(\lambda)_{nonres}$,
$$\cal{B}_1(\lambda)=\{\vec{\nu} \in S_1 :
k \vec{\nu} \in \cal{D}_0(\lambda)_{nonres}\}.$$ It is easy to see
that $\cal{B}_1(\lambda )$   is a unit circle with holes, centered
at the origin. We denote by $\Theta _1(\lambda)$ the set of angles $\varphi$,
 corresponding to $\cal{B}_1(\lambda)$:
%%%%%%%\begin{equation}
$$\Theta _1(\lambda)=\{\varphi \in [0,2\pi ):\ (\cos \varphi ,\sin \varphi
)\in \cal{B}_1(\lambda)\}. $$
%%%%%\label{May17a}
%%%%%%% \end{equation}
%%%%%By part 4 of the Geometric Lemma, the length of each connected
%%%%%%component of the non-resonance set $\chi _1(\lambda, s_1, \delta )$
%%%%%is at least $k^{-1-2s_1-\delta }$. Obviously, it is true also for
%%%%%$\cal{D}_0(\lambda)_{nonres}$. Hence, the length of each connected
%%%%%component of $B_1$ is at least $k^{-2-2s_1-\delta }$ and $\Theta
%%%%%_1 $ consists of a finite number of intervals of the length at
%%%%%least $k^{-2-2s_1-\delta }$.

Let $\vec{\varkappa } \in \cal{D}_0(k)_{nonres}$.\footnote{Usually the
vector $\vec p_j(t)$ is denoted by $\vec k$, the corresponding
plane wave being $e^{ \langle \vec k, \vec x \rangle}$. We use
less common notation $\vec{\varkappa }$, since we already have
other $k$'s in the text.} Then, there exists
$j\in \Z^2$, $t\in \chi _1(\lambda ,\delta )$
such that $\vec{\varkappa }=\vec p_j(t)$. Obviously, $t={\cal K}_1\vec{\varkappa }$
and, by \eqref{p}, $t \in \chi_1(\lambda, \delta)$.
According to Theorem \ref{T:2.3}, for sufficiently large $k$,
there exists an eigenvalue of the operator $H^{(1)}_{\alpha}(t)$,
$t={\cal K_1} \vec \varkappa $, $0 \leq \alpha \leq 1$, given by
(\ref{2.16}). It is convenient here to denote $\lambda
_j^{(1)}(\alpha, t)$ by $\lambda^{(1)}(\alpha,\vec{\varkappa})$;
we can do this since there is a one-to-one correspondence between
$\vec{\varkappa}$ and the pair $(t,j)$. We rewrite (\ref{2.16}) and \eqref{2.45} in
the forms
    \begin{equation}\label{2.66}
    \lambda^{(1)}(\alpha,\vec{\varkappa})=\varkappa ^{2l}+f_1(\alpha,\vec{\varkappa}),
    \ \ \ \ \ \varkappa =|\vec \varkappa |,
    \end{equation}
  %%%%%%%%  \begin{equation}\label{2.67}
  %%%%%%%%  |f_1(\alpha,\vec \varkappa)|\leq 2 \alpha^2k^{2l-2-4s_1-2\gamma_0-\delta},
  %%%%%%%%  \end{equation}
    \begin{equation}\label{2.67a}
    |T(m)f_1(\alpha,\vec \varkappa)|\leq 2 m!
    \alpha^2\varkappa^{2l-2-4s_1-2\gamma_0-\delta+|m|(1+4s_1+2\delta)}.
\end{equation}
where
%%%%%%%%$\gamma_0=2l-2-4s_1-2\delta $,
$f_1(\alpha,\vec{\varkappa})=\sum _{r=2}^{\infty}
\alpha^{r}g_r^{(1)}(\vec{\varkappa})$,
%%%%%%%%%\begin{equation}f_1(\alpha,\vec{\varkappa})=\sum _{r=2}^{\infty}
%%%%%%%%%\alpha^{r}g_r^{(1)}(\vec{\varkappa}),\label{july5b}
%%%%%%%%%\end{equation}
$g_r^{(1)}(\vec{\varkappa})$ being defined by (\ref{2.11}) with $j$
and $t$ such that $\vec{p}_j(t)=\vec{\varkappa}$. By Theorem
\ref{T:2.3}, the formulas (\ref{2.66}), (\ref{2.67a}) hold in
$(2k^{-1-4s_1-2\delta})$-neighborhood of $\cal{D}_0(\lambda
)_{nonres}$, i.e., they hold for any $\varkappa \vec{\nu}$ such that
$\vec{\nu} \in \cal{B}_1(\lambda),\ |\varkappa
-k|<2k^{-1-4s_1-2\delta}$. We define $\cal{D}_1(\lambda )$ as the
level set of the function $\lambda^{(1)}(\alpha,\vec{\varkappa})$ in
this neighborhood:
    \begin{equation}\label{2.68}
    \cal{D}_1(\lambda ):=\{\vec{\varkappa _1 }=\varkappa _1 \vec{\nu}:
    \vec{\nu}\in
    \cal{B}_1(\lambda), \  |\varkappa _1-k|<2k^{-1-4s_1-2\delta},\
    \lambda^{(1)}(\alpha,\vec{\varkappa} _1)=\lambda \}.
    \end{equation}

\begin{Lem}\label{L:2.13}
\begin{enumerate}

\item For sufficiently large $\lambda$, the set $\cal{D}_1(\lambda )$ is a distorted circle with
holes which is strictly inside the circle of the radius $k$ (see Fig. \ref{F:1});
it can be described by the formula
\begin{equation}
\cal{D}_1(\lambda )=\bigl\{\vec \varkappa _1\in \R^2: \vec
\varkappa _1=\varkappa _1(\varphi)\vec \nu,\ \ \vec{\nu}=(\cos \varphi,\sin \varphi)
\in \cal{B}_1(\lambda)\bigr\},\label{May20}
\end{equation} where
    $ \varkappa_1 (\varphi)=k+h_1(\varphi) $
and $h_1(\varphi)$ obeys the inequalities
    \begin{equation}\label{2.75}
    |h_1|<k^{-1-4s_1-2\gamma_0-\delta},
%    \end{equation}
%    \begin{equation}\label{2.76}
    \left|\frac{\partial h_1}{\partial \varphi} \right| \leq
    k^{-2\gamma_0+1+\delta},
    \end{equation}
$h_1(\varphi) <0$ when $W_1 \neq 0$.

\item The total length of $\cal B _1(\lambda)$ satisfies the estimate
    \begin{equation}\label{B_1}
    L(\cal B_1)=2\pi (1+O(k^{-\delta /2})).
    \end{equation}

\item The function $h_1(\varphi)$ can be extended as a
holomorphic function of $\varphi $ to the complex
$(2k^{-2-4s_1-2\delta })$-neighborhood of each connected component of
$\Theta _1(\lambda)$ and  estimates (\ref{2.75}) hold.

\item The curve $\cal{D}_1(\lambda)$ has a length which is
asymptotically close to that of the whole circle in the
sense that
    \begin{equation}\label{2.77}
     L\bigl(\cal{D}_1(\lambda)\bigr)\underset{\lambda \rightarrow
     \infty}{=}2\pi k\bigl(1+O(k^{-\delta/2})\bigr),\quad \lambda=k^{2l}.
     \end{equation}
     \end{enumerate}
%where $\omega_n$ is the surface area of a unit sphere in $\R^n$.
\end{Lem}
%It can be easily shown that $\lambda _1(\vec \varkappa )>|\vec
%\varkappa |^{2l}$ (since $g_2^{(1)}(\vec{\varkappa})>0$).
%Therefore, $\cal{D}_1(\lambda)$ is strictly inside the perfect
%circle.

\begin{figure}
\centering
\includegraphics[totalheight=.2\textheight]{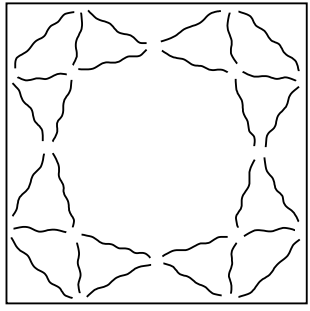}
\caption{The  set $\chi ^*_1(\lambda)$}\label{F:5}
\end{figure}

Next, we define the non-resonance subset $\chi_1^*(\lambda )$ of
 isoenergetic set  $S_1(\lambda )$ as the parallel shift of
 $\cal{D}_1(\lambda)$ into $K_1$ (Fig. \ref{F:5}):
    \begin{equation}\label{2.81}
    \chi_1^*(\lambda ):=\cal{K}_1\cal{D}_1(\lambda ).
    \end{equation}

\begin{Lem} \label{Apr4}
The set $\chi_1^*(\lambda )$ belongs to the
$(k^{-1-4s_1-2\gamma _0-\delta })$-neighborhood of $\chi_1(\lambda )$
in $K_1$. If $t\in \chi_1^*(\lambda )$, then the operator
$H^{(1)}_{\alpha }(t)$ has a simple eigenvalue $\lambda _n^{(1)}(\alpha ,t)$, $n\in \N$,
equal to $\lambda $, no other eigenvalues being in the interval
$\varepsilon_1 (k,\delta )$, $\varepsilon_1 (k,\delta ):=
(k^{2l}-k^{2l-2-4s_1-\delta }, k^{2l}+k^{2l-2-4s_1-\delta })$. This
eigenvalue is given by the perturbation series (\ref{2.16}), where
$j $ is uniquely defined by $t$ from the relation
$p_j^{2l}(t)\in \varepsilon_1 (k,\delta )$.
\end{Lem}

\begin{Lem}\label{L:May10a}
The formula (\ref{2.81}) establishes
a one-to-one correspondence between $\chi_1^*(\lambda )$ and $\cal{D}_1(\lambda )$.
\end{Lem}

From the geometric point of view, this means that $\chi_1^*(\lambda )$ does not have self-intersections.

\subsection{The Second Step of Approximation.}
\subsubsection{The Operator $H_{\alpha}^{(2)}$.}

%%%%%%%%From now on we denote $H_1^{(1)}$ ($\alpha=1$) just by $H_1$.
%%%%Choosing $s_2$ such that
%%%%%.
 %%%%%   \begin{equation}\label{3.0}
%%%    2s_1 \leq s_2<2l-11-14s_1
%%%5    \end{equation}
Choosing $s_2=2s_1$, we define the second operator
$H_{\alpha}^{(2)}$ by the formula
    \begin{equation}\label{3.1}
     H_{\alpha }^{(2)}=H^{(1)}+\alpha W_2,\quad     ( 0\leq \alpha \leq
     1), \qquad
     W_2=\sum_{r=M_1+1}^{M_2}V_r,
     \end{equation}
where $H^{(1)} $ is defined by (\ref{2.1}) and $M_2$ is chosen in such
a way that $2^{M_2} \approx k^{s_2}$. Obviously, the periods of
$W_2$ are $2^{M_2-1} (\beta _1,0)$ and $2^{M_2-1} (0,\beta _2)$. We
write them in the form $N_1(a_1,0)$ and $N_1(0,a_2)$, where
$(a_1,0),\ (0,a_2)$ are the periods of $W_1$ and $N_1=2^{M_2-M_1},\ \frac{1}{4}k^{s_2-s_1}<N_1<
4k^{s_2-s_1}$. Note that
    \begin{equation}\|W_2\|_{\infty} \leq \sum_{n=M_1+1}^{M_2}
    \|V_n\|_{\infty} \leq \sum_{n=M_1+1}^{M_2}\exp(-2^{\eta n})
    <\exp(-k^{\eta s_1}). \label{W2}
    \end{equation}

\subsubsection{Multiple Periods of $W_1(x)$. }\label{S:2.2}

 The operator $H^{(1)}=H_0 +W_1(x)$
%
 %   \begin{equation}\label{3.2}
 %   H^{(1)}=H_0 + W_1(x),
 %   \end{equation}
%
has the periods $a_1,\ a_2$. The corresponding family of operators,
$\{H^{(1)}(t)\}_{t \in K_1}$, acts in $L_2(Q_1)$, where $Q_1=[0,a_1]
\times [0,a_2]$ and $K_1=[0, 2\pi/a_1)\times [0, 2\pi/a_2)$.
The eigenvalues of $H^{(1)}(t)$ are denoted by $\lambda_n^{(1)}(t)$, $n
\in \N$, and its spectrum by $\Lambda ^{(1)}(t)$. Now let us
consider the same $W_1(x)$ as a periodic function with the periods
$N_1a_1,\ N_1a_2$. Obviously, the definition of the operator
$H^{(1)}$ does not depend on how we define the periods of
$W_1$. However, the family of operators $\{H^{(1)}(t)\}_{t \in K_1}
$ does change when we replace the periods $a_1,\ a_2$
by $N_1a_1,\ N_1a_2$. The family of operators $\{H^{(1)}(t)\}_{t \in K_1}$ has to
be  replaced by a family of operators $\{ \tilde{H}_1(\tau) \}_{\tau
\in K_2}$ acting in $L_2(Q_2)$, where $Q_2=[0,N_1a_1] \times
[0,N_1a_2]$ and $K_2=[0, 2\pi/N_1a_1)\times [0, 2\pi/N_1a_2)$. We
denote the eigenvalues of $\tilde{H}_1(\tau)$ by
$\tilde{\lambda}_n^{(1)}(\tau)$, $n \in \N$, and its spectrum by
$\tilde{\Lambda}^{(1)}(\tau)$. The next lemma establishes a
connection between spectra of the operators $H^{(1)}(t)$ and
$\tilde{H}_1(\tau)$. It follows easily from Bloch theory (see, e.g.,
\cite{RS}).
\begin{Lem}\label{L:3.1}
For any $\tau \in K_2$,
    \begin{equation}\label{3.3}
    \tilde{\Lambda}^{(1)}(\tau)=\bigcup_{p \in P} \Lambda^{(1)}(t_p),
    \end{equation}
 where \begin{equation}
  P=\{p=(p_1,p_2) \in \Z^2 : 0 \leq p_1 \leq N_1-1,\ 0 \leq p_2
\leq N_1-1\} \label{May10a}
\end{equation}
 and $t_p=(t_{p,1}, t_{p,2})=(\tau _1 +2\pi
p_1/N_1a_1, \tau_2+2\pi p_2/N_1a_2) \in K_1$. See Fig. 6.
\end{Lem}
    \begin{figure}\label{F:6}
\centering
    \psfrag{2p/a1}{\hspace{2mm}\small{$\frac{2\pi}{a_1}$}} \psfrag{2p/a2}{\small{$\frac{2\pi}{a_2}$}}
    \psfrag{2p/Na1}{\small{$\frac{2\pi}{N_1a_1}$}}
    \psfrag{2p/Na2}{\hspace{2mm}\small{$\frac{2\pi}{N_1a_2}$}}
\includegraphics[totalheight=.2\textheight]{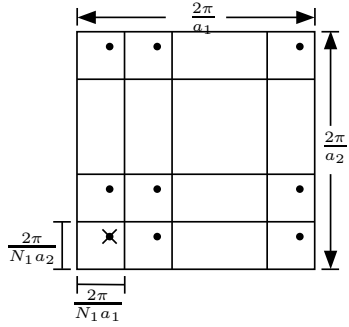}
 \caption{Relation between $\tau$ and $t_p$}
\end{figure}

%proof of lemma 3.1
We defined the isoenergetic set $S_1(\lambda ) \subset K_1$ of
$H^{(1)}$ by formula (\ref{2.65.1}). Obviously, this definition is
directly associated with the family of operators $H^{(1)}(t)$ and,
therefore, with the periods $a_1, a_2$, which we assigned to
$W_1(x)$. Now, assuming that the periods are equal to $N_1a_1,
N_1a_2$, we give an analogous definition of the isoenergetic set
$\tilde{S}_1(\lambda )$ in $K_2$:
    %\begin{equation}\label{3.4}
    $$\tilde{S}_1(\lambda ):=\{\tau \in K_2: \exists n \in \N:\ \
    \tilde{\lambda}_n^{(1)}(\tau)=\lambda \}.$$
    %\end{equation}
By Lemma \ref{L:3.1}, $\tilde{S}_1(\lambda )$ can be expressed as
follows:
    %\begin{equation}\label{3.5}
    $$\tilde{S}_1(\lambda )=\bigl\{\tau \in K_2: \exists n \in \N,\ p \in P:\
    \lambda_n^{(1)}\bigl(\tau+2\pi p/N_1a\bigr)=\lambda \bigr\},
  %  \end{equation}
  %%%  \begin{equation}
 \ \ \    2\pi
    p/N_1a=\left(\frac{2\pi p_1}{N_1a_1}, \frac{2\pi
    p_2}{N_1a_2}\right). $$
    %\end{equation}
The relation between $S_1(\lambda)$ and $\tilde{S}_1(\lambda)$ can be easily
understood from the geometric point of view as
    $\tilde{S}_1=\cal{K}_2 S_1, $
where $\cal{K}_2$ is the parallel shift into $K_2$, i.e.,
    %\begin{equation}\label{3.5.2}
    $$\cal{K}_2:\R^2 \rightarrow K_2,\   \cal{K}_2(\tau+2\pi
    m/N_1a)=\tau,\ m \in \Z^2,\ \tau \in K_2.$$
    %\end{equation}
Thus, $\tilde{S}_1(\lambda )$ is obtained from $S_1(\lambda )$ by
cutting $S_1(\lambda )$ into pieces of the size $K_2$ and shifting them
together in $K_2$.

\begin{Def} \label{D:May6}
We say that $\tau $ is a point of
self-intersection of $\tilde{S}_1(\lambda )$,  if there is a
pair $n,\hat n\in \N$, $n\neq \hat n$ such that $\tilde{\lambda}_n^{(1)}(\tau)=
\tilde{\lambda}_{\hat n}^{(1)}(\tau)=\lambda $.
\end{Def}

\begin{Rem} \label{R:May6}
By Lemma \ref{L:3.1}, $\tau $ is a point of
self-intersection of $\tilde{S}_1(\lambda )$,  if there is a
pair $p,\hat p\in P$ and a pair $n,\hat n \in \N$ such that
$|p-\hat p|+|n-\hat n|\neq 0$ and $\lambda_n^{(1)}(\tau+2\pi p/N_1a)=
\lambda_{\hat n}^{(1)}(\tau+2\pi \hat p/N_1a)=\lambda $.
\end{Rem}

Now let us recall that the isoenergetic set $S_1(\lambda )$ consists
of two parts: $\chi_1^*(\lambda )$ and
$S_1(\lambda)\setminus~\chi_1^*(\lambda )$,
   %%%%%%%% $$S_1(\lambda )=\chi_1^*(\lambda )\cup \bigl(S_1(\lambda )
   %%%%%%%% \setminus \chi_1^*(\lambda ) \bigr),$$
where $\chi_1^*(\lambda )$ is the first non-resonance set given by
(\ref{2.81}). Obviously $\cal{K}_2\chi_1^*(\lambda )\subset
\cal{K}_2S_1(\lambda)=\tilde S_1(\lambda )$ and can be described
by the formula: \begin{equation}\cal{K}_2\chi_1^*(\lambda
)=\left\{\tau \in K_2:\ \exists p\in P: \tau +2\pi p/N_1a \in
\chi_1^*(\lambda )\right\}.\label{May6}
\end{equation}
Let us  consider only those self-intersections of $\tilde{S}_1$
which belong to $\cal{K}_2\chi_1^*(\lambda )$, i.e., we consider the
points of intersection of $\cal{K}_2\chi_1^*(\lambda )$ both with
itself and with $\tilde S_1(\lambda )\setminus
\cal{K}_2\chi_1^*(\lambda )$.%%%% Combining Remark \ref{R:May6} of
 %%%%with  formula (\ref{May6}) and Lemma \ref{Apr4}, we
%%%%5obtain the following result.
%%%%\begin{Lem}\label{L:May6} A self-intersection $\tau $ of $\tilde S_1(\lambda )$
%%%%belongs to $\cal{K}_2\chi_1^*(\lambda )$ if and only if there is a
%%%%pair $p,\hat p\in P$, $p\neq \hat p$ and a pair $n,\hat n\in N$ such
%%%%that $\tau +2\pi p/N_1a \in \chi_1^*(\lambda )$ and $\lambda
%%%%_n^{(1)}(\tau +2\pi p/N_1a)=\lambda _{\hat n}^{(1)}(\tau +2\pi \hat
%%%%p/N_1a)=\lambda$, the eigenvalue $\lambda _n^{(1)}(\tau +2\pi
%%%%p/N_1a)$ being given by the series (\ref{2.16}) with $t=\tau +2\pi
%%%%p/N_1a $ and
%%%%    $j $  uniquely defined by $t$ from the relation
%%%%    $p_j^{2l}(t)\in \varepsilon _1$, $\varepsilon _1=
%%%%    (k^{2l}-k^{2l-2-4s_1-\delta }, k^{2l}+k^{2l-2-4s_1-\delta })$.
%%%%    \end{Lem}

To obtain a new non-resonance set $\chi_2(\lambda )$, we remove from
$\cal{K}_2\chi_1^*(\lambda )$ a  neighborhood of its
self-intersections with$\tilde{S}_1(\lambda )$. More precisely, we
remove from $\cal{K}_2\chi_1^*(\lambda )$ the set
    \begin{equation}\label{omega}
    \begin{split}
     \Omega_1(\lambda )=\bigl\{\tau \in \cal{K}_2\chi_1^*(\lambda ): \exists n,\hat{n} \in
    \N,\ p,\hat{p} \in P,\  p\neq \hat{p}:\ \lambda_{n}^{(1)}(\tau+2\pi
    p/N_1a)=\lambda ,\\
    \tau+2\pi p/N_1a \in \chi_1^*(\lambda ),\
     \left|\lambda_{n}^{(1)}(\tau+2\pi p/N_1a)-\lambda_{\hat{n}}^{(1)}(\tau+2\pi
    \hat{p}/N_1a)\right| \leq \epsilon_1  \bigr\},\quad \epsilon_1 =e^{-\frac{1}{4}k^{\eta
    s_1}}.
    \end{split}
    \end{equation}
We define $ \chi_2(\lambda )$  by the formula
    \begin{equation}\label{3.5.1}
     \chi_2(\lambda )=\cal{K}_2 \chi _1^*(\lambda )\setminus \Omega _1(\lambda ).
    \end{equation}

\subsubsection{Perturbation Formulas.}\label{S:3.3}

%%%%%%%Before proving the main result, we formulate the Geometric Lemma:
%
\begin{Lem}[Geometric Lemma]\label{L:3.2}
For an arbitrarily small positive $\delta $, $9\delta <2l-11-16s_1$
and sufficiently large $\lambda $, $\lambda
>\lambda _1(V,\delta)$, there exists a non-resonance set $\chi
_2(\lambda ,\delta ) \subset \cal{K}_2\chi_1^*$ such that the following hold.
\begin{enumerate}
\item For any $\tau \in \chi _2$,
    \begin{enumerate}
    \item there exists a unique $p \in P$
    such that $\tau + 2\pi p/N_1a \in \chi_1^*$;
    \footnote{From the geometric point of view, this means that $\chi _2(\lambda)$
    does not have self-intersections.}
    \item
    $ \lambda _j^{(1)}(\tau + 2\pi p/N_1a)=k^{2l},$
    where $\lambda _j^{(1)}(\tau + 2\pi p/N_1a)$ is
      given  by the perturbation series
    (\ref{2.16}) with $\alpha
    =1$, $j$ being uniquely defined by $t=\tau + 2\pi p/N_1a $ as is described
    in Part 2 of
    Lemma \ref{L:2.1}.
    \item The eigenvalue $\lambda _j^{(1)}(\tau + 2\pi p/N_1a)$ is
    a simple eigenvalue of $\tilde H^{(1)}(\tau )$,
    whose distance from all other eigenvalues $\lambda _{\hat n}^{(1)}(\tau + 2\pi \hat{p}/N_1a),\
    {\hat n} \in
    \N$, of $\tilde{H}_1(\tau )$ is greater than $\epsilon_1 = e^{-\frac{1}{4}k^{\eta s_1}}$:
    \begin{equation}\label{3.6}
    |\lambda _j^{(1)}(\tau + 2\pi p/N_1a)-\lambda _{\hat n}^{(1)}(\tau + 2\pi
    \hat{p}/N_1a)|>\epsilon_1.
    \end{equation}
    \end{enumerate}

\item For any $\tau$ in the $(\epsilon_1 k^{-2l+1-\delta })$-neighborhood
in $\C^2$ of $\chi_2$, there exists a unique $p \in
P$ such that $\tau +2\pi p/N_1a$ is in the $(\epsilon_1
k^{-2l+1-\delta })$-neighborhood in $\C^2$ of $\chi_1^*$ and
    \begin{equation}\label{3.7}
    | \lambda_j^{(1)}(\tau +2\pi p/N_1a)-k^{2l}| < \epsilon_1 k^{-\delta
    },
    \end{equation}
$j$ being uniquely defined by $\tau + 2\pi p/N_1a $ as is described
in Part 2 of Lemma \ref{L:2.1}.

\item The second non-resonance set $\chi_2$ has asymptotically
full measure in $\chi_1^*$ in the sense that
\begin{equation}\label{3.9}
\frac{L(\cal{K}_2\chi_1^* \setminus \chi
_2))}{L(\chi_1^*)}<k^{-2-2s_1}.
%%%%%%O(e^{-k^{s_1}}).
\end{equation}
%%%%%\item The length of each connected component in the second
%%%%%non-resonance set $\chi_2(k,s_1,\delta,\eta)$ is at least
%%%%%$k^{-1-2s_2-3\delta}$.
%
\end{enumerate}
\end{Lem}

\begin{Rem} \label{R:11}
The dual lattice $2\pi m/N_1a$ ($m\in \Z^2$), corresponding to
larger periods $N_1a_1,\ N_1a_2$, is finer than the dual lattice
$2\pi j/a$ ($j\in \Z^2$), corresponding to $a_1,\ a_2$.
Every point $2\pi m/N_1a$ of a dual lattice corresponding to the periods
$N_1a_1,\ N_1a_2$ can be uniquely represented in the form $2\pi m/N_1a=2\pi j/a+2\pi
p/N_1a$, where $m=N_1j+p$ and $2\pi j/a$ is a point of the dual
lattice for periods $a_1$, $a_2$, while $p\in P$ is responsible
for refining the lattice.

 Let us consider a normalized eigenfunction
$\psi_n(t,x)$ of $H^{(1)}(t)$ in $L_2(Q_1)$. We extended it
quasiperiodically to $Q_2$, renormalize in $L_2(Q_2)$ and denote the
new function by $\tilde{\psi}_n(\tau,x)$, $\tau ={\cal K}_2t$. The
Fourier representations of $\psi_n(t,x)$ in $L_2(Q_1)$ and
$\tilde{\psi}_n(\tau,x)$ in $L_2(Q_2)$ are simply related. If we
denote Fourier coefficients of $\psi_n(t,x)$ with respect to the
basis of exponential functions
    $|Q_1|^{-1/2}e^{i \langle t+2\pi j/a,x \rangle}$, $j \in \Z^2,$
in $L_2(Q_1)$  by $C_{nj}$, then,  the Fourier coefficients
$\tilde{C}_{nm}$ of $\tilde{\psi}_n(\tau,x)$ with respect to the
basis of exponential functions
    $|Q_2|^{-1/2}e^{i(\tau + 2\pi m/N_1a, x)}$, $m \in \Z^2,$
in $L_2(Q_2)$ are given by the formula
    \begin{equation*}
    \tilde{C}_{nm}=
        \begin{cases}
        C_{nj},  &\text{if $m=jN_1+p$;}\\
        0,       &\text{otherwise},
        \end{cases}
    \end{equation*}
    $p$ being defined from the relation
$t=\tau+2\pi p/N_1a,\ p \in P$. Hence, the matrices of the projections
on $\psi_n(t,x)$ and $\tilde{\psi}_n(\tau,x)$ with respect to the
above bases are simply related by
    \begin{equation*}
    (\tilde{E}_n)_{j\hat{j}}=
        \begin{cases}
        (E_n)_{m\hat{m}},  &\text{if $m=jN_1+p,\ \hat{m}=\hat{j}N_1+p$;}\\
        0,       &\text{otherwise},
        \end{cases}
    \end{equation*}
$\tilde{E}_n$ and $E_n$ being the projections in $L_2(Q_2)$ and
$L_2(Q_1)$, respectively.

Let us denote by $\tilde{E}_j^{(1)}\bigl(\tau+2\pi p/N_1a \bigr)$
the spectral projection ${E}_j^{(1)}(\alpha, t)$ (see (\ref{2.17}))
with $\alpha =1$ and $t=\tau+2\pi p/N_1a $, ``extended" from
$L_2(Q_1)$ to $L_2(Q_2)$.
\end{Rem}

\medskip

 By analogy with (\ref{2.11}), (\ref{2.12}), we define functions $g_r^{(2)}(k,\tau)$ and
operator-valued functions $G_r^{(2)}(k,\tau)$, $r=1, 2, \cdots$,
%%%%%%for $ \tau \in \chi _2$
as follows:
\begin{equation}\label{3.13}
 g_r^{(2)}(k,\tau)=\frac{(-1)^r}{2\pi ir}\mbox{Tr}\oint _{C_2}
 \bigl(\bigl(\tilde{H}_1(\tau)-z\bigr)^{-1}W_2\bigr)^rdz,
 \end{equation}
\begin{equation}\label{3.14}
G_r^{(2)}(k,\tau)=\frac{(-1)^{r+1}}{2\pi i}\oint
_{C_2}\bigl(\bigl(\tilde{H}_1(\tau)-z\bigr)^{-1}W_2\bigr)^r
\bigl(\tilde{H}_1(\tau)-z\bigr)^{-1}dz.
 \end{equation}
We consider the operators $H_{\alpha}^{(2)}=H^{(1)}+\alpha W_2$ and
the family $H_{\alpha}^{(2)}(\tau)$, $\tau \in K_2$, acting in
$L_2(Q_2)$.
\begin{Thm}\label{T:3.4}
 Suppose $\tau$ belongs to the
$(\epsilon_1 k^{-2l+1-\delta })$-neighborhood in $K_2$ of the
 second non-resonance set $\chi _2(\lambda ,\delta)$, $0<9\delta
<2l-11-16s_1$, $\epsilon_1=e^{-\frac{1}{4}k^{\eta s_1}}$. Then, for
sufficiently large $\lambda $, $\lambda
>\lambda _1(V,\delta)$ and for all $\alpha $, $0 \leq \alpha \leq 1$,
there exists a single eigenvalue of the operator
$H_{\alpha}^{(2)}(\tau)$ in
the interval $\varepsilon_2 (k,\delta ):= (k^{2l}-\epsilon_1 /2,
k^{2l}+\epsilon_1 /2)$. It is given by the series
\begin{equation}\label{3.15}
\lambda_{\tilde{j}}^{(2)} (\alpha ,\tau)=\lambda_j^{(1)}\bigl(
\tau+2\pi p/N_1a \bigr)+\sum _{r=1}^{\infty }\alpha ^r
g_r^{(2)}(k,\tau),\ \ \ \ \tilde j=j+p/N_1,
\end{equation}
converging absolutely in the disk $|\alpha|  \leq 1$, where $p\in P$
and $j\in Z^2$ are as in Lemma \ref{L:3.2}. The
spectral projection corresponding to $\lambda_{\tilde{j}}^{(2)}
(\alpha ,\tau)$ is given by the series
\begin{equation}\label{3.16}
E_{\tilde{j}}^{(2)} (\alpha ,\tau)=\tilde{E}_j^{(1)}\bigl(\tau+2\pi
p/N_1a \bigr)+\sum _{r=1}^{\infty }\alpha ^rG_r^{(2)}(k,\tau),
\end{equation}
which converges in the trace class $\mathbf{S_1}$ uniformly with
respect to $\alpha $ in the disk  $| \alpha | \leq 1$.

The following estimates hold for coefficients $g_r^{(2)}(k,\tau)$,
$G_r^{(2)}(k,\tau)$, $r\geq 1$:
%\footnote{Here and below $\| \cdot \|_r$,  $r\in N$ is the norm in the class
%$\cal{D}_r$.}
%
\begin{equation}\label{3.17}
\bigl| g_r^{(2)}(k,\tau)
\bigr|<\frac{3\epsilon_1}{2}(4\epsilon_1^3)^r, \ \ \ \ \bigl\|
G_r^{(2)}(k,\tau)\bigr\| _1< 6r (4 \epsilon_1^3)^r.
\end{equation}
%
%%%%%%%%\begin{equation}\label{3.18}
%%%%%%%%\bigl\| G_r^{(2)}(k,\tau)\bigr\| _1< 6r (4 \epsilon_1^3)^r.
%%%%%%%%\end{equation}
%

%The operator $G_r^{(1)}$ is
%nonzero only on the finite-dimensional subspace $(\sum_{i:| i-j|
%<rR_0}E_i)l_2^2.$
\end{Thm}
\begin{Cor}\label{C:3.5}
The following estimates hold for the perturbed eigenvalue and its
spectral projection:
\begin{equation}\label{3.19}
\bigl| \lambda_{\tilde{j}}^{(2)} (\alpha ,\tau)-\lambda_j^{(1)}
\bigl(\tau+2\pi p/N_1a\bigr)\bigr| \leq  12 \alpha \epsilon_1 ^4,
\end{equation}
\begin{equation}\label{3.20}
\bigl\|E_{\tilde{j}}^{(2)} (\alpha
,\tau)-\tilde{E}_j^{(1)}\bigl(\tau+2\pi p/N_1a\bigr) \bigr\|_1\leq
48 \alpha \epsilon_1^3 .
\end{equation}
\end{Cor}
%
%%%%%%%\begin{Rem} \label{R:May10a} The theorem states that
%%%%%%%$\lambda_{\tilde{j}}^{(2)} (\alpha ,\tau )$ is a single eigenvalue
%%%%%%%in the interval $\varepsilon _2 (k,\delta )$. This means that
%%%%%%%$\bigl|\lambda_j^{(1)} (\alpha ,t)-k^{2l}\bigr|<\epsilon _1/2$.
%%%%%%%Formula (\ref{3.19}) provides a stronger estimate on the location of
%%%%%%%$\lambda_j^{(1)} (\alpha ,t)$.\end{Rem}
%{\em Remark 2.1.} Here we discuss only the selfadjoint case, and
%hence all arguments are carried out only for real $\alpha , -1\leq
%\alpha \leq 1$. Actually, the majority of the formulas are
%preserved also for complex $\alpha $, although completely new
%spectral phenomena arise here which we shall discuss in detail
%elsewhere.
The proof of Theorem \ref{T:3.4} is analogous to that of Theorem \ref{T:2.3}
and is based on expanding the resolvent
$(H_{\alpha}^{(2)}(\tau)-z)^{-1}$ in a perturbation series for $z
\in C_2 $, $C_2$ being the contour around the unperturbed eigenvalue
$k^{2l}$:
    $C_2=\{ z: |z-k^{2l}| =\frac{\epsilon_1}{2} \}. $
Integrating the resolvent yields the formulas for an eigenvalue of
$H_{\alpha}^{(2)}$ and its spectral projection.
%%%%\begin{Lem}\label{T:3.6} The following estimates
%%%%hold for the coefficients $g_r^{(2)}(k,\tau)$ and
%%%%$G_r^{(2)}(k,\tau)$ in the complex $(\frac{1}{2}\epsilon_1
%%%%k^{-2l+1-\delta })$-neighborhood  of the non-resonance set:
%%%%$\chi_2$,:
%
%%%%    \begin{align}
%%%%    |T(m)g_r^{(2)}(k,\tau)| &< m!\cdot 3 \cdot 2^{2r-1+|m|}
%%%%    \epsilon_1^{3r+1-|m|}k^{|m|(2l-1+\delta)},\label{3.39}\\
%%%%    \| T(m)G_r^{(2)}(k,\tau)\|_1 &<
%%%%    m!\cdot 3r \cdot 2^{2r+1+|m|} \epsilon_1^{3r-|m|} k^{|m|(2l-1+\delta)}.\label{3.40}
%%%%    \end{align}
%%%%\end{Lem}
\begin{Thm}\label{T:3.5a}
%%%%%%Under the conditions of Theorem \ref{T:2.3}
Under the conditions of Theorem \ref{T:3.4}, the
series~(\ref{3.15}),~(\ref{3.16})  can be extended as holomorphic
functions of $\tau$ in the complex
$(\frac{1}{2}\epsilon_1k^{-2l+1-\delta})$-neighborhood of the non-resonance set $\chi_2$
 and the following
estimates hold in the complex neighborhood:
%%%%%\end{Thm}
%
%%%%%%\begin{Cor}\label{C:2.6}
%%%%%%The following estimates hold  for the perturbed eigenvalue and its
%%%%%%spectral projection:
\begin{align}
    \left| T(m)\left(\lambda_{\tilde{j}}^{(2)} (\alpha ,\tau)
    -\lambda_j^{(1)}(\tau+2\pi p/N_1a)\right)\right| &<
   \alpha C_m\epsilon_1^{4-|m|}k^{|m|(2l-1+\delta)},
    \label{3.41}\\
   \left \| T(m)\left(E_{\tilde{j}}^{(2)} (\alpha ,\tau)-\tilde{E}_j^{(1)}(\tau+2\pi p/N_1a)
   \right)\right\|_1
    &< \alpha C_m\epsilon_1^{3-|m|}k^{|m|(2l-1+\delta)}, \ \ \
    C_m=48m!2^{|m|}.
    \label{3.42}
    \end{align}
\end{Thm}
%%%%\begin{Cor}
%%%%\begin{equation}\label{2.20**}
%%%%\nabla \lambda_j^{(2)} (\alpha ,\tau )=p_j^{2l-1}(\tau +2\pi
%%%%p/N_1a)\vec p_j(\tau +2\pi p/N_1a)\bigl(1+o(1)\bigr).
%%%%\end{equation}
%%%%\begin{equation}
%%%%T(m)\lambda_j^{(2)}(\alpha ,\tau )=O\bigl(k^{2l-2}\bigr)\ \ \ \
%%%%\mbox{if}\ \ \ |m|=2. \label{2.20**a}
%%%%\end{equation}
%%%%\end{Cor}
\subsubsection{Sketch of the Proof of the Geometric Lemma \ref{L:3.2}.}\label{S:3.2}
Parts 1 and 2 of Geometric Lemma \ref{L:3.2} easily follow from the definition of the
non-resonance set. The main problem is to prove that the
non-resonance set exists and is rather extensive, i.e., Part 3. We
outline a proof of Part 3 below.

 {\bf Determinants. Intersections
and Quasi-intersections. Description of the set $\Omega _1$ in
terms of determinants.} We have considered self-intersections of $\tilde
S_1(\lambda)$ belonging to ${\cal K}_2\chi _1^*$.
 %%%%%   (see
%%%%%      Definition \ref{D:May6} and Lemma \ref{L:May6}).
We describe self-intersections as  zeros of
determinants of  operators of the type $I+A$, $A \in \mathbf{S_1}$. (see, e.g., \cite{RS}).
Let us represent  the
operator $(H^{(1)}(t)-\lambda )(H_0(t)+\lambda )^{-1}$ in the form
$I+A_1,\ A_1 \in \mathbf{S_1}$:
    \begin{equation}
    (H^{(1)}(t)-\lambda )(H_0(t)+\lambda )^{-1}=I+A_1(t),\ \ \ A_1(t)=(W_1-2\lambda )
    (H_0(t)+\lambda )^{-1}. \label{july9a}
    \end{equation}
%
%
%    $$\prod_{n=1}^{N(A(k^{2l},t))}(1+\lambda_n(A(k^{2l},t)))=0.$$
%
%
Obviously, $A_1(t) \in \mathbf{S_1}$.
%%%%%%%%\begin{Rem}\label{R:May22}
From properties of determinants and  the definition of $S_1(\lambda
)$ it follows easily that the isoenergetic set $S_1(\lambda)$ of
$H^{(1)}$  is the zero set of $\det\bigl(I+A_1(t)\bigr)$ in $K_1$.
%%%%%%%%%\end{Rem}
%%%%% \begin{Rem} \label{R:May20} Matrices $H_0(t)$, $H^{(1)}(t)$ and $A_1(t)$ can
%%%%% be extended as holomorphic functions of two variables to a vicinity
%%%%% of $R^2$. If $\vec y=p_j(0)+t$, then matrices $H_0(\vec y)$,
%%%%% $H^{(1)}(\vec y)$ and $A_1(\vec y)$ differ from $H_0(t)$, $H^{(1)}(t)$ and
%%%%% $A_1(t)$ only by a shift of indices: $H_0(\vec
%%%%% y)_{mm}=H_0(t)_{m+j,m+j}$, etc. Hence, $\left\|(H_{0,1}(\vec
%%%%% y)-z)^{-1}\right\|=\left\|(H_{0,1}(t)-z)^{-1}\right\|$ and $\det
%%%%% \bigl(I+A_1(\vec y)\bigr)=\det \bigl(I+A_1(t)\bigr).$ \end{Rem}

Now recall that the set ${\cal D}_1(\lambda )$ can be
 described in terms of vectors $\vec \varkappa_1 (\varphi ),\ \varphi \in
 \Theta _1(\lambda)$; see Lemma \ref{L:2.13}.  By definition, $\chi
_1^*(\lambda )={\cal K}_1{\cal D}_1(\lambda )$. Lemma
\ref{L:May10a} shows that $\chi _1^*(\lambda )$ does not have
self-intersections (Fig.\ref{F:5}), i.e., for every $t\in \chi _1^*(\lambda )$,
there is a single $\vec \varkappa_1 (\varphi )\in {\cal
D}_1(\lambda )$ such that $t={\cal K}_1\vec \varkappa_1 (\varphi
)$. Next, if $\tau \in {\cal K}_2\chi _1^*(\lambda )$, then there is $p\in P$
  such that $\tau +2\pi p/N_1a \in \chi _1^*(\lambda
 )$.  Note that $p$  is not uniquely defined by $\tau $, since ${\cal K}_2\chi _1^*(\lambda
 )$
 may have self-intersections. Hence, every $\tau \in {\cal K}_2\chi _1^*(\lambda
 )$ can be represented  as $\tau ={\cal K}_2\vec \varkappa_1 (\varphi
 )$, where $\vec \varkappa_1 (\varphi
 )$ is not necessary uniquely defined.
The next lemma describes self-intersections  of  $\tilde S_1$ belonging to ${\cal K}_2\chi
_1^*(\lambda
 )$ as
 zeros of a group of determinants.
 \begin{Lem} \label{L:May24} If $\tau $ is a point of
self-intersection of $\tilde S_1$ (Definition \ref{D:May6}),
belonging to ${\cal K}_2\chi _1^*(\lambda )$,
 then $\tau = {\cal K}_2\vec \varkappa_1 (\varphi
 )$, where $\varphi \in
\Theta _1(\lambda)$ and satisfies the equation
\begin{equation}
\det \bigl(I+A_1\left(\vec y(\varphi)\right)\bigr)=0, \ \ \ \vec
y(\varphi)=\vec \varkappa_1 (\varphi )+\vec b,\  \ \vec b=2\pi
p/N_1a, \label{May18c*}
\end{equation}
%%%%with $$ \vec y(\varphi)=\vec \varkappa_1 (\varphi )+\vec b,\  \ \vec
%%%%b=2\pi p/N_1a,
%%%%5$$
 for some $p\in P\setminus \{0\}$. Conversely, if (\ref{May18c*})
is satisfied for  some $p\in P\setminus \{0\}$, then $\tau = {\cal
K}_2\vec \varkappa_1 (\varphi )$ is a point of self-intersection.
\end{Lem}

\begin{Def} \label{D:July3a} Let
$\varPhi _1$ be the complex $\bigl(k^{-2-4s_1-2\delta
}\bigr)$-neighborhood of $\Theta _1$.
\end{Def}

By Lemma
\ref{L:2.13}, $\vec \varkappa_1 (\varphi )$ is an analytic
function in $\varPhi _1$, and hence
    $$\det \bigl(I+A\bigl(\vec y (\varphi \bigr)\bigr),\quad \vec y (\varphi)= \vec
    \varkappa_1 (\varphi )+\vec b \quad \vec b\in K_1,$$
is analytic too.
\begin{Def}
We say that  $\varphi
\in \varPhi _1$ is a quasi-intersection of ${\cal K}_2\chi _1^*$
with $\tilde S_1(\lambda )$ if  (\ref{May18c*})
%%%%%%%%%\begin{equation} \det
%%%%%%%%%\bigl(I+A_1\bigl(\vec y(\varphi)\bigr)\bigr)=0, \vec y(\varphi)=\vec
%%%%%%%%%\varkappa_1 (\varphi )+\vec b,\ \ \ \vec b=2\pi p/N_1a,
%%%%%%%%%\label{May18c*}
%%%%%%%%%\end{equation}
%%%%% with
%%%%% $$ \vec y(\varphi)=\vec \varkappa_1 (\lambda, \vec \nu )+\vec b,\ \
%%%% \ \ \vec \nu =(\cos \varphi ,\sin \varphi ),\ \
%%%% \vec b=2\pi p/N_1a,
 holds for some $p\in P\setminus \{0\}.$
\end{Def}

Thus, real intersections correspond to real zeros of the
determinant, while quasi-intersections may have a small imaginary
part (quasi-intersections include intersections).

Next we describe the resonance set $\Omega_1$ (defined in \eqref{omega})
in terms of determinants.

\begin{Lem}\label{U}
If  $\tau \in \Omega _1$, then $\tau = {\cal K}_2\vec \varkappa_1 (\varphi
 )$ where $\varphi \in
\Theta _1$ satisfies the equation
    \begin{equation}
    \det \left(\dfrac{H^{(1)}\bigl(\vec y(\varphi)\bigr)-k^{2l}-\epsilon }
    {H_0\bigl(\vec  y(\varphi)\bigr)+k^{2l}}\right)=0,\ \ \  \vec y(\varphi)
    =\vec\varkappa_1 (\varphi )+\vec b,\ \ \vec b=2\pi p/N_1a,
    \label{May18c+}
    \end{equation}
for some $p\in P\setminus \{0\}$ and $|\epsilon |<\epsilon _1$.
Conversely, if (\ref{May18c+})
is satisfied for  some $p\in P\setminus \{0\}$ and $|\epsilon
|<\epsilon _1$, then $\tau = {\cal K}_2\vec \varkappa_1 (\varphi )$
belongs to $\Omega _{1}$.
\end{Lem}

We denote by $\omega _1$ the set of $\varphi \in \Theta _1$
corresponding to $\Omega _1$ , i.e., $\omega
_1=\{\varphi \in \Theta _1(\lambda): {\cal K}_2\vec \varkappa_1
(\varphi
 )\in \Omega _1\}\subset [0,2\pi )$.
%%%%%%%%Let us discuss separately the case $W_1=0$. If $\varphi $ is a
%%%%%%%%point of intersection, then (\ref{May18c}) means that
%%%%%%%%\begin{equation}
%%%%%%%%\left|\vec \varkappa_1 (\varphi )+2\pi j/a+2\pi p/N_1a\right|^2=k^2
%%%%%%%%\label{july25} \end{equation} for some $j\in Z^2$ and $p\in P$, and
%%%%%%%%vise versa. Thus, (\ref{May18c}) means that $\vec \varkappa_1 (\varphi
%%%%%%%%)$ is a point of intersection of the circle of the radius $k$
%%%%%%%%centered at the origin ($\left|\vec \varkappa_1 (\varphi
%%%%%%%%)\right|^2=k^2$ when $W_1=0$) and the circle of the same radius
%%%%%%%%centered at a point $2\pi j/a+2\pi p/N_1a$ of a dual lattice
%%%%%%%%corresponding to the periods $N_1a_1$, $N_1a_2$. If there is a
%%%%%%%%quasii-ntersection, then (\ref{july25}) holds for some $\varphi \in
%%%%%%%%\varPhi _1$. We will refer to such $\varphi $ as a point of
%%%%%%%%quasi-intersection of two circles. Geometrically, such terminology is
%%%%%%%%justified, since  validity of (\ref{july25}) for some $\varphi $
%%%%%%%%with a small imaginary part means, that two circles come close
%%%%%%%%together, even, if they do not intersect.

{\bf Complex resonant set.}\label{CRS} Further we consider a
complex resonance set $\omega _1^*(\lambda )$, which is the set of
zeros of the determinants (\ref{May18c+}) in $\varPhi _1$ ($p\in
P\setminus \{0\}$, $|\epsilon |<\epsilon _1$). By Lemma \ref{U},
$\omega _1=\omega _1^*\cap \Theta _1$. We prefer to
consider quasi-intersections instead of intersections and the
complex resonance set instead of just the real one, for the
following reason: the determinants (\ref{May18c*}) and
(\ref{May18c+}), involved in the definitions of quasi-intersections
and the complex resonance set $\omega _1^*$, are holomorphic
functions of $\varphi $ in $\varPhi _1$. Thus we can apply theorems of complex
analysis to these determinants. Rouch\'{e}'s theorem is
particularly important here, since it implies the stability of  zeros of
a holomorphic function with respect to  small perturbations of the
function. We take the determinant (\ref{May18c*}) as a holomorphic
function, its zeros being quasi-intersections: the initial
determinant corresponds to the case $W_1=0$,  the perturbation
obtained by ``switching on" a potential $W_1$. Since there is no analogue of Rouch\'{e}'s
theorem for real functions on the real axis, introducing the
region $\varPhi _1$ and analytic extension of the determinants into
this region is in the core of our considerations. We also use the
well-known inequality for the determinants (see \cite{RS})
    \begin{equation}\label{3.2.27}
    \bigl|\det(I+A)-\det(I+B)\bigr| \leq \|A-B\|_1
    \text{exp}\bigl(\|A\|_1+\|B\|_1+1\bigr),\ A,B \in \mathbf{S}_1.
    \end{equation}
    Note that $\omega _1^*=\bigcup _{p\in P\setminus \{0\}}\omega
    _{1,p}^*$, where $\omega
    _{1,p}^*$ corresponds to a fixed $p$ in (\ref{May18c+}); and
    similarly, $\omega _1=\bigcup _{p\in P\setminus \{0\}}\omega
    _{1,p}$.
    We fix $p\in P$ and study $\omega
    _{1,p}^*$ separately.
We start by the case $W_1=0$. The
    corresponding determinant (\ref{May18c*}) is
    \begin{equation}
    \det\bigl(I+A_0\bigl(\vec
    y_0(\varphi)\bigr)\bigr), \ I+ A_0\bigl(\vec
    y_0(\varphi)\bigr)=\bigl(H_0\bigl(\vec y_0(\varphi)\bigr)-\lambda \bigr)\bigl(H_0\bigl(\vec
    y_0(\varphi)\bigr)+\lambda \bigr)^{-1},\ \label{eee} \end{equation}
$\vec
    y_0(\varphi)=k(\cos \varphi ,\sin
\varphi )+\vec b.$ This determinant  can be investigated by
elementary means. We easily check that the number of zeros of the
determinant in $\varPhi _1$ does not exceed $c_0k^{2+2s_1},\
c_0=32\beta_1\beta_2$. The resolvent $ \bigl(H_0\bigl( \vec
y_0(\varphi)\bigr)-\lambda \bigr)^{-1}$ has poles at zeros of the
determinant. The resolvent norm at $\varphi \in \varPhi _1$ can be
easily estimated by the distance from $\varphi $ to the nearest zero
of the determinant. Next, we introduce the union ${\cal O}(\vec b)$
of all disks of radius $r=k^{-4-6s_1-3\delta }$ surrounding zeros of
the determinant (\ref{eee}) in $\varPhi _1$. Obviously, any $\varphi
\in \varPhi _1\setminus {\cal O}(\vec b)$ is separated from zeros of
the determinant (\ref{eee}) by the distance no less than $r$. This
estimate on the distance yields an estimate for the norm of the
resolvent $\bigl(H_0\bigl(\vec y_0(\varphi)\bigr)-\lambda
\bigr)^{-1},$ when $\varphi \in \varPhi _1\setminus {\cal O}(\vec
b)$. Further, we introduce the potential $W_1$. It is shown in
\cite{KL} that the number of zeros of each determinant
(\ref{May18c+}) is preserved in each connected component $\Gamma
(\vec b)$ of ${\cal O}(\vec b)$ when we switch from the case $W_1=0,
A_1=A_0$ to the case of non-zero $W_1$ and from $\epsilon =0$ to
$|\epsilon |<\epsilon _1$. We also  show in \cite{KL} that estimates
for the resolvent are stable under such change when $\varphi \in
\varPhi _1\setminus {\cal O}(\vec b)$. We ``switch on" the potential
$W_1$ in two steps. First, we replace $\vec y_0(\varphi )$ by $\vec
y(\varphi )$ and consider $\det\bigl(I+A_0\bigl(\vec
y(\varphi)\bigr)\bigr)$ and $\bigl(H_0\bigl(\vec
y(\varphi)\bigr)-k^{2l}\bigr)^{-1}$ in $\varPhi _1$. We take into
account that $\vec y(\varphi)-\vec y_0(\varphi)$ is small and
holomorphic in $\varPhi _1$ (Lemma \ref{L:2.13}),  use
(\ref{3.2.27}) on the boundary of $\Gamma$, and apply Rouch\'{e}'s
theorem. This enables us to conclude  that the number of zeros of
the determinant in $\Gamma (\vec b)$ is preserved when we replace
$\vec y_0(\varphi)$ by $\vec y(\varphi)$. Applying Hilbert relation
for resolvents, we show that the estimates for the resolvent in
$\varPhi _1\setminus {\cal O}(\vec b)$ are also stable under such
change. In the second step we replace $H_0\bigl(\vec
y(\varphi)\bigr)$ by $H^{(1)}\bigl(\vec y(\varphi)\bigr)+\epsilon I$
and prove similar results.
   %%%%%%% Further
  %%%%%%%   we consider the determinant (\ref{May18c+}). Obviously it is
  %%%%%%%   equal to (\ref{May18c*}) when $\epsilon =0$. Since, $\epsilon
 %%%%%%%    $ is very small, its properties are similar to those of
 %%%%%%%    (\ref{May18c*}).
%%%%%%%     In particular, it has the same number of zeros (for a fixed
%%%%%%%     $\epsilon $) as (\ref{May18c*}) at each connected component
 %%%%%%%    $\Gamma (\vec b)$ of ${\cal O}(\vec b)$.
    %%%%%%%%%(Lemma \ref{L:3.17}).
    %%%%%Moreover
    %%%%%we prove that the zeros of the determinants (\ref{May18c*}) and (\ref{May18c+})
    %%%%%are close to each other. For each zero of (\ref{May18c+})
    %%%%%there is a zero of (\ref{May18c*}) at the distance less than
    %%%%%$d=\epsilon ^{1/J}\cdot
    %%%%%k^{-2-4s_1-3\delta}$, $J<ck^{2+2s_1}$ (Lemma \ref{L:3.18}).
    %%%%%hence the whole set $\Omega ^*_1$ is in the $d$ neighborhood
    %%%%%of the zeros of (\ref{May18c*}).
    %%%%%Since the number of zeros
    %%%%%does not exceed $ck^{2+2s_1}$, the total width of $\Omega
    %%%%%^*_{1,p}$ and hence of $\Omega
    %%%%%^_{1,p}$ does not exceed $cdk^{2+2s_1}$. Substituting the
    %%%%%value of $d$ in the last estimate, we prove (\ref{.}).
From this, we see that  $\omega^*_{1,p}\subset {\cal O}(\vec b)$,
$\vec b=2\pi p/N_1a$ and
    \begin{equation}
    \omega^*_{1}\subset {\cal O}^{}_*:=\bigcup _{p^{}\in P^{}\setminus \{0\}}{\cal
    O}^{}\bigl(2\pi p^{}/N_1a\bigr).\label{respect}
    \end{equation}
Considering ${\cal O}(\vec b)$ is formed by no more than $c_0k^{2+2s_1}$  disks and
the set $P$ contains no more than $4k^{2s_2-2s_1}$ elements,
$s_2=2s_1$, we easily obtain that ${\cal O}^{}_*$ contains no
more than $4c_0k^{2+4s_1}$ disks.
Taking the real parts of the sets, we conclude
$\omega_{1}\subset {\cal O}^{}_*\cap \Theta _1(\lambda)$.
Noting ${\cal O}^{}_*$ is formed by   disks of the radius
$r=k^{-4-6s_1-3\delta }$ and using the estimate for the number of disks,
we obtain that the total length of
$\omega_{1}$ does not exceed $k^{-2-2s_1-3\delta}$
and hence the length of $\Omega_{1}$ does not exceed $k^{-1-2s_1-3\delta }$.
    %%%%%%%%% This estimate is proven in Section \ref{St3}.

We introduce the new notation
    \begin{equation}
    %%%%%%{\cal O}^{}_*=\cup _{p^{}\in P^{}\setminus \{0\}}{\cal
    %%%%%%O}^{}\bigl(2\pi p^{}/N_1a\bigr), \ \ \ \
    \varPhi
    _2=\varPhi _1\setminus {\cal O}^{}_*,
     \label{ut1}
    \end{equation}
where $\varPhi _1$ is given by Definition \ref{D:July3a}.
    \begin{figure}
    \centering
\psfrag{Phi_2}{$\Phi_2$}
\includegraphics[totalheight=.2\textheight]{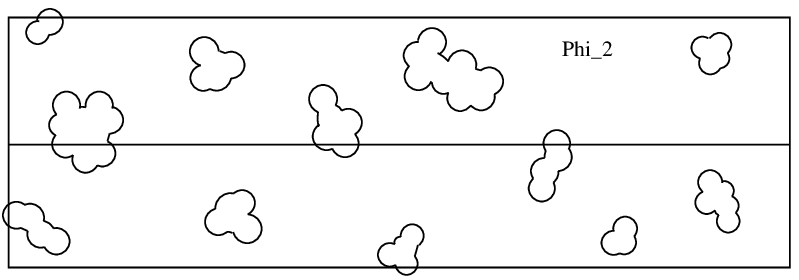}
\caption{The set $\varPhi
    _2$.}\label{F:7}
\end{figure}
 %%%%%%%    Note that the set ${\cal O}_*$  is formed by small disks
 %%%%%%% centered at  points $\varphi _{m,p}^{\pm }\in \varPhi _1$, which are
%%%%%%%  quasi-intersections of the circle of the
%%%%%%%  radius $k$
%%%%%%% around the origin   with the circles of the same radius $k$ around
%%%%%%%  points $\vec p_m(0)+2\pi p/N_1a $, $p\neq 0$, of the dual lattice corresponding to the
%%%%%%% periods $N_1a_1, N_1a_2$. We will show (Corollary \ref{C:ut}) that
%%%%%%% $\omega _1^*\subset {\cal
 %%%%%%%    O}^{}_*$ and $\omega _1 \subset {\cal
%%%%%%%     O}^{}_*\cap \Theta _1.$ From now on we call ${\cal
%%%%%%%     O}^{}_*$ the first complex non-resonance set in $\varPhi
%%%%%%%     _1$ and ${\cal
%%%%%%%     O}^{}_*\cap \Theta _1$ the first non-resonance set in
%%%%%%%     $\Theta _1$.
Obviously, to obtain $\varPhi _2$, we produce round holes in each connected
component of $\varPhi _1$. The set $\varPhi _2$ has a
structure of Swiss cheese (Fig. \ref{F:7}); we  add more holes of a smaller
size at each  step of approximation.

Basing on the perturbation formulas (\ref{3.15}), (\ref{3.16}),
we construct $\cal{B}_2(\lambda)$,  $\cal{D}_2(\lambda )$
(see (\ref{Dn*}), (\ref{Dn}) for $n=2$ and Fig. \ref{F:2}) and
$\chi_2^*(\lambda )$ in the way analogous to the first step. In
particular,
\begin{equation}
|\varkappa _2(\varphi )-\varkappa _1(\varphi )|< 2\epsilon
_1^4k^{-2l+1}\label{add}
\end{equation}
\begin{equation}
\left|\frac{\partial \varkappa _2 }{\partial \varphi }(\varphi
)-\frac{\partial \varkappa _1}{\partial \varphi }(\varphi )\right|<
4\epsilon _1^3k^{1+\delta }\label{add"}
\end{equation}
for $\vec \nu =(\cos \varphi , \sin \varphi )\in
\cal{B}_2(\lambda)$.

\subsection{Next Steps of Approximation.}

 On the $n$-th step, $n\geq 3$, we choose
$s_{n}=2s_{n-1}$ and define the operator $H_{\alpha}^{(n)}$ by the
formula
 \begin{equation}\label{4.1}
     H_{\alpha }^{(n)}=H^{(n-1)}+\alpha W_{n},\quad      (0\leq \alpha \leq
     1),\qquad
     W_{n}=\sum_{r=M_{n-1}+1}^{M_n}V_r, \notag
     \end{equation}
where $M_{n}$ is chosen in such a way that $2^{M_{n}}\approx
k^{s_{n}}$. Obviously, the periods of $W_{n}$ are $2^{M_{n}-1}
(\beta_1,0)$ and $2^{M_{n}-1} (0,\beta_2)$. We  write the periods in the
form: $N_{n-1}\cdots N_1(a_1,0)$ and $N_{n-1}\cdots N_1(0,a_2)$,
where $N_{n-1}$ is of order of $k^{s_n-s_{n-1}}$,
namely, $N_{n-1}=2^{M_{n}-M_{n-1}}$.
%%%%%%%$\frac{1}{4}k^{s_{n}-s_{n-1}}<N_{n-1}< 4k^{s_n-s_{n-1}}$.
Note that
    $\|W_n\|_{\infty} \leq \sum_{r=M_{n-1}+1}^{M_n}
    \|V_r\|_{\infty} \leq
    \exp(-k^{\eta s_{n-1}}).$

Let us start by establishing a lower bound for $k$.
 Since $\eta s_1>2+2s_1$, there is a number $k_*>e$ such that
\begin{equation}\label{kstar}
 C_*(1+s_1)k^{2+2s_1}\ln k <k^{\eta s_1},\quad  C_*=400l(c_0+1)^2,
 \end{equation}
  for any $k>k_*$. Assume also that $k_*$ is sufficiently
 large to ensure validity of all estimates in the first two steps
 for any $k>k_*$. Further we consider $\lambda =k^{2l}$, where
 $k>k_*$.

 %%%%%%%   We note that  $k$ satisfying (\ref{k}), obeys the analogous
 %%%%%%%   condition for the n-th step:
%%%%%%%    \begin{equation}
%%%%%%% \hat C(1+s_{n-2})k^{2+2s_{n-2}}\ln k <k^{\eta s_{n-2}}\label{k*}
%%%%%%% \end{equation}
%%%%%%%with the same constant $\hat C$. The inequality (\ref{*k}) can be
%%%%%%%derived from (\ref{k}) by the method of mathematical induction. This
%%%%%%%is a very important fact. It means that the lower bound for $k$ does
%%%%%%%not grow with $n$, i.e. all steps hold uniformly in $k$ for $k>k_*$.
%%%%%%%Further we assume $k>k_*$.

The geometric lemma for $n$-th step is the same as  that for Step 2
up to shift of indices. Note only that we need an inductive
procedure to define the set $\chi_{n-1}^*(\lambda)$, which is defined by
(\ref{2.81}) for $n=1$ and in the analogous way for $n\geq 2
$.
 The estimate (\ref{3.9}) for $n$-th step takes the form
\begin{equation}\label{4.9*}
\frac{L\left(\cal{K}_n \chi_{n-1}^* \setminus \chi
_n)\right)}{L\left(\chi_{n-1}^*\right)}<k^{-\cal{S}_n},\ \ \ \
\cal{S}_n=2\sum _{i=1}^{n-1}(1+s_i).
\end{equation}
It is easy to see that $\cal{S}_n=2(n-1)+\left(2^{n}-2\right)s_1$
and $\cal{S}_n\approx 2^{n}s_1 \approx s_n$. The formulation of the main
results (perturbation formulas) for $n$-th step is the same as for
the second step up to shift of indices. The formula for the
resonance set $\Omega _{n-1}$ and non-resonance set $\chi _n$ are
analogous to those for $\Omega _1$, $\chi _2$ (see (\ref{3.5.1})).
The proof of the first and second statements of Geometric Lemma follows
from the definition of the non-resonance set. Now we describe
shortly a proof of the third statement.

 In the second step, we defined the union ${\cal O}(\vec b)$
of all disks of the radius $r=k^{-4-6s_1-3\delta }$ surrounding
zeros of the determinant (\ref{eee}) in $\varPhi _1$. Let us change
the notation: ${\cal O}(\vec b)\equiv {\cal O}^{(1)}(\vec b^{(1)})$
.
 Now we define ${\cal O}^{(n-1)}(\vec b^{(n-1)})$,  $\vec b^{(n-1)}\in K_{n-1}$, $n\geq 3$, by the
formula
    \begin{equation}
    {\cal O}^{(n-1)}(\vec b^{(n-1)})=\bigcup _{p^{(n-2)}\in P^{(n-2)}}{\cal
    O}^{(n-2)}_s\left(\vec b^{(n-1)}+2\pi
    p^{(n-2)}/\hat N_{n-2}a \right), \label{july8a*}
    \end{equation}
here and below, $\hat N_{n-2}\equiv N_{n-2}\cdots N_1$ and
$P^{(m)}=\bigl\{p^{(m)}=(p_1^{(m)},p_2^{(m)}),\ 0\leq
p_1^{(m)}<N_m-1,\ 0\leq p_2^{(m)}<N_m-1\bigr\}$. The set ${\cal
O}^{(m)}_s(\vec b^{(m)})$, $m\geq 1$, is a collection of disks of
the radius $r^{(m+1)}=r^{(m)}k^{-2-4s_{m+1}-\delta }$,
$r^{(1)}=r=k^{-4-6s_1-3\delta }$ around zeros of the determinant
$\det\bigl(I+A_{m}\left(\vec y^{(m)}(\varphi )\right)\bigr)$ in
$\varPhi _m$, $\vec y^{(m)}=\vec \varkappa _{m}(\varphi )+\vec
b^{(m)}$, the set $\varPhi _m$ being defined  earlier for $m=1,2$
(Definition \ref{D:July3a}, \eqref{ut1}) and by the formula below
for $m\geq 3$.
    \begin{equation}
    {\cal O}^{(n-1)}_*=\bigcup _{p^{(n-1)}\in P^{(n-1)}\setminus \{0\}}{\cal
    O}^{(n-1)}\bigl(2\pi p^{(n-1)}/\hat N_{n-1}a\bigr), \ \ \ \ \varPhi
    _n=\varPhi _{n-1}\setminus {\cal O}^{(n-1)}_*.
     \label{ut1**}
    \end{equation}
If $n=2$, then   (\ref{ut1**}) gives
us ${\cal O}_*$, see (\ref{respect}).
Note that the complex non-resonance set $\varPhi _n$ is defined by
the  recurrent formula analogous to (\ref{ut1}).

\begin{Lem} \label{L:new2}
The set ${\cal O}^{(m)}_s(\vec b^{(m)})$, $\vec b^{(m)}\in K_m$
contains no more than $4^{m-1}c_ok^{2+2s_{m}}$ disks.
\end{Lem}

\begin{Cor}\label{C:c1}
The set ${\cal O}^{(n-1)}(\vec b^{(n-1)})$ contains no more than
$4^{n-2}c_ok^{2+2s_{n-1}}$ disks.
\end{Cor}

\begin{Cor} \label{C:c2}
The set ${\cal O}^{(n-1)}_*$ contains no more than
$4^{n-1}c_ok^{2+2s_{n}}$ disks.
\end{Cor}

The lemma is proved by an induction procedure.  Corollaries
\ref{C:c1} and \ref{C:c2} are based on the fact that $P^{(n-1)}$
contains no more than $4k^{2(s_n-s_{n-1})}$ elements and a
similar estimate holds for $P^{(n-2)}$.

\begin{figure}
\centering
 \psfrag{Phi_3}{$\Phi_3$}
\includegraphics[totalheight=.2\textheight]{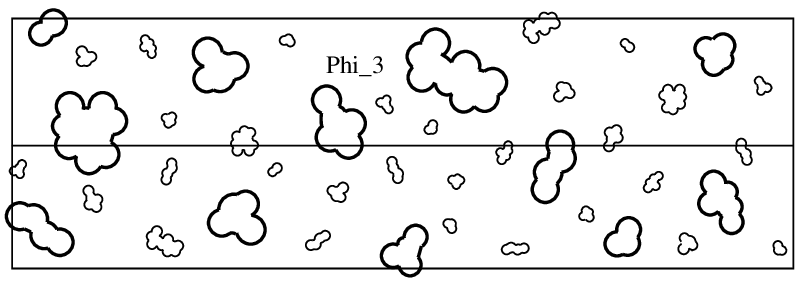}
\caption{The set $\varPhi
    _3$.}\label{F:8}
\end{figure}

Obviously, $\varPhi _{n}$ has the structure of Swiss cheese,
more and more holes of smaller and smaller radii appear at each
step of approximation (Fig. \ref{F:8}). Note that the disks are more
and more precisely ``targeted" at each step of approximation.
At the $n$-th step the disks of ${\cal O}^{(n-1)}_*$ are centered around the zeros of the
determinants
     $$\det\bigl(I+A_{n-2}(\vec \varkappa _{n-2}(\varphi )+2\pi p^{(n-2)}/\hat N_{n-2}a
     +2\pi p^{(n-1)}/\hat N_{n-1}a )\bigr),$$
where $ p^{(n-2)}\in P^{(n-2)},\ p^{(n-1)}\in P^{(n-1)},$
$\vec \varkappa _{n-2}(\varphi) \in \cal{D}_{n-2}$, the
corresponding operator $H^{(n-2)}$ being closer and closer to
the operator $H$. Here, $\lambda^{(n-2)}(\vec \varkappa_{n-2}(\varphi))=\lambda$.
If $W_1=W_2=...=W_{n-2}=0$, then ${\cal O}^{(n-1)}_*$ is just
the union of disks centered at quasi-intersections of the ``unperturbed" circle
$\vec k =k(\cos \varphi ,\sin \varphi )$, $k=\lambda ^{\frac{1}{2l}}$, $\varphi \in [0,2\pi )$
with circles of the same
radius centered at points $2\pi j/a+2\pi p^{(1)}/N_1a+....+2\pi p^{(n-1)}/\hat N_{n-1}a$, these
points being nodes of the dual lattice corresponding  to
the periods $\hat N_{n-1}a_1,\hat N_{n-1}a_2$.
After constructing $\chi _n(\lambda )$ as the real part of
$\varPhi _n$, we define the non-resonance subset  $\chi _n^*(\lambda )$ of the
isoenergetic set $S_n(\lambda )$ of  $H_{\alpha }^{(n)}$, $S_n(\lambda )\subset K_n.$
It corresponds to  the non-resonance eigenvalues  given by  perturbation series.
The sets $\chi _1^*(\lambda )$, $\chi _2^*(\lambda )$
are defined in the previous steps as well
as the non-resonance sets $\chi _1(\lambda )$, $\chi _2(\lambda  )$.
Recall that we started by the
definition of $\chi _1(\lambda )$ (Fig. \ref{F:4}) and used it to define
${\cal D}_1(\lambda )$ (Fig. \ref{F:1}) and $\chi _1^*(\lambda )$,
$\chi _1^*={\cal K}_1{\cal D}_1$ (Fig. \ref{F:5}).
In the second step, we constructed $\chi _2(\lambda )$,
using $\chi _1^*(\lambda )$.
Next,  we defined ${\cal D}_2(\lambda )$ (Fig. \ref{F:2}) and $\chi _2^*(\lambda )$,
$\chi _2^*={\cal K}_2{\cal D}_2$.
Thus, the process looks like
$\chi _1\to {\cal D}_1\to \chi _1^*\to \chi_2\to {\cal D}_2
\to \chi _2^*\to \chi _3\to {\cal D}_3 \to \chi _3^*\to ...$.
At Every step, the set $\chi _n$ is constructed using $\chi _{n-1}^*$
by a formula analogous to (\ref{3.5.1}).
Using perturbation formulas, we show that the ``radius" $\varkappa _n(\varphi )$
of $ {\cal D}_n$ satisfies the estimates
    \begin{equation}
    |\varkappa _n(\varphi )-\varkappa _{n-1}(\varphi )|< 2\epsilon _{n-1}^4k^{-2l+1},\ \ n\geq 2,
    \label{add*}
    \end{equation}
    \begin{equation}
    \left|\frac{\partial \varkappa _n }{\partial \varphi }(\varphi )
    -\frac{\partial \varkappa _{n-1}}{\partial \varphi }(\varphi)\right|
    < 4\epsilon _{n-1}^3k^{1+\delta }, \ \ n\geq 2,\label{add"*}
    \end{equation}
where
    \begin{equation}\label{rrr}
    \epsilon _n=e^{-\frac{1}{4}k^{\eta s_n}}
    \end{equation}
and  $\eta $ is the parameter in (\ref{r}).
Note that $\epsilon _n$ decays super exponentially with $n$.

\section{Limit-Isoenergetic Set and Eigenfunctions} \label{chapt7}

\subsection{Limit-Isoenergetic Set and Proof of Bethe-Sommerfeld Conjecture}
At each step $n$, we have constructed a set
$\cal{B}_n(\lambda)$, $\cal{B}_{n}(\lambda)\subset
\cal{B}_{n-1}(\lambda)\subset S_1(\lambda)$,  and a function
$\varkappa _n(\lambda,\vec{\nu})$, $\vec{\nu} \in
\cal{B}_n(\lambda)$, with the following properties. The set
$\cal{D}_{n}(\lambda )$ of vectors $\vec{\varkappa}=\varkappa
_n(\lambda ,\vec{\nu})\vec{\nu}$,
$\vec{\nu} \in \cal{B}_{n}(\lambda )$,
is a slightly distorted circle with holes;
see Figs.\ref{F:1}, \ref{F:2}, formula (\ref{Dn}) and Lemma \ref{L:2.13}.
For any $\vec \varkappa _n(\lambda,\vec{\nu})\in
\cal{D}_{n}(\lambda )$, there is a single eigenvalue of
 $H^{(n)}(\vec \varkappa _n)$
equal to $\lambda $ and given by a perturbation
series analogous to (\ref{3.15}). Let
    $\cal{B}_{\infty}(\lambda)=\bigcap_{n=1}^{\infty}\cal{B}_n(\lambda).$
Since $\cal{B}_{n+1} \subset \cal{B}_n$ for every $n$,
$\cal{B}_{\infty}(\lambda)$ is the unit circle with an infinite number of
holes, more and more holes of smaller and smaller size appearing at
each step. \begin{Lem} \label{L:Dec9} The length of
$\cal{B}_{\infty}(\lambda)$ satisfies estimate (\ref{B}) with
$\gamma _3=\delta /2$.
\end{Lem}

\begin{proof}
 Using  (\ref{4.9*}) and noting that $S_n\approx 2^ns_1$, we
easily conclude that
$L\left(\cal{B}_n\right)=\left(1+O(k^{-\delta/2})\right)$,
$k=\lambda ^{1/2l}$, uniformly in $n$. Since $\cal{B}_n$ is a
decreasing sequence of sets, (\ref{B}) holds.
\end{proof}

Let us consider
$\varkappa _{\infty}(\lambda, \vec{\nu})=\lim_{n \to \infty}\varkappa _n(\lambda,\vec{\nu}),\quad
\vec{\nu} \in \cal{B}_{\infty}(\lambda ).$

\begin{Lem}
The limit $\varkappa _{\infty}(\lambda, \vec{\nu})$
exists for any $\vec{\nu} \in \cal{B}_{\infty}(\lambda )$.
The following estimates hold when $n\geq 1$:
    \begin{equation}\label{6.1}
    \left|\varkappa _{\infty}(\lambda, \vec{\nu})-\varkappa
    _n(\lambda,\vec{\nu})\right|<4\epsilon_{n} ^4k^{-2l+1},\ \
    \epsilon _{n}=\exp(-\frac{1}{4}k^{\eta
    s_{n}}),\ \
    s_n=2^{n-1}s_1.
    \end{equation}
\end{Lem}

\begin{Cor}\label{Dec18}
For every $\vec{\nu} \in \cal{B}_{\infty}(\lambda)$, estimate (\ref{h}) holds,
where $\gamma _4=(4l-3-4s_1-3\delta)/2l>0$.
\end{Cor}

The lemma follows easily from (\ref{add*}). To obtain the corollary, we use
(\ref{2.75}) and take into account that $\gamma _0=2l-2-4s_1-2\delta $.

The estimate (\ref{add"*}) justifies convergence of the sequence
$\frac{\partial \varkappa _n}{\partial \varphi}$.
We denote the limit of this sequence by $\frac{\partial \varkappa
_{\infty}}{\partial \varphi }.$

\begin{Lem}
The  estimate (\ref{Dec9a*}) with $\gamma _5=(4l-5-8s_1-4\delta)/2l >0$
holds for any $\vec \nu \in \cal{B}_{\infty}(\lambda)$.
\end{Lem}

We define $\cal{D}_{\infty}(\lambda )$ by (\ref{D}). Clearly,
$\cal{D}_{\infty}(\lambda )$ is a slightly distorted circle of
radius $k$ with infinite number of holes. We can assign a tangent
vector $\frac{\partial \varkappa }{\partial \varphi }\vec \nu
+\varkappa \vec \mu $, $\vec \mu =(-\sin \varphi ,\cos \varphi )$ to
the curve $\cal{D}_{\infty}(\lambda )$, this tangent vector being
the limit of corresponding tangent vectors for  curves
$\cal{D}_{n}(\lambda )$ at points $\vec \varkappa _n(\lambda ,\vec
\nu )$ as $n\to \infty $.

\begin{Rem} \label{R:Dec9}
We see easily from (\ref{6.1}) that any $\vec{\varkappa}
\in \cal{D}_{\infty}(\lambda )$ belongs to the $\left(4\epsilon_{n}
^4k^{-2l+1}\right)$-neighborhood of $\cal{D}_n(\lambda )$. Applying
perturbation formulas for $n$-th step, we conclude that  there
is an eigenvalue  $\lambda^{(n)}(\vec \varkappa )$ of $H^{(n)}(\vec \varkappa )$
satisfying the estimate $\lambda^{(n)}(\vec \varkappa )=\lambda
+\delta _n$, $\delta _n=O\left(\epsilon _{n}^4\right)$, where the
eigenvalue $\lambda^{(n)}(\vec \varkappa )$ is given by a perturbation
series of the type (\ref{3.15}). Hence, for every $\vec{\varkappa}
\in \cal{D}_{\infty}(\lambda)$, one has the limit
\begin{equation} \lim _{n\to \infty }\lambda^{(n)}(\vec \varkappa
)=\lambda,\label{6.2}
\end{equation}
\begin{equation} \left|\lambda^{(n)}(\vec \varkappa
)-\lambda \right|<\delta _n, \ \ \ \delta _n=24\epsilon
_n^4.\label{6.2a}
\end{equation}
\end{Rem}

\begin{Thm}[Bethe-Sommerfeld Conjecture]
The spectrum of  operator $H$ contains a semi-axis.
\end{Thm}

\begin{proof}
By Remark \ref{R:Dec9}, there is a point of the spectrum of $H_n$ in
the $\delta _n$-neighborhood of $\lambda $ for every $\lambda
>k_*^{2l}$, where $k_*$ is given by \eqref{kstar}. Since
$\|H_n-H\|<\epsilon_{n}^4$, there is a point of the spectrum of $H$
in the $\delta _n^*$-neighborhood of $\lambda $, $\delta _n^*=\delta
_n+\epsilon_{n}^4$. Since this is true for every $n$ and the
spectrum of $H$ is closed, $\lambda $ is in the spectrum of $H$.
\end{proof}

\subsection{Generalized Eigenfunctions of $H$}
A plane wave is usually written by $e^{i\langle \vec k, x \rangle}$,
$\vec k \in \R^2$. Here we  use $\vec \varkappa $ instead of
$\vec k$ to conform to our previous notations. We show that for
every  $\vec \varkappa $ in the set
    \begin{equation}
    \cal{G} _{\infty }=\bigcup _{\lambda >\lambda _*}\cal{D}_{\infty}(\lambda ),
    \ \ \lambda _*=k_*^{2l},\label{ac3}
    \end{equation}
$k_*$ being given in \eqref{kstar}, there is a solution $\Psi _{\infty }(\vec \varkappa , x)$ of the
equation for eigenfunction equation
    \begin{equation}
    (-\Delta)^{2l}\Psi _{\infty}(\vec \varkappa , x)+V(x)\Psi _{\infty }(\vec \varkappa ,
    x)=\lambda _{\infty}(\vec \varkappa )\Psi _{\infty }(\vec \varkappa , x)
    \label{6.2.1}
    \end{equation}
which can be represented in the form
    \begin{equation}
    \Psi _{\infty }(\vec \varkappa , x)=e^{i\langle \vec \varkappa , x
    \rangle}\bigl(1+u_{\infty}(\vec \varkappa , x)\bigr),   \label{6.2.1a}
    \end{equation}
where $u_{\infty}(\vec \varkappa , x)$ is a limit-periodic function satisfying the estimate
    \begin{equation}\label{toyota}
    \bigl\|u_{\infty}(\vec \varkappa , x))\bigr\| _{L_{\infty }(\R^2)}<10|\vec \varkappa |^{-\gamma_1},
    \end{equation}

 $\gamma _1=2l-4-7s_1-2\delta >0$; the eigenvalue $\lambda _{\infty}(\vec \varkappa )$
 satisfies the asymptotic
formula
    \begin{equation}\lambda _{\infty}(\vec \varkappa )=|\vec \varkappa |^{2l}+O(|\vec
    \varkappa |^{-\gamma _2}), \ \ \ \gamma _2=2l-2-4s_1-3\delta
    >0.\label{6.2.4}
    \end{equation}
We also show that the set $\cal{G} _{\infty }$ satisfies
(\ref{full}).

 In fact,   by (\ref{6.1}), any $\vec{\varkappa} \in \cal{D}_{\infty}(\lambda
)$ belongs to the $(\epsilon_n k^{-2l+1-\delta})$-neighborhood of
$\cal{D}_n(\lambda )$. Applying the perturbation formulas proved in
the previous sections, we obtain the inequalities
    \begin{equation}
    \bigl\|E^{(1)}(\vec{\varkappa})-{E}^{(0)}(\vec{\varkappa})\bigr\|_1<2k^{-\gamma_0},
    \quad \gamma_0=2l-2-4s_1-2\delta,\label{6.2.2}
    \end{equation}
    \begin{equation}
    \bigl\|E^{(n+1)}(\vec{\varkappa})-\tilde{E}^{(n)}(\vec{\varkappa})\bigr\|_1<
    48\epsilon_n^3, \quad n \geq 1,\label{6.2.2s}
    \end{equation}
    \begin{equation}\bigl|\lambda ^{(1)}(\vec{\varkappa})-|\vec \varkappa |^{2l}
     \bigr|
     <2k^{-\gamma _2}, \label{6.2.3}
     \end{equation}
     \begin{equation}
    \bigl|\lambda ^{(n+1)}(\vec \varkappa )-\lambda ^{(n)}(\vec \varkappa )\bigr|<12\epsilon _n^4,
     \quad n \geq 1,
    \label{6.2.3s}
    \end{equation}
where $E^{(n+1)},\ \tilde{E}^{(n)}$ are one-dimensional spectral
projectors in $L_2(Q_{n+1})$ corresponding to the potentials $W_{n+1}$
and $W_n$, respectively; $\lambda ^{(n+1)}(\vec \varkappa )$ is the
eigenvalue corresponding to $E^{(n+1)}(\vec{\varkappa})$; and
${E}^{(0)}(\vec{\varkappa})$ corresponds to $V=0$ and the periods
$a_1$, $a_2$. The estimate (\ref{6.2.3s}) means that for every
$\vec \varkappa \in \cal G_{\infty} $ there is a limit
$\lambda _{\infty}(\vec \varkappa )$ of $\lambda ^{(n)}(\vec \varkappa )$ as $n\to
\infty $:
    \begin{equation}
    \lambda _{\infty }(\vec \varkappa )=\lim _{n\to \infty }\lambda ^{(n)}(\vec \varkappa ),
    \label{June1}
    \end{equation}

    \begin{equation}
    \left|\lambda _{\infty }(\vec \varkappa )-\lambda ^{(n)}(\vec
    \varkappa )\right|<24\epsilon _n^4, \ \ n \geq 2.
    \label{June2}
    \end{equation}
The estimates (\ref{6.2.2}), \eqref{6.2.2s} mean that for
properly chosen eigenfunctions $\Psi _{n+1}(\vec \varkappa ,x)$,
    \begin{equation}\label{psi1}
    \|\Psi _{1}-\Psi _0\|_{L_2(Q_{1})}<4k^{-\gamma_0}|Q_1|^{1/2},\ \ \
    \Psi _0(x)=e^{i \langle \vec \varkappa ,x
    \rangle},
    \end{equation}
    \begin{equation}
    \|\Psi _{n+1}-\tilde \Psi _n\|_{L_2(Q_{n+1})} < 100\epsilon_n^{3}|Q_{n+1}|
    ^{1/2},\label{Dec9c}
    \end{equation}
where $\tilde \Psi _n $ is $\Psi _n $ extended quasi-periodically from $Q_n$ to
$Q_{n+1}$. The eigenfunctions $\Psi _{n}$, $n\geq 1$, are chosen to obey
two conditions: $\|\Psi _{n}\|_{L_2(Q_{n})}=|Q_{n}|^{1/2}$;
    \footnote{The condition
    $\|\Psi _{n}\|_{L_2(Q_{n})}=|Q_{n}|^{1/2}$ implies
    $\|\tilde \Psi _{n}\|_{L_2(Q_{n+1})}=|Q_{n+1}|^{1/2}$.}
and $\text{Im}(\Psi _{n},\tilde \Psi _{n-1})=0$;
here    $(\cdot ,\cdot)$ is an inner product in $L_2(Q_{n})$. These two
conditions obviously determine a unique choice of each $\Psi _n$.
Noting $\Psi _{n+1}$ and $\tilde \Psi _n$ satisfy  eigenfunction equations
and taking into account (\ref{6.2.3s}),
(\ref{Dec9c}), we obtain
    \begin{equation}\|\Psi _{n+1}-\tilde \Psi _n\|_{W_2^{2l}(Q_{n+1})} <
    ck^{2l}\epsilon_n^{3}|Q_{n+1}|^{1/2},\ \ \ n \geq 1,
    \label{s++}
    \end{equation}
and hence
$\|\Psi _{n+1}-\tilde \Psi _n\|_{L_{\infty}(Q_{n+1})}
< ck^{2l}\epsilon_n^{3}|Q_{n+1}|^{1/2}. $ Since $\Psi _{n+1}$ and $\tilde \Psi _n$ obey the same
quasiperiodic conditions, the same inequality holds in all of $\R^2$:
    \begin{equation}
    \|\Psi _{n+1}-\Psi _n\|_{L_{\infty}(\R^2)}
    < c_l k^{2l}\epsilon_n^{3}|Q_{n+1}|^{1/2}, \ \ n \geq 1, \label{Dec10}
    \end{equation}
where $\Psi _{n+1},\Psi _n$ are quasiperiodically extended to
$\R^2$. Obviously, we have a Cauchy sequence in $L_{\infty}(\R^2)$.
Let \begin{equation} \Psi _{\infty }(\vec \varkappa ,x)=\lim_{n \to
\infty}\Psi _n(\vec \varkappa, x).\label{June3} \end{equation}
This limit is defined pointwise uniformly in $x$ and in
$W_{2,loc}^{2l}(\R^2)$. From the estimate (\ref{Dec10}), we easily
obtain
\begin{equation}
    \|\Psi _{\infty}-\Psi _n\|_{L_{\infty}(\R^2)}
    < c)l k^{2l}\epsilon_n^{3}|Q_{n+1}|^{1/2}, \ \ n \geq 2. \label{Dec10*}
    \end{equation}

\begin{Thm} \label{T:Dec10}
For every sufficiently large $\lambda,\ \lambda >\lambda_*(V,\delta)$
 and $\vec{\varkappa} \in
\cal{D}_{\infty}(\lambda)$, the sequence of functions $\Psi
_n(\vec{\varkappa},x)$ converges  in $L_{\infty}(\R^2)$ and
$W_{2,loc}^{2l}(\R^2)$. The limit function $\Psi _{\infty
}(\vec{\varkappa},x)$ satisfies the equation
    \begin{equation}\label{6.7}
     (-\Delta)^{2l}\Psi _{\infty }(\vec{\varkappa}, x)+V(x)\Psi _{\infty }(\vec{\varkappa},
    x)= \lambda \Psi _{\infty }(\vec{\varkappa}, x).
    \end{equation}
It can be represented in the form (\ref{6.2.1a}),
where $u_{\infty}(\vec{\varkappa}, x)$ is the limit-periodic function
    \begin{equation}\label{6.5}
    u_{\infty}(\vec{\varkappa}, x)=\sum_{n=1}^{\infty}\tilde u_n(\vec{\varkappa},
    x),
    \end{equation}
and $\tilde u_n(\vec{\varkappa}, x)$ are periodic function with the periods
$2^{M_n-1}\beta _1, 2^{M_n-1}\beta _2$, $2^{M_n}\approx
k^{2^{n-1}s_1}$,
\begin{equation}
    \|\tilde{u}_1\|_{L_{\infty}(\R^2)} <9k^{-\gamma_1}, \ \ \
    \gamma _1=2l-4-7s_1-2\delta >0, \label{6.6a}
    \end{equation}
    \begin{equation}\label{6.6}
    \|\tilde{u}_n\|_{L_{\infty}(\R^2)} <c_l k^{2l}\epsilon_{n-1}^{3}|Q_{n}|^{1/2}, \ \ \ n\geq
    2.\end{equation}
    The eigenvalue $\lambda $ in (\ref{6.7}) is equal to
    $\lambda _{\infty }(\varkappa )$, defined by (\ref{June1}),
    (\ref{June2}), and the estimate (\ref{6.2.4}) holds.
\end{Thm}
\begin{Cor}
The function $u_{\infty}(\vec{\varkappa}, x)$ satisfies the
 estimate (\ref{toyota}).
\end{Cor}
\begin{Rem}
    If $V$ is sufficiently smooth, say $V\in C^1(R)$,
    then  estimate (\ref{6.6a}) and hence (\ref{6.2.1a}) can be
    improved by replacing $\gamma _1$ by $\gamma _0$.
\end{Rem}

\begin{proof}
 Let us show that $\Psi _{\infty }$ is a
limit-periodic function. Obviously,
    $ \Psi _{\infty } =\Psi _0+\sum_{n=0}^{\infty}(\Psi_{n+1}-\Psi _n)
$, the series converging in $L_{\infty}(\R^2)$ by (\ref{Dec10}).
Writing $u_{n+1}=e^{-i \langle \vec \varkappa
,x\rangle}(\Psi_{n+1}-\Psi _n)$, we arrive at (\ref{6.2.1a}),
(\ref{6.5}). Note that
 $\tilde{u}_n$ is periodic with  the periods $2^{M_n-1}\beta_1,
 2^{M_n-1}\beta_2$. Estimate (\ref{6.6})  follows from (\ref{Dec10}). We check
 (\ref{6.6a}). Indeed,  by (\ref{psi1}), the Fourier coefficients
 $(u_1)_j$, $j\in \Z^2$, satisfy the estimate
 $\left|(\tilde{u}_1)_j\right|<4k^{-\gamma_0}|Q_1|^{1/2}<8k^{-\gamma_0+s_1}$. This estimate is
 easily improved for $j$ such that $p_j(0)>2k$: $\left|(\tilde{u}_1)_j\right|<c|j|^{-2l}$.
 Summarizing these
 inequalities and taking into account that the number of $j:p_j(0)\leq 2k$
 does not exceed $c_0k^{2+2s_1}$, we conclude that
 (\ref{6.6a}) holds for sufficiently large $k$, $k_0(\sum_{r=1}^{\infty}\|V_r\|, l,\delta)$.
It remains to prove (\ref{6.7}). Indeed, $\Psi _n(\vec \varkappa,
x)$, $n\geq 1$, satisfy the eigenfunction equations: $H^{(n)}\Psi
_n=\lambda ^{(n)}(\vec \varkappa )\Psi _n $. Since $\Psi
_n(\vec \varkappa,x)$ converges to $\Psi (\vec \varkappa, x)$ in
$W_{2l,loc}^2$ and relation (\ref{6.2}) holds, we arrive at (\ref{6.7}).
The estimate (\ref{6.2.4}) follows from (\ref{6.2.3}) -- (\ref{June1}).
\end{proof}

\begin{Rem}
    Theorem \ref{T:Dec10} holds for $\vec{\varkappa} \in
    \cal{D}_{\infty}(\lambda)$ and all $\lambda>\lambda_*$.
    Hence it holds in $\cal{G} _{\infty }=\cup_{\lambda >\lambda*}\cal D_{\infty}(\lambda)$.
\end{Rem}

\section{Absolute Continuity of the Spectrum}\label{chapt8}

\subsection{Sets $\cal{G}_n$ and Projections $E_n(\cal{G}_n')$, $\cal{G}_n'\subset \cal{G}_n$.}

Let us consider the sets $\cal{G}_n$ given by
 \begin{equation}\cal{G}_n=\bigcup _{\lambda
>\lambda _*}\cal{D}_n(\lambda ),\label{tt1}
\end{equation}
where $\lambda_*=k_*^{2l}$ and $k_*$ is introduced in \eqref{kstar}.
Since the perturbation formulas hold in a small neighborhood of each
point of $\cal{G}_n$, we consider, with slightly abused notations,
that $\cal{G}_n$ is open. The function $\lambda ^{(1)}(\vec
\varkappa )$ is differentiable in a neighborhood of each $\vec
\varkappa \in \cal{G}_1$, estimates (\ref{2.66}), (\ref{2.67a})
being valid. Similar results hold for all $\lambda ^{(n)}(\vec
\varkappa )$ and $\cal{G}_n$, $n=1,2,...$.

 There is a family of Bloch eigenfunctions $\Psi _n(\vec
    \varkappa ,x)$, $\vec \varkappa \in \cal{G}_n$, of the operator
    $H^{(n)}$, which are described by the perturbation formulas.
    Let   $\cal{G}_n'$ be a Lebesgue measurable subset of  $\cal{G}_n$.
    We consider the spectral projection
    $E_n\left(\cal{G}_n'\right)$ of $H^{(n)}$ corresponding to
    functions $\Psi _n(\vec
    \varkappa ,x)$, $\vec \varkappa \in \cal{G}_n'.$
Note that, as in \cite{6r}, $E_n\left( \cal{G}'_n\right):
L_2(\R^2)\to L_2(\R^2)$ can be written as
    \begin{equation} E_n\left( \cal{G}'_n\right)F=\frac{1}{4\pi ^2}\int
    _{ \cal{G}'_n}\bigl( F,\Psi _n(\vec
    \varkappa )\bigr) \Psi _n(\vec
    \varkappa ) d\vec \varkappa \label{s}
    \end{equation}
    for any $F\in C_0^{\infty}(\R^2)$, here and below $\bigl( \cdot ,\cdot \bigr)$
    is the canonical scalar product in $L_2(\R^2)$, i.e.,
    $$\bigl( F,\Psi _n(\vec
    \varkappa )\bigr)=\int _{\R^2}F(x)\overline{\Psi _n(\vec
    \varkappa ,x)}dx.$$
More precisely, we write
    \begin{equation} E_n\left(\cal{G}'_n\right)=S_n\left(\cal{G}'_n\right)T_n \left(
    \cal{G}'_n\right), \label{ST}
    \end{equation}
    $$T_n: L_2(\R^2)\to L_2\left(  \cal{G}'_n\right), \ \
    \ \ S_n:L_2\left( \cal{G}'_n\right)\to L_2(\R^2),$$
    \begin{equation}
    T_nF=\bigl( F,\Psi _n(\vec
    \varkappa )\bigr) \mbox{\ \ for any $F\in C_0^{\infty}(\R^2)$},
    \label{eq2}
    \end{equation}
    $T_nF$ being in $L_{\infty }\left(  \cal{G}'_n\right)$, and
    \begin{equation}S_n\varphi = \int _{  \cal{G}'_n}\varphi (\vec \varkappa)\Psi _n(\vec
    \varkappa ,x)d\vec \varkappa  \mbox{\ \ for any $\varphi \in L_{\infty }\left(
    \cal{G}'_n\right)$.} \label{ev}
    \end{equation}
It is
    easy to show that $T_nF\in L_{\infty }(\cal G_n)$, when $F\in C_0^{\infty }(\R^2)$.
    Hence $E_n\left( \cal{G}'_n\right)$ can be described by formula
    (\ref{s}) for $F\in C_0^{\infty }(\R^2)$.
Moreover, as in  \cite{6r}, $\|T_n\|\leq 1$ on $C_0^{\infty }(\R^2)$
and $\|S_n\|\leq 1$ on $L_{\infty }(\cal G'_n)$
    and hence $T_n$, $S_n$ can be extended by
    continuity from $C_0^{\infty }(\R^2)$, $L_{\infty }\left(  \cal{G}'_n\right)$
    to $L_2(\R^2)$ and $L_2\left( \cal{G}'_n\right)$,
    respectively. Thus the operator $E_n\left( \cal{G}'_n\right)$ is
    described by (\ref{ST}) in the whole space $L_2(\R^2)$.

Let us introduce new coordinates in $\cal{G}_n$, $(\lambda _n
,\varphi )$, $\lambda _n=\lambda ^{(n)}(\vec \varkappa )$, $(\cos
\varphi ,\sin \varphi )=\frac{\vec \varkappa }{|\vec \varkappa |}$.
\begin{Lem}
Every point  $\vec \varkappa $ in $\cal{G}_n$ is represented by a
unique pair $(\lambda _n,\varphi )$, $\lambda _n>\lambda _*$,
$\varphi \in [0,2\pi )$, where $\lambda _*=k_*^{2l}$.
\end{Lem}
\begin{proof}
Obviously, to every $\vec \varkappa $ in $\cal{G}_n$, there exists a
pair $(\lambda _n ,\varphi )$ such that $\lambda _n =\lambda
^{(n)}(\vec \varkappa )$ and that $(\cos \varphi ,\sin \varphi
)=\frac{\vec \varkappa }{|\vec \varkappa |}$. For uniqueness,
suppose there are two points $\vec \varkappa _1, \vec \varkappa _2$
corresponding to $(\lambda _n,\varphi )$, i.e., $\lambda ^{(n)}(\vec
\varkappa _1)=\lambda ^{(n)}(\vec \varkappa _2) =\lambda _n$ and
$\frac{\vec \varkappa_1 }{|\vec \varkappa_1 |}=\frac{\vec
\varkappa_2 }{|\vec \varkappa_2 |}=\varphi$. Since both $\vec
\varkappa_1$ and $\vec \varkappa_2$ belong to $\cal{D}_n(\lambda
_n)$ which is parameterized by $\varphi $, $\vec \varkappa _1=\vec
\varkappa _2$.
\end{proof}
For any function $f(\vec \varkappa)$ integrable on $\cal{G}_n$, we
use the new coordinates and write
    \begin{align*}
    \int_{\cal{G}_n} f(\vec \varkappa)d\vec \varkappa &= \int_{\R^2}\chi
    \left(\cal{G}_n,\vec \varkappa\right)f(\vec \varkappa)d\vec \varkappa\\
    &=\int_0^{2\pi} \int_{\lambda _*}^{\infty }
    \chi \left( \cal{G}_n, \vec \varkappa (\lambda_n,\varphi )\right)
    f\left(\vec \varkappa (\lambda_n,\varphi )\right)
    \frac{\varkappa (\lambda_n,\varphi )}{\frac{\partial \lambda_n}{\partial \varkappa }
    }d\lambda_n d\varphi,
    \end{align*}
where    $\chi \left(\cal{G}_n,\vec \varkappa\right)$ is the
characteristic function on $\cal{G}_n$.

Let
\begin{equation} \cal{G}_{n, \lambda}=\{ \vec \varkappa \in
{\cal{G}}_n: \lambda_n(\vec \varkappa) < \lambda\}. \label{d}
\end{equation}
 This set is Lebesgue measurable since ${\cal{G}}_n $ is
open and $\lambda_n(\vec \varkappa)$ is continuous on $
{\cal{G}}_n$.

\begin{Lem}\label{L:abs.6}
$\left|{\cal{G}}_{n,\lambda+\varepsilon} \setminus
{\cal{G}}_{n,\lambda}\right| \leq 2\pi \lambda ^{-(l-1)/l}
\varepsilon $ when $0\leq \varepsilon \leq 1$.
\end{Lem}
\begin{proof}
Considering that
    $ {\cal{G}}_{n,\lambda+\varepsilon} \setminus
{\cal{G}}_{n,\lambda}=\{\vec \varkappa \in {\cal{G}}_n: \lambda \leq
\lambda_n(\vec \varkappa) < \lambda+\epsilon \}, $ we get
    \begin{align*}
    \left|{\cal{G}}_{n,\lambda+\varepsilon} \setminus
    {\cal{G}}_{n,\lambda}\right|
    &=\int_{\cal{G}_n} \chi \left({\cal{G}}_{n,\lambda+\varepsilon} \setminus
{\cal{G}}_{n,\lambda},\vec \varkappa \right) d\vec \varkappa
    &= \int_\lambda^{\lambda+\varepsilon}\int_{\Theta _n(\lambda _n)}
    \frac{\varkappa (\lambda_n, \varphi )}{\frac{\partial \lambda_n}{\partial
    \varkappa }}d\varphi d\lambda_n,
   %%%%%%%%%55 &=C[(\lambda+\epsilon)^{1/l}-\lambda^{1/l}](1+o(1))\\
%%%%%5<C(\lambda ) \epsilon .
    \end{align*}
where $\Theta _n(\lambda _n)\subset [0,2\pi)$ is the set of $\varphi
$
    corresponding to ${\cal D}_n(\lambda _n)$. By perturbation
    formulas (e.g., \eqref{2.67a}, \eqref{3.41}), we have  $\frac{\partial \lambda_n}{\partial
    \varkappa }=2l\varkappa ^{2l-1}(1+o(1))$ and easily
    arrive at the inequality in the lemma.
\end{proof}
By (\ref{s}),
$E_n\left({\cal{G}}_{n,\lambda+\varepsilon}\right)-E_n\left({\cal{G}}_{n,\lambda}\right)
=E_n\left({\cal{G}}_{n,\lambda+\varepsilon}\setminus
    {\cal{G}}_{n,\lambda}\right)$. Let us obtain an estimate for
    this projection.
\begin{Lem}\label{L:abs.7}
For any $F \in C_0^{\infty}(\R^2)$ and $0\leq \varepsilon \leq 1$,
\begin{equation}
\left\|\bigl(E_n({\cal{G}}_{n,\lambda+\varepsilon})-E_n({\cal{G}}_{n,\lambda})\bigr)F\right\|^2_{L_2(\R^2)}
 \leq C( F) \lambda ^{-\frac{l-1}{l}}\epsilon , \label{tootoo1}
 \end{equation}
 where $C(F)$ is uniform with respect to $n$ and $\lambda$.
\end{Lem}
%%%%%%%%\begin{Cor}\label{add1}
%%%%%%$$\left\|\left(E_n({\cal{G}}_{n,\lambda+\epsilon})-E_n({\cal{G}}_{n,\lambda})
%%%\right)F\right\|_{L_2(R^2)}\leq C(F)\lambda ^{\frac{2l}{2l-1}} \epsilon
%%%%%.$$
%%%%%%%%\end{Cor}
\begin{proof}
Considering formula (\ref{s}), we easily see that
    \begin{align*}
    \bigl( \bigl(E_n({\cal{G}}_{n,\lambda+\varepsilon})-E_n({\cal{G}}_{n,\lambda})\bigr)F,
    F \bigr)
   %%%%%%%%%% & \frac{1}{|K_n|}
   %%%%%%%%% \int_{K_n}\sum_{j_n \in P({\cal{G}}_n, \lambda+\epsilon, t_n)
   %%%%\setminus P({\cal{G}}_n, \lambda, t_n)}
   %%%%%%%%% \\bigl(| \bigl( F(\cdot), \psi_{n, j_n}(t_n, \cdot) \bigr) _{L_2(\R^2)}\bigr)|^2 dt_n\\
    =\int_{{\cal{G}}_{n,\lambda+\varepsilon} \setminus {\cal{G}}_{n,\lambda}}
    \left| \bigl( F, \Psi_n(\vec \varkappa ) \bigr) \right|^2
    d\vec \varkappa .
    \end{align*}
Using estimates (\ref{psi1}), (\ref{Dec9c}) for every cell of
periods covering the support  of $F$
    and summarizing over such
    cells, we readily obtain
 $$\left| \bigl( F, \Psi_n(\vec \varkappa )
    \bigr)\right|^2< C(F).$$
 Hence, by Lemma \ref{L:abs.6},
$$ \bigl( \left(E_n({\cal{G}}_{n,\lambda+\varepsilon})-E_n({\cal{G}}_{n,\lambda})\right)F,
    F \bigr) \leq C( F) \left|{\cal{G}}_{n,\lambda+\varepsilon} \setminus
    {\cal{G}}_{n,\lambda}\right|
    \leq C( F) \lambda ^{-\frac{l-1}{l}}\varepsilon .$$
Estimate (\ref{tootoo1}) follows since
    $E_n({\cal{G}}_{n,\lambda+\varepsilon})-E_n({\cal{G}}_{n,\lambda})$
    is a projection.
\end{proof}

\subsection{ Sets ${\cal{G}}_{\infty}$ and ${\cal{G}}_{\infty ,\lambda }$.} \label{S:8.1}

Recall, from (\ref{ac3}) and (\ref{tt1}), that
$$\cal{G}_{\infty}=\bigcup_{\lambda > \lambda_*}\cal D _{\infty}(\lambda) \text{ and }
\cal{G}_{n}=\bigcup _{\lambda >\lambda_*} \cal D_n(\lambda).$$

\begin{Lem} The relation
    \begin{equation}\label{tt}
    \cal{G}_{\infty}=\bigcap_{n=1}^{\infty}{\cal{G}}_n
    \end{equation}
holds and $\cal{G}_{\infty}$ satisfies (\ref{full}) with $\gamma _3=\delta/2 $.
\end{Lem}

\begin{Cor}
The perturbation formulas for $\lambda ^{(n)}(\vec \varkappa )$ and $\Psi _n(\vec
\varkappa )$ hold in $\cal{G}_{\infty}$ for all $n$. Moreover, Coordinates
$(\lambda_n, \varphi )$ can be used in  $\cal{G}_{\infty}$  for every $n$.
\end{Cor}

\begin{proof}
%%%%%%%%%%%%%5Using (\ref{h}), which is proven in Corollary~\ref{Dec18}, with
%%%%%%%5$\varkappa _{\infty }=|\vec \varkappa |$, we easily obtain
%%%%%%%5(\ref{6.2.4}). It remains to prove (\ref{full}).
We start by considering a small region $U_n(\lambda _0)=\bigcup _{|\lambda -\lambda _0|<r_n}
\cal{D}_{n}(\lambda )$, $r_n=\epsilon _{n-1}k^{-2\delta}$,
$k=\lambda _0^{1/2l}$ around the isoenergetic surface $\cal{D}_n(\lambda _0)$
for $\lambda _0>\lambda_*$.
Taking into account that the estimate $\nabla
\lambda ^{(n)}(\vec \varkappa )= 2l|\vec \varkappa |^{2l-2}\vec
\varkappa +o(1)$ holds in the
$\left(\epsilon_{n-1}k^{-2l+1-2\delta}\right)$-neighborhood of
$\cal{D}_n(\lambda _0)$, we conclude that $U_n(\lambda _0)$ is an
open set (a distorted ring with holes), and the width of the ring is
of order $\epsilon_{n-1}k^{-2l+1-2\delta}$. Hence, $|U_n(\lambda
_0)|=2\pi k r_n\left(1+o(k^{-\delta /2})\right)$. It follows easily
from the relations $\cal B_{n+1}\subset \cal B_n$ and (\ref{add*}) that
$U_{n+1}\subset U_n$. The definition of $\cal{D}_{\infty }(\lambda
_0)$ yields $\cal{D}_{\infty }(\lambda _0)=\cap _{n=1}^{\infty
}U_n(\lambda _0)$. Hence,
 $$
\cal{G}_{\infty }\subset \bigcap _{n=1}^{\infty }\cal{G}_n^+,\ \ \
\cal{G}_n^+=\bigcup _{\lambda
>\lambda _*-\delta _n}\cal{D}_n(\lambda ).$$
The set $\cal{G}_n$ differs from $\cal{G}_n^+$ only in the region
near $\cal{D}_n(\lambda _*)$. Since $\lambda _*$ is  not strictly
fixed, this difference is not essential. With a slightly abused
notations, we replace $\bigcup _{\lambda >\lambda _*-\delta
_n}\cal{D}_n(\lambda )$ by $\cal{G}_{n}$. Thus,
$\cal{G}_{\infty}\subset \bigcap _{n=1}\cal{G}_n$. If $\vec
\varkappa \in \bigcap _{n=1}\cal{G}_n$, then $\lambda _n(\vec
\varkappa )$ exists for every $n$ and satisfies (\ref{6.2.3}),
(\ref{6.2.3s}). Hence, $\lambda _n(\vec \varkappa )$ has a limit
$\lambda _{\infty }(\vec \varkappa )\equiv \lambda _0$, i.e.,
$\varkappa \in \cal{D}_{\infty }(\lambda _0)$, $\lambda _0\geq
\lambda _*$. This means $\bigcap_{n=1}^{\infty}{\cal{G}}_n\subset
\cal{G}_{\infty }$. The formula (\ref{tt}) is proved.

Now let us estimate the Lebesgue measure of $\cal G_\infty$.
Since $U_{n+1}\subset U_n$ for every $\lambda
_0>\lambda _*$,
    \begin{equation}
    \cal{G}_{n+1}\subset \cal{G}_n. \label{ac2}
    \end{equation}
Hence $\left|\cal{G} _{\infty}\cap \bf B_R\right|=\lim _{n\to \infty }\left|\cal{G} _{n}\cap
\bf B_R\right|$. Summing the volumes of the regions $U_{n}$, we conclude that
    \begin{equation}
    \left|\cal{G} _{n}\cap
    \bf B_R\right|=|{\bf B_R}|\left(1+O(R^{-\delta /2})\right)\label{+++}
    \end{equation}
uniformly in $n$. Thus, we have obtained (\ref{full}) with $\gamma _3=\delta/2 $.
\end{proof}

Let
     \begin{equation}
     \cal{G}_{\infty, \lambda }=\left\{\vec \varkappa \in
    \cal{G}_{\infty }, \lambda _{\infty }(\vec \varkappa )<\lambda
    \right\}. \label{dd}
    \end{equation}
The function $\lambda _{\infty }(\vec \varkappa )$ is a Lebesgue measurable function
since it is a limit of the sequence of measurable functions.
Hence, the set  $\cal{G}_{\infty, \lambda }$ is measurable.

\begin{Lem}\label{add6}
The measure of the symmetric difference of two sets
$\cal{G}_{\infty, \lambda }$ and $\cal{G}_{n, \lambda}$ converges uniformly in $\lambda$
to zero as $n \to \infty$:
    $$\lim _{n\to \infty }\left|\cal{G}_{\infty, \lambda }\Delta \cal{G}_{n, \lambda
    }\right|=0,$$
where $A\Delta B=(A \setminus B)\cup(B \setminus A)$.
\end{Lem}

\begin{proof}
Using the relation $\cal{G}_{\infty }\subset \cal{G}_{n}$ and
estimate (\ref{June2}), we readily check  that $\cal{G}_{\infty,
\lambda }\subset \cal{G}_{n, \lambda +\delta _n},$ $\delta
_n=24\epsilon _n^4$. Therefore,
    $$\cal{G}_{\infty, \lambda}\setminus \cal{G}_{n, \lambda
    }\subset \cal{G}_{n, \lambda +\delta _n}\setminus \cal{G}_{n, \lambda
    }.$$
Since $\cal{G}_{\infty, \lambda }\supset \cal{G}_{n, \lambda -\delta
    _n}\cap \cal{G}_{\infty }$,
    $$\cal{G}_{n, \lambda}\setminus \cal{G}_{\infty, \lambda }\subset \cal{G}_{n, \lambda
    }\cap \bigl(\cal{G}_{n, \lambda -\delta
    _n}\cap \cal{G}_{\infty }\bigr)^c\subset \left(\cal{G}_{n,
    \lambda
    }\setminus \cal{G}_{n,
    \lambda -\delta _n
    }\right)\cup \left( \cal{G}_{n}\setminus \cal{G}_{\infty }\right).$$
    %%%%%%%%%%%\ \ \ \ \ \ \delta _n=\sup _{\cal{G}_{\infty }}|\lambda _{\infty }(\vec \varkappa
   %%%%%%%%%%% )-\lambda _{n }(\vec \varkappa )|.$$
Combining the two, we get
    $$\cal{G}_{\infty, \lambda}\Delta \cal{G}_{n, \lambda }\subset \left(\cal{G}_{n,
    \lambda +\delta _n
    }\setminus \cal{G}_{n,
    \lambda -\delta _n
    }\right)\cup \left( \cal{G}_{n}\setminus \cal{G}_{\infty
    }\right),$$
hence,
    $$\left| \cal{G}_{\infty, \lambda }\Delta \cal{G}_{n, \lambda
    }\right|\leq \left|\cal{G}_{n,
    \lambda -\delta _n
    }\setminus \cal{G}_{n,
    \lambda +\delta _n
    }\right|+\left|\cal{G}_{n}\setminus \cal{G}_{\infty
    }\right|.$$
Let us consider the first term of the right hand side. Using Lemma
\ref{L:abs.6} with $\varepsilon =2\delta _n$, we obtain
$\left|\cal{G}_{n, \lambda -\delta _n}\setminus \cal{G}_{n, \lambda
+\delta _n }\right|<48\pi\lambda ^{-(l-1)/l}\epsilon _n^4$. By the
definition (\ref{6.1}) of $\epsilon _n $, we conclude easily
    that the first term goes to zero uniformly in $\lambda $. By \eqref{tt} and (\ref{ac2}),
    the second term  goes to zero too.
    \end{proof}

    \subsection{Spectral Projections $E(\cal{G}_{\infty , \lambda })$.}

In this section, we show that spectral projections
$E_n(\cal{G}_{\infty , \lambda })$ have a strong limit $E_{\infty
}(\cal{G}_{\infty , \lambda })$ in $L_2(\R^2)$ as $n$ tends to
infinity. The operator $E_{\infty }(\cal{G}_{\infty , \lambda })$ is
a spectral projection of $H$. It can be represented in the form
$E_{\infty }(\cal{G}_{\infty , \lambda })=S_\infty T_{\infty }$,
where $S_{\infty }$ and $T_{\infty }$ are strong limits of
$S_n(\cal{G}_{\infty , \lambda })$ and $T_n(\cal{G}_{\infty ,
\lambda })$, respectively.  For any $F\in C_0^{\infty }(\R^2)$, we
show
\begin{equation} E_{\infty }\left(
\cal{G}_{\infty , \lambda }\right)F=\frac{1}{4\pi ^2}\int
    _{ \cal{G}_{\infty , \lambda }}\bigl( F,\Psi _{\infty }(\vec
    \varkappa )\bigr) \Psi _{\infty }(\vec
    \varkappa ) d\vec \varkappa ,\label{s1}
    \end{equation}
    \begin{equation} HE_{\infty }\left(
\cal{G}_{\infty , \lambda }\right)F=\frac{1}{4\pi ^2}\int
    _{ \cal{G}_{\infty , \lambda }}\lambda _{\infty }(\vec
    \varkappa )\bigl( F,\Psi _{\infty }(\vec
    \varkappa )\bigr) \Psi _{\infty }(\vec
    \varkappa ) d\vec \varkappa .\label{s1uu}
    \end{equation}
   %%%%%%% $\bigl( F,\Psi _{\infty }(\vec
 %%%%%%%   \varkappa )\bigr) \Psi _{\infty }$ being an integral analogous
 %%%%%%%%   to the dot product in $L_2(R^2)$:
 %%%%%%%%    \begin{equation}\bigl( F,\Psi _{\infty}(\vec
%%%%%%%%     \varkappa )\bigr)=\int _{R^2}F(x)\overline{\Psi _{\infty }(\vec
%%%%%%%%     \varkappa ,x)}dx. \label{eq1}
%%%%%%%%     \end{equation}
Using properties of $E_{\infty }\left( \cal{G}_{\infty , \lambda
}\right)$, we  prove absolute continuity of the branch of the
spectrum corresponding to functions $\Psi _{\infty }(\vec
    \varkappa )$.

Now we consider the sequence of operators $T_n(\cal{G}_{\infty ,
\lambda
    })$ which are given by (\ref{eq2}) with $\cal G_n'=\cal{G}_{\infty , \lambda
    }$ and act from $L_2(\R^2)$ to $L_2(\cal{G}_{\infty , \lambda
    })$. We prove that the sequence has a strong limit and
    describe its properties.

\begin{Lem} \label{Lem1}
The sequence $T_n(\cal{G}_{\infty , \lambda
    })$  has a strong limit $T_{\infty }(\cal{G}_{\infty , \lambda
    })$. The operator $T_{\infty }(\cal{G}_{\infty , \lambda
    })$ satisfies $\|T_{\infty }\|\leq 1$ and can be described by the
    formula
    $T_{\infty }F=\bigl( F,\Psi _{\infty }(\vec
    \varkappa )\bigr) $ for any $F\in C_0^{\infty }(\R^2)$.
    The convergence of $T_n(\cal{G}_{\infty , \lambda
    })F$ to $T_{\infty }(\cal{G}_{\infty , \lambda
    })F$ is uniform in $\lambda $ for every $F\in L_2(\R^2)$.
\end{Lem}
\begin{proof}
Let $F\in C_0^{\infty }(\R^2)$. We consider $T_{\infty }F=\bigl(
F,\Psi _{\infty }(\vec
    \varkappa )\bigr) $. It  follows from (\ref{Dec10*}) and  (\ref{eq2})
     that
     $$\bigl|(T_{\infty }-T_n)F(\vec \varkappa )\bigr|<C(F)g_n(\vec \varkappa ),\ \
     \ g_n(\vec \varkappa )=\varkappa ^{2l}\epsilon_n^{3}|Q_{n+1}|^{1/2}, \ \ \
     \epsilon_n=\exp(-\frac{1}{4}\varkappa ^{\eta
    s_{n}}).$$
It is easy to see that $g_n(\vec \varkappa )\in L_2(\cal G_{\infty
})$ for all $n$ and $g_n(\vec \varkappa )$
    tends to zero in  $L_2(\cal G_{\infty })$ as $n \to \infty$.
    Therefore, $g_n(\vec \varkappa )$ tends to zero in  $L_2(\cal G_{\infty ,\lambda })$ uniformly in $\lambda
    $. Hence, $\bigl\|(T_{\infty }-T_n)F\bigr\|_{L_2(\cal G_{\infty ,\lambda
    })}$ tends to zero  uniformly in $\lambda $
    for every $F\in C_0^{\infty }(\R^2)$ as $n \to \infty$.
Considering $\|T_n\|\leq 1$, we obtain that $T_nF$ has a limit for
every $F\in L_2(\R^2)$ uniformly in $\lambda $. The estimate
$\|T_{\infty }\|\leq 1$ is now obvious.
\end{proof}

Now we consider the sequence of operators $S_n(\cal{G}_{\infty ,
\lambda })$ which are given by (\ref{ev}) with $\cal
G_n'=\cal{G}_{\infty , \lambda
    }$ and act from $L_2(\cal{G}_{\infty , \lambda
    })$ to $L_2(\R^2)$. We prove that the sequence has a strong limit and
    describe its properties.

\begin{Lem} \label{Lem2}
The sequence of operators $S_n(\cal{G}_{\infty , \lambda
    })$ has a strong limit $S_{\infty }(\cal{G}_{\infty , \lambda
    })$. The operator $S_{\infty }(\cal{G}_{\infty , \lambda
    })$ satisfies $\|S_{\infty }\|\leq 1$ and can be described by the
    formula
    \begin{equation}
    (S_{\infty }\varphi) (x)= \int _{\cal{G}_{\infty , \lambda
    }}\varphi (\vec \varkappa)\Psi _{\infty }(\vec
    \varkappa ,x) d\vec \varkappa  \label{ev1}
    \end{equation}
for any $\varphi \in L_{\infty }\left( \cal{G}_{\infty , \lambda
}\right)$. The convergence of $S_n(\cal{G}_{\infty , \lambda
    })\varphi $ to $S_{\infty }(\cal{G}_{\infty , \lambda
    })\varphi $  is uniform in $\lambda $ for every $\varphi \in L_{2}\left(
    \cal{G}_{\infty }\right)$.
\end{Lem}
\begin{proof}
We start by proving that $S_n(\cal{G}_{\infty , \lambda
    })\varphi $ is a Cauchy sequence in $L_2(\R^2)$ for every $\varphi \in L_{\infty}\left(
    \cal{G}_{\infty ,\lambda
    }\right)$. The function $\Psi _n(\vec \varkappa,x)$ is quasiperiodic in $Q_n$ and hence can be
    represented as a  combination of plane waves:
    \begin{equation}\Psi _n(\vec \varkappa,x)=
    \frac{1}{2\pi}\sum _{r\in \Z^2}c_r^{(n)}(\vec \varkappa)
    \exp i\langle \vec \varkappa+\vec p_r(0)/\hat
    N_{n-1},x\rangle,\label{++}
    \end{equation}
where $c_r^{(n)}(\vec \varkappa)$ are Fourier coefficients, $\hat
N_{n-1}=N_{n-1}\cdots N_1\approx 2^{s_n}$ and $\vec
p_r(0)=(\frac{2\pi r_1}{a_1},\frac{2\pi r_2}{a_2})$. The Fourier
transform of $\widehat \Psi_n$ is a combination of $\delta
$-functions
    $$\widehat \Psi _n(\vec \varkappa,\xi)=\sum _{r\in \Z^2}c_r^{(n)}(\vec \varkappa)
    \delta \bigl(\xi +\vec \varkappa+\vec p_r(0)/\hat N_{n-1}\bigr).$$
From this, we compute easily the Fourier
    transform of $S_n\varphi $
    $$ (\widehat{S_n\varphi})(\xi)=\sum _{r\in \Z^2}c_r^{(n)}\bigl(-\xi-
    \vec p_r(0)/\hat N_{n-1}\bigr)\varphi \bigl(-\xi-
    \vec p_r(0)/\hat N_{n-1}\bigr)\chi \bigl(\cal G_{\infty ,\lambda},-\xi-
    \vec p_r(0)/\hat N_{n-1}\bigr),$$
where $\chi (\cal G_{\infty ,\lambda},\cdot ) $ is the
characteristic function on $\cal G_{\infty
    ,\lambda }$. Since $\cal G_{\infty ,\lambda }$ is bounded, the series contains only a
    finite number of non-zero terms for every $\xi $. By Parseval's identity, triangle
    inequality and a parallel shift of the variable,
    \begin{multline*}
    \|S_n\varphi\|_{L_2(\R^2)}=\|\widehat{S_n\varphi}\|_{L_2(\R^2)} \\
    \leq  \sum _{r\in \Z^2}\left \|c_r^{(n)}\bigl(-\xi-
    \vec p_r(0)/\hat N_{n-1}\bigr)\varphi \bigl(-\xi-
    \vec p_r(0)/\hat N_{n-1}\bigr)\chi \bigl(\cal G_{\infty , \lambda},-\xi-
    \vec p_r(0)/\hat N_{n-1}\bigr)\right\|_{L_2(\R^2)}=\\ \sum _{r\in \Z^2}\|c_r^{(n)}\varphi \|_{L_2(\cal G_{\infty ,
    \lambda})}\leq  \|\varphi \|_{L_{\infty}(\cal G_{\infty ,
    \lambda})}\sum _{r\in \Z^2}\|c_r^{(n)}\|_{L_2(\cal G_{\infty ,
    \lambda})} \leq \\
     \|\varphi \|_{L_{\infty}(\cal G_{\infty ,
    \lambda})}\left(\sum _{r\in \Z^2}p_r^{2l}(0)\|c_r^{(n)}\|^2_{L_2(\cal G_{\infty ,
    \lambda})}\right)^{1/2}\left(\sum_{r\in \Z^2} p_r^{-2l}(0)\right)^{1/2}.
    \end{multline*}
%%%%%%%Obviously, each term of the series can be bounded by
 %%%%%%%    $\|c_r^{(n)}\|_{L_{\infty}(\cal G_{\infty ,\lambda})}\|\varphi\|_{L_{\infty
%%%%%%%    }(\cal G_{\infty ,\lambda})}|\cal G_{\infty,\lambda }|^{1/2}.$
By (\ref{++}),  Fourier coefficients $c_r^{(n)}(\vec \varkappa)$ can
be estimated as follows:
    $$\sum_{r \in \Z^2}p_r^{2l}(0)|c_r^{(n)}(\vec \varkappa)|^2
    \leq \|\Psi _n(\vec \varkappa,\cdot )\exp -i\langle \vec \varkappa,\cdot \rangle\|^2_{W^{2l}_2(Q_n)}
    |Q_n|^{-1}\hat N_{n-1}^{2l}<$$ $$2|\vec \varkappa|^{2l}\|\Psi _n(\vec \varkappa,\cdot)\|^2_{W^{2l}_2(Q_n)}
    |Q_n|^{-1}\hat N_{n-1}^{2l}
    .$$
    Integrating the last inequality over $\cal G_{\infty ,
    \lambda}$, we arrive at
    $$\sum_{r \in \Z^2}p_r^{2l}(0)\|c_r^{(n)}\|^2_{L_2(\cal G_{\infty ,
    \lambda})}
    \leq 2|\cal G_{\infty ,\lambda }||Q_n|^{-1}\hat N_{n-1}^{2l}\sup _{\vec \varkappa \in \cal G_{\infty ,\lambda}}|\vec \varkappa|^{2l}
    \|\Psi _n(\vec \varkappa,\cdot )\|^2_{W^{2l}_2(Q_n)}
    $$
    %%%%%%%%%$$\sup _{\vec \varkappa \in \cal G_{\infty ,\lambda}}2|\vec \varkappa|^{2l}\|\Psi _n(\vec \varkappa,\cdot)\|^2_{W^{2l}_2(Q_n)}
%%%%%%%%    |Q_n|^{-1}\hat N_{n-1}^{2l}
%%%%%%%    .$$
    %%%%%%%%Considering also that the sum in $r$ contains no more than
   %%%%%%%% $|G_{\infty ,\lambda }||K_n|^{-1}$ nonzero terms,
Considering that $\sum_r p_r^{-2l}(0)<ck^{2s_1}$,
%%%%%%%55we estimate
%%%%%%%$\sum_{r \in \Z^2}\|c_r^{(n)}\|_{L_{\infty}(\cal G_{\infty
%%%%%%%,\lambda})}$ by H\"{o}lder inequality, and, hence,
we obtain
   %%%%%%%%  estimate:
     $$ \|S_n\varphi\|_{L_2(\R^2)}<ck^{s_1}|\cal G_{\infty ,\lambda }|^{1/2}\|\varphi\|_{L_{\infty
   }(\cal G_{\infty ,\lambda})}
    |Q_{n}|^{-1/2}\hat N_{n-1}^{l}\sup _{\vec \varkappa \in \cal G_{\infty ,\lambda}}|\vec \varkappa|^{l}\|\Psi _n(\vec \varkappa,\cdot)\|_{W^{2l}_2(Q_n)}
  .$$
Similarly,
    \begin{multline*}
     \|(S_{n+1}-S_n)\varphi\|_{L_2(\R^2)}\\
    <ck^{s_1}|\cal G_{\infty ,\lambda }|^{1/2}\|\varphi\|_{L_{\infty }(\cal G_{\infty ,\lambda})}
    |Q_{n+1}|^{-1/2}\hat N_{n}^{l}
    \sup _{\vec \varkappa\in \cal G_{\infty ,\lambda}}
    |\vec \varkappa|^{l}\|\bigl(\Psi _{n+1}(\vec \varkappa, \cdot)-\tilde \Psi _n(\vec \varkappa,\cdot)\bigr)\|_{W^{2l}_2
    (Q_{n+1})}.
    \end{multline*}
Now, using (\ref{s++}) and taking into account that $|\vec
\varkappa|^{2l}<\lambda +o(1)$, $k^{2l}=\lambda $, we obtain
    $$ \|(S_n-S_{n+1})\varphi\|_{L_2(\R^2)}
    \leq c|\cal G_{\infty,\lambda}|^{1/2}\|\varphi\|_{L_{\infty}(\cal G_{\infty ,\lambda })}
    \hat N_{n}^{l}k^{3l+s_1} \epsilon_n^{3}.$$
Considering that $\epsilon_n$ decays super exponentially with $n$
(see (\ref{rrr})) and the estimates $|\cal G_{\infty ,\lambda }|<
\pi \lambda (1+o(1))$, $\hat N_{n}\approx k^{s_n}$, we conclude that
$S_n\varphi $ is a Cauchy sequence in $L_2(\R^2)$ for every $\varphi
\in L_{\infty}\left(\cal{G}_{\infty ,\lambda}\right)$. It is easy to
see that  convergence is uniform in $\lambda $ for every $\varphi
\in L_{\infty}(\cal G_{\infty })$.  We denote the limit of $S_n(\cal
G_{\infty ,\lambda })\varphi $ by $S_{\infty }(\cal G_{\infty
,\lambda })\varphi $.

We see from formula (\ref{ev}) and estimate (\ref{Dec10*}) that
    $$\lim _{n\to \infty}\bigl(S_n(\cal G_{\infty ,\lambda })\varphi \bigr)(x)
    =\int _{\cal G_{\infty ,\lambda}}\varphi (\vec \varkappa)\Psi _{\infty }(\vec \varkappa,x)d\vec \varkappa ,
    $$
for all $x \in \R^2$ when $\varphi \in L_{\infty}(\cal G_{\infty
,\lambda}).$ Hence, (\ref{ev1}) holds.

Since $\|S_n\|\leq 1$, the limit $S_{\infty }(\cal G_{\infty ,\lambda })\varphi $ exists
for all $\varphi \in L_2(\cal G_{\infty ,\lambda })$,
the convergence being uniform in $\lambda $ for every $\varphi  \in L_2(\cal G_{\infty })$.
It is obvious now that $\|S_{\infty }\| \leq 1$.
\end{proof}
\begin{Lem}\label{May8}
Spectral projections $E_n(\cal{G}_{\infty , \lambda })$ have a
strong limit $E_{\infty }(\cal{G}_{\infty , \lambda })$ in
$L_2(\R^2)$, the convergence being uniform in $\lambda $ for every
element. The operator $E_{\infty }(\cal{G}_{\infty , \lambda })$ is
a projection given by the formula (\ref{s1}) for any $F\in
C_0^{\infty }(\R^2)$. The formula (\ref{s1uu}) holds for $HE_{\infty
}(\cal{G}_{\infty , \lambda })$.
\end{Lem}
\begin{proof}
By (\ref{ST}), $E_n=S_nT_n$. Both $S_n$ and
$T_n$ have strong limits $S_{\infty }$, $T_{\infty }$ and
$\|S_n\|\leq 1,$ $\|T_n\|\leq 1$. It follows easily that $E_n$ has
the strong limit $E_{\infty }=S_{\infty }T_{\infty }$. Since $E_n$
is a sequence of projections, its strong limit satisfies the
relations: $E_{\infty }=E_{\infty }^*$, $E_{\infty }^2=E_{\infty }$.
Hence $E_{\infty }$ is a projection \cite{RN}. Using last two lemmas
and considering that $T_{\infty }(\cal{G}_{\infty , \lambda })F_0\in
L_{\infty}(\cal G_{\infty ,\lambda })$ for any $F\in C_0^{\infty }(\R^2)$,
we arrive at (\ref{s1}). Applying  equation (\ref{6.2.1}) for $\Psi
_{\infty }$, we obtain (\ref{s1uu}). It remains to prove that
convergence of $E_nF$ is uniform in $\lambda$ for every $F\in
L_{2}(\R^2)$. First, let $F\in C_0^{\infty }(\R^2) $. By the triangle
inequality,
$$\|(E_{\infty }-E_n)F\|\leq \|(S_{\infty }-S_n)T_{\infty
}F\|+\|S_n(T_{\infty }-T_n)F\|.$$ Since  $T_nF$ converges to
$T_{\infty }F$ uniformly in $\lambda $  and $\|S_n\|\leq 1$, the
second term goes to zero uniformly in $\lambda $. We see easily from
(\ref{s1}) that $T_{\infty }F\in L_{\infty }(\cal G_{\infty})$. Then, by
Lemma \ref{Lem2}, $S_nT_{\infty }F$ converges to $E_{\infty
}(\cal{G}_{\infty , \lambda })F$ uniformly in $\lambda $. This mean
that $E_n(\cal{G}_{\infty , \lambda })F$ converges to $E_{\infty
}(\cal{G}_{\infty , \lambda })F$ uniformly in $\lambda $ for $F\in
C_0^{\infty }(\R^2) $. Using $\|E_n\|=1$, we obtain that uniform
convergence holds for all $F\in L_{2}(\R^2)$.

\end{proof}

\begin{Lem}\label{onemore}
There is a strong limit $E_\infty(\cal{G}_{\infty})$ of the projections
$E(\cal{G}_{\infty,\lambda })$ as $\lambda $ goes to infinity.
\end{Lem}
\begin{Cor}\label{onemore1} The operator $E(\cal{G}_{\infty})$ is a
projection.
\end{Cor}
\begin{proof}
Considering that $\lim _{n  \to \infty }E_n(\cal{G}_{\infty,\lambda})= E_\infty(\cal{G}_{\infty,\lambda })$
and $E_n(\cal{G}_{\infty,\lambda })$ is a monotone in $\lambda $,
we conclude that $E_\infty(\cal{G}_{\infty,\lambda })$ is monotone too.
It is well-known that a monotone sequence of projections has a strong limit.
\end{proof}
\begin{Lem}\label{L:abs.9}
Projections
$E_\infty(\cal{G}_{\infty},\lambda )$, $\lambda \in \R$, and
$E_\infty(\cal{G}_{\infty})$ reduce the operator $H$.
\end{Lem}

%%%%%%%%%%\begin{Cor} \label{May10} The operator $E(G_{\infty})$ is a projection
%%%%%%%%%%reducing $H$.
%%%%%%%%%%\end{Cor}

%%%%%%5To proof the corollary, note that $E_n(\cal{G}_{\infty},\lambda
%%%%%%)=E_n(\cal{G}_{\infty}$, when $\lambda > 2k^{2l}+1$. Hence,
%%%%%%%$E(\cal{G}_{\infty},\lambda )=E(\cal{G}_{\infty}$, when $\lambda $
%%%%%%%%5satisfies the last inequality. Now the corollary immediately follows
%%%%%%%%from the lemma.

\begin{proof}
Let us show $E_\infty(\cal{G}_{\infty},\lambda )$ reduces $H$, i.e.,
$E_\infty(\cal{G}_{\infty,\lambda }) Dom (H)\subset Dom (H)$ and
$E_\infty(\cal{G}_{\infty,\lambda})H=HE_\infty(\cal{G}_{\infty,\lambda })$ on $Dom (H)$
(e.g., see Theorem 40.2 in \cite{AG}).
For any $F,\ G \in Dom(H)=Dom(H^{(n)})$,
    \begin{multline*}
    \bigl( F, E_\infty(\cal{G}_{\infty,\lambda })HG \bigr) =
    \bigl( E_\infty(\cal{G}_{\infty,\lambda })F, HG \bigr) =
    \lim_{n \to \infty} \bigl( E_n(\cal{G}_{\infty,\lambda })F, H_n G\bigr)\\
    =\lim_{n \to \infty} \bigl( H^{(n)} E_n(\cal{G}_{\infty,\lambda })F,  G \bigr)
    =\lim_{n \to \infty} \bigl( E_n(\cal{G}_{\infty,\lambda }) H^{(n)} F,  G \bigr)\\
    =\lim_{n \to \infty} \bigl( H^{(n)} F,  E_n(\cal{G}_{\infty,\lambda }) G \bigr)
    =\bigl( HF, E_\infty(\cal{G}_{\infty,\lambda })G \bigr)
    =\bigl( E_\infty(\cal{G}_{\infty,\lambda })HF, G \bigr)
    \end{multline*}
Hence, $E_\infty(\cal{G}_{\infty,\lambda })H$
is symmetric. Since $E_\infty(\cal{G}_{\infty,\lambda })$ is bounded,
$(E_\infty(\cal{G}_{\infty,\lambda })H)^{*}=HE_\infty(\cal{G}_{\infty,\lambda })$
 (e.g., see $\S$115 in \cite{RN}). Therefore, $E_\infty(\cal{G}_{\infty,\lambda })H
\subset HE_\infty(\cal{G}_{\infty,\lambda })$ which means that for every
$F \in Dom(H)$, $E_\infty(\cal{G}_{\infty,\lambda })F \in Dom(H)$ and
$E_\infty(\cal{G}_{\infty,\lambda })HF=HE_\infty(\cal{G}_{\infty,\lambda })F$.
%Thus, $E_\infty(\cal{G}_{\infty},\lambda )$ reduces $H$.

Now we show that $E_\infty(\cal{G}_{\infty})$ reduces $H$. Noting that
$E_\infty(\cal{G}_{\infty})$ is the strong limit of
$E_\infty(\cal{G}_{\infty,\lambda })$ as $\lambda \to \infty $,  for any
$F,\ G \in Dom(H)$,
    \begin{multline*}
    \bigl( F, E_\infty(\cal{G}_{\infty})HG \bigr) =
     \lim_{\lambda  \to \infty} \bigl( F, E_\infty (\cal{G}_{\infty,\lambda })H G
    \bigr)
    =\lim_{\lambda \to \infty} \bigl( H E_\infty(\cal{G}_{\infty,\lambda })F,  G \bigr) \\
    =\lim_{\lambda \to \infty} \bigl( E_\infty(\cal{G}_{\infty,\lambda }) H F,  G \bigr)
     =\bigl( E_\infty(\cal{G}_{\infty})HF, G \bigr),
    \end{multline*}
 %%%%%%5   $$ \bigl( F, E(\cal{G}_{\infty})HG \bigr) =
  %%%%%%5    \bigl( E(\cal{G}_{\infty })F, HG \bigr) =\lim_{\lambda  \to \infty}
  %%%%\bigl( E (\cal{G}_{\infty,\lambda })F, H G
  %%%%%%5    \bigr)
 %%%%%%5     =\lim_{\lambda \to \infty} \bigl( H E(\cal{G}_{\infty,\lambda })F,  G \bigr) $$
 %%%%%%5     $$=\lim_{\lambda \to \infty} \bigl( E(\cal{G}_{\infty,\lambda }) H F,  G \bigr)
 %%%%%%5     =\lim_{\lambda \to \infty} \bigl( H F,  E(\cal{G}_{\infty,\lambda }) G \bigr)
  %%%%%%5    =\bigl( HF, E(\cal{G}_{\infty})G \bigr) =\bigl( E(\cal{G}_{\infty})HF, G \bigr)
  %%%%%%5    $$
i.e.,  $E_\infty(\cal{G}_{\infty })H$
is symmetric. Considering $(E_\infty(\cal{G}_{\infty})H)^{*}=HE_\infty(\cal{G}_{\infty})$ as before,
we obtain $E_\infty(\cal{G}_{\infty })H\subset HE_\infty(\cal{G}_{\infty })$
which means that for every $F \in
Dom(H)$, $E_\infty(\cal{G}_{\infty })F \in Dom(H)$ and $E_\infty(\cal{G}_{\infty
})HF=HE(\cal{G}_{\infty })F$. Thus, $E_\infty(\cal{G}_{\infty})$ reduces $H$.
\end{proof}

\begin{Lem}
The family of projections $E_\infty(\cal{G}_{\infty},\lambda )$
is the resolution of identity belonging to the operator $HE_\infty(\cal{G}_{\infty})$.
\end{Lem}

\begin{proof}
First, we show that $\lim _{\lambda \to -\infty}E_\infty(\cal{G}_{\infty,\lambda }) =0$.
It is enough to check that
$\cal{G}_{\infty,\lambda }=\emptyset $ for every $\lambda < \lambda
_*$. We see from  the definition (\ref{tt1}) of $\cal{G}_{n}$ and
the definition (\ref{d}) of $\cal{G}_{n,\lambda }$
 that $\cal{G}_{n,\lambda _*}=\emptyset$. It follows from
(\ref{June2}) and (\ref{dd}) that $\cal{G}_{\infty,\lambda _*-\delta
_n}\subset \cal{G}_{n,\lambda _*} $, here $\delta _n =24\epsilon
_n^4$, $n\geq 2$. Hence, $\cal{G}_{\infty,\lambda }=\emptyset $ for
every $\lambda < \lambda _*$.

Second, $\lim _{\lambda \to \infty }E_\infty(\cal{G}_{\infty,\lambda }
)=E_\infty(\cal{G}_{\infty})$ by Lemma \ref{onemore}.

Third, the family $E_\infty(\cal{G}_{\infty,\lambda })$ is left-continuous
since each $E_n(\cal{G}_{\infty,\lambda })$ is left-continuous and
$E_n(\cal{G}_{\infty,\lambda })F$ converges to
$E_\infty(\cal{G}_{\infty,\lambda })F$  uniformly in $\lambda$  for every
$F$ (Lemma \ref{May8}).

Fourth, let $\lambda >\mu $. Then,
    \begin{multline*}
    \bigl( E_\infty(\cal{G}_{\infty,\lambda } )E_\infty(\cal{G}_{\infty,\mu } )F,G\bigr)
    =\bigl(  E_\infty(\cal{G}_{\infty,\mu} )F,E_\infty(\cal{G}_{\infty,\lambda } )G\bigr)
    =\lim_{n \to \infty} \bigl( E_n(\cal{G}_{\infty,\mu} ) F, E_n(\cal{G}_{\infty,\lambda } )G\bigr)\\
    =\lim_{n \to \infty} \bigl( E_n(\cal{G}_{\infty,\lambda } )E_n(\cal{G}_{\infty,\mu} )F,G\bigr)
    =\lim_{n \to \infty} \bigl( E_n(\cal{G}_{\infty,\mu})F,G\bigr)
    =\bigl( E_\infty(\cal{G}_{\infty,\mu} )F,G\bigr) .
    \end{multline*}
This means that $E_\infty(\cal{G}_{\infty,\lambda } )E_\infty(\cal{G}_{\infty,\mu
} )= E_\infty(\cal{G}_{\infty,\mu} )$.

Last, we check that for any $f \in
[E_\infty(\cal{G}_{\infty,\lambda } )-E_\infty(\cal{G}_{\infty,\mu} )]D(H),\
\lambda>\mu$,
    \begin{equation}\label{abs.7}
     \mu \|f\|^2 \leq \bigl( Hf,f \bigr) \leq
    \lambda \|f\|^2.
    \end{equation}
In fact, let
    \begin{equation}
    f=[E_\infty(\cal{G}_{\infty,\lambda } )-E_\infty(\cal{G}_{\infty,\mu}
    )]F, \ \ \ F\in C_0^{\infty }(\R^2).  \label{here}
    \end{equation}
By (\ref{s1}), (\ref{s1uu}),
    \begin{align}
    f(x)&=\frac{1}{4\pi ^2}\int_{ \cal{G}_{\infty , \lambda }\setminus \cal{G}_{\infty , \mu }}
    \bigl( F,\Psi _{\infty }(\vec\varkappa )\bigr)\Psi _{\infty }(x)  d\vec \varkappa ,\notag\\
    Hf(x)&=\frac{1}{4\pi ^2}\int_{ \cal{G}_{\infty , \lambda }\setminus \cal{G}_{\infty , \mu }}
    \lambda _{\infty }(\vec \varkappa )\bigl( F,\Psi _{\infty }(\vec
    \varkappa )\bigr)\Psi _{\infty }(x)  d\vec \varkappa ,\notag\\
    \|f\|_{L_2(R^2)}^2&=\bigl(f,F\bigr)=\frac{1}{4\pi ^2}\int
    _{ \cal{G}_{\infty , \lambda }\setminus \cal{G}_{\infty , \mu }}\left|\bigl( F,\Psi _{\infty }
    (\vec
    \varkappa )\bigr) \right|^2 d\vec \varkappa ,\label{s1*c}\\
    \bigl(Hf,f\bigr)&=\bigl(Hf,F\bigr)=\frac{1}{4\pi ^2}\int
    _{ \cal{G}_{\infty , \lambda }\setminus \cal{G}_{\infty , \mu }}\lambda _{\infty }(\vec \varkappa )
    \left|\bigl( F,\Psi _{\infty }(\vec
    \varkappa )\bigr) \right|^2 d\vec \varkappa .\label{s1*}
    \end{align}
By the definitions of $\cal{G}_{\infty , \mu }$ and $\cal{G}_{\infty
, \lambda }$,  the inequality $\mu \leq \lambda _{\infty }(\vec
\varkappa )<\lambda $ holds, when $\vec \varkappa  \in
\cal{G}_{\infty , \lambda }\setminus \cal{G}_{\infty , \mu }$. Using
the last equality in (\ref{s1*}) and considering (\ref{s1*c}), we
obtain (\ref{abs.7}) for all $f$ given by (\ref{here}). Since
$C_0^{\infty }(\R^2)$ is dense in $Dom(H)$ with respect to
$\|F\|_{L_2(\R^2)}+\|HF\|_{L_2(\R^2)}$ norm, inequality
(\ref{abs.7}) can be extended to all $f=[E_\infty(\cal{G}_{\infty,\lambda }
)-E_\infty(\cal{G}_{\infty,\mu})]F$, $F\in Dom(H)$.

From five properties of $E_\infty(\cal{G}_{\infty,\lambda } )$ proved
above, it follows that $E_\infty(\cal{G}_{\infty,\lambda } )$ is the
resolution of identity belonging to $HE_{\infty}(\cal{G}_{\infty})$ \cite{AG}.
\end{proof}

\subsection{Proof of Absolute Continuity.}

Now we show that the  branch of spectrum (semi-axis) corresponding
to $\cal G_{\infty }$  is absolutely continuous.

\begin{Thm}\label{T:abs}
For any $F\in C_0^{\infty }(\R^2)$ and $0\leq \varepsilon \leq 1$,
    \begin{equation}
    | \bigl((E_\infty(\cal{G}_{\infty,\lambda+\varepsilon})-E_\infty(\cal{G}_{\infty,\lambda}))F,F
    \bigr) | \leq C_F \varepsilon .\label{May10*}
    \end{equation}
\end{Thm}
\begin{Cor}
The spectrum of the operator $HE_\infty(\cal G_{\infty})$ is
absolutely continuous.
\end{Cor}
\begin{proof}
By formula (\ref{s1}),
    $$ | \bigl((E_\infty(\cal{G}_{\infty,\lambda+\varepsilon})-E(\cal{G}_{\infty,\lambda}))F,F
    \bigr) | \leq C_F\left| \cal{G}_{\infty , \lambda +\varepsilon
    }\setminus \cal{G}_{\infty , \lambda } \right| .$$
Applying Lemmas \ref{L:abs.6} and \ref{add6}, we immediately get (\ref{May10*}).

\end{proof}

\end{document}